\documentclass[twocolumn]{emulateapj}

\usepackage{graphicx}
\usepackage{epsfig}

\newcommand{\etal}{et al.}

\newcommand{\atlas}{{\sc Atlas$^{3D}$}}

\def \arcmin{{\em $^\prime$}}
\def \arcsec{{\em $^{\prime\prime}$}}

\def \apjs{{ApJs}}
\def \aap{{AAP}}
\def \apjl{{ApJl}}

\shorttitle{HI in CO-Rich Early-Types}
\shortauthors{Lucero \etal}

\begin{document}


\title{A High Resolution Study of the Atomic Hydrogen in CO-Rich Early-Type Galaxies}
\author{D. M. Lucero\altaffilmark{1,2} and L. M. Young\altaffilmark{1}}
\altaffiltext{1}{Physics Department, New Mexico Institute of Mining and Technology,
    Socorro, NM 87801, USA}
\altaffiltext{2}{Department of Astronomy, University of Cape Town,
    Rondebosch 7701, Republic of South Africa}
\begin{abstract}
We present an analysis of new and archival VLA HI observations of a sample of eleven early-type galaxies rich in CO, with detailed comparisons of CO and HI distributions and kinematics.  The early-type sample consists of both lenticular and elliptical galaxies in a variety of environments. A range of morphologies and environments were selected in order to give a broader understanding of the origins, distribution, and fate of the cold gas in early-type galaxies.  Six of the eleven galaxies in the sample are detected in both HI and CO.  The H$_{2}$ to HI mass ratios for this sample range from 0.2-120.  The HI morphologies of the sample are consistent with that of recent HI surveys of early-type galaxies which also find a mix of HI morphologies and masses, low HI peak surface densities, and a lack of HI in early-type galaxies which reside in high density environments. The HI-detected galaxies have a wide range of HI masses (1.4$\times$10$^{6}$ to 1.1$\times$10$^{10}$ M$_{\odot}$). There does not appear to be any correlation between the HI mass and morphology (E versus S0).  When HI is detected, it is centrally peaked - there are no central kpc-scale central HI depressions like those observed for early-type spiral galaxies at similar spatial resolutions and scales. A kinematic comparison between the HI and CO indicates that both cold gas components share the same origin.  The primary goal of this and a series of future papers is to better understand the relationship between the atomic and molecular gas in early-type galaxies, and to compare the observed relationships with those of spiral galaxies where this relationship has been studied in depth.  
\end{abstract}
\section{Introduction}
It is well known that many early-type galaxies (E and S0) contain significant amounts of cold gas (Wiklind \etal\ 1995; Young 2002; Welch \& Sage 2003; Sage \& Welch 2006; Morganti \etal\ 2006; Sage \etal\ 2007; Combes \etal\ 2007; Osterloo \etal\ 2007; di Serego Alighieri \etal\ 2007; Welch \etal\ 2010; Young \etal\ 2011). The evolutionary pathway of early-type galaxies is thought to be driven in part by the acquisition and transformation of this cold gas into new stars. To the best of our knowledge stars form only from the molecular gas phase.  Clearly, knowledge of the amount of cold gas in the molecular phase versus the atomic phase is an important constraint for theoretical models of star formation and galaxy evolution. 

Currently, the physical processes which determine the balance of the atomic and molecular gas are poorly understood in early-type galaxies.  The question of molecule formation has been thoroughly studied in the literature mostly through two different methods.  The first is an empirical study using the CO and HI maps of nearby disk galaxies (Blitz \& Rosolowsky 2004, 2006; Leroy \etal\ 2008).  These studies infer that the molecular to atomic surface density ratio in disks is entirely a function of the hydrostatic midplane pressure which is in turn a function of the stellar and gas volume densities (hereafter BR06 model).  The second method also utilizes gas maps and a first principles approach which models the chemical and physical processes that regulate the balance between the formation and dissociation of H$_{2}$ molecules (Elmegreen 1993; Krumholz \etal\ 2008, 2009; McKee \& Krumholz 2010).  Interestingly, both of these approaches predict values of the molecular fraction, F$_{mol}=\Sigma_{H_{2}}/\Sigma_{g}$, with $\Sigma_{g}=$$\Sigma_{H_{2}}+\Sigma_{HI}$, which are roughly consistent with molecular and atomic observations in nearby disk galaxies (Krumholz \etal\ 2009).  Fumagalli \etal\ (2010) suggest that the two approaches yield similar values of the molecular fraction, because Blitz \& Rosolowsky (2006) use a fixed stellar density typical of nearby disk galaxies and in effect both formalisms become a sole function of the total gas column density. 

Testing whether photo-dissociation and star formation laws (empirical and theoretical) hold in early-type galaxies requires detailed information about the spatial distribution, column density, and kinematics of both the atomic and molecular gas components.  There has been a recent explosion in the number of available HI and CO maps of early-type galaxies (e.g. Serra \etal\ 2012; di Serego \etal\ 2007; Oosterloo \etal\ 2007; Alatalo \etal\ 2012).  However, most of these HI maps have resolutions $\geq$45\arcsec and so are not at high enough resolution to make an adequate comparison with existing CO interferometric observations which typically have resolutions $\leq$7\arcsec.  In this paper we present the first comparison of high-resolution HI and CO observations in a sample of eleven early-type galaxies.  The new HI observations have 2 to 16 times better resolution than the current HI surveys.  

Higher resolution observations of the HI can also help us to better constrain the origins of the cold gas.  There are two basic models for the origin of the cold gas in early-type galaxies: internal and external.  The internal origin means mass loss from the stars in the galaxy itself.  Estimates of the stellar mass loss rate are certainly more than enough to reproduce the cold gas contents in early-types over a Hubble time (Faber \& Gallagher 1976; Ciotti \etal\ 1991; Brighenti \& Mathews 1997; Athey \etal\ 2002).   However, recent observational studies of the cold gas in nearby early-type galaxies show that on average the total observed gas masses add up to only about 10$\%$ of what is expected from mass loss after the first 0.5 Gyr regardless of luminosity (e.g. Sage \& Welch 2006; Welch \etal\ 2010).  It has been speculated that most of the returned stellar mass is either quickly depleted in star formation, used to fuel a central AGN, stripped, or heated into an ionized phase.  Possible external origins include gas accretion from mergers or accretion of primordial gas from the intergalactic medium (IGM).   

A large amount of evidence is accumulating which suggests that the bulk of the cold gas in local early-type galaxies is likely supplied via external processes (e.g.\ Davis \etal\ 2011; Morganti \etal\ 2006; Serra \etal\ 2012; Emsellem \etal\ 2004; McDermid 2006a).  Arguments for an external origin are supported by the fact that the cold gas mass in early-type galaxies only weakly correlate with other optical properties such as stellar luminosity and stellar velocity dispersion (e.g.\ Davis \etal\ 2011), the existence of tidal features in HI maps (Morganti \etal\ 2006; Serra \etal\ 2012), kinematic mismatch between cold gas and stars (Davis \etal\ 2011), and the existence of kinematically decoupled stellar cores (Emsellem \etal\ 2004; McDermid \etal\ 2006a).  Some early-type galaxies do sometimes contain stellar discs which correlate with nuclear and global galaxy properties suggesting that the gas which formed these stars was produced via internal mass loss (Davis \etal\ 2011).  However, in these cases it may be that galaxy wide processes in the most massive early-types or the early-types host environment are effectively destroying any signatures of an external origin (Davis \etal\ 2011).  

Recent single dish observations of HI and CO in a volume limited (dec $\leq$10$^o$, and distances $<$20 Mpc) sample of early-type galaxies indicate that the atomic and molecular cold gas phases themselves could have separate origins (Welch \& Sage 2003; Sage \& Welch 2006; Sage \etal\ 2007; Welch \etal\ 2010). The key piece of evidence which support their hypothesis is that the HI and CO kinematics (line profiles) for many of their lenticular galaxies do not match. Sage \& Welch (2006) speculate that the two cold gas phases in early-type galaxies are separate, because the molecular gas originates from stellar mass loss while the atomic gas has been acquired from an outside source (primordial or through mergers).  This single dish survey compares HI data with angular resolutions $\geq$45\arcsec to CO data with angular resolutions of 55\arcsec.  A more recent \atlas\ volume limited multiwavelength survey of 260 early-type galaxies compares 45\arcsec resolution HI data to CO data with resolutions ranging from 3\arcsec to 10\arcsec.  Their data show that the ionized, atomic,  and molecular gas in local early-type galaxies always have similar kinematics and so likely have a common origin (Davis \etal\ 2011; Alatalo \etal\ 2012). Resolving the discrepancy between these two studies requires closely matched high resolution HI and CO data.

This is the first in a series of papers that aim to: (1) Use the cold gas kinematics and gas morphologies of our sample to look for indications of separate origins for the HI and CO in early-type galaxies.  (2) Test theoretical models of star formation known for spiral galaxies.  Is there evidence of star formation where star formation models predict it to occur? (3) Test whether the theoretical H$_{2}$ molecule formation models of Elmegreen (1993), Krumholz \etal\ (2008, 2009), McKee \& Krumholz (2010) and the empirical models of Blitz \& Rosolowsky (2004) can predict the observed molecular fraction in early-type galaxies. 

In Paper 1 we present the results of the new HI observations.  Paper I is organized as follows:  In Section \ref{sampsel} the sample description is presented.  In Section \ref{obsre} we present the 21-cm HI observations and data reduction.  In sections \ref{res1} and \ref{res2} we present the results of the observations, and the HI fluxes and masses.  In section \ref{morph} we compare the HI and CO morphologies.  In section \ref{kin} we present the HI kinematics and the dynamical masses.  In section \ref{disc} we discuss the role of the environment on the HI morphology and the origin of the cold gas in early-type galaxies.  In section \ref{concl} we present the conclusions.
\section{Selection and Galaxy Properties \label{sampsel}}\label{sampsel}
\begin{deluxetable*}{lccccccc} 
\tabletypesize{\scriptsize}
\tablecaption{\bfseries Sample Galaxies \label{hih2samp}}
\tablewidth{0pt}
\tablehead{
\colhead{Galaxy}    &  \colhead{Type}  &  \colhead{RA}         &  \colhead{DEC}        &  \colhead{V$_{helio}$}    &  \colhead{D}      &  \colhead{M$_K$}  &  \colhead{Environ}\\
\colhead{}          &  \colhead{}      &  \colhead{(J2000.0)}  &  \colhead{(J2000.0)}  &  \colhead{(km s$^{-1}$)} &  \colhead{(Mpc)}  &  \colhead{(mag)}  &  \colhead{}}
\startdata
N83              &E0       &00h21m22.4s    &22d26m01s    &6359 (27)   &85.1(6.1)    &$-$25.35      &Group\\
U1503            &E        &02h01m19.8s    &33d19m46s    &5086(6)     &71(5)        &$-$23.69      &field\\
N807             &E        &02h04m55.7s    &28d59m15    &4764(12)    &66(5)        &$-$24.74      &field\\
N2320            &E        &07h05m42.0s    &50d34m52s    &5944 (15)   &84(7)        &$-$25.77      &Abell 569\\
N3032            &S0       &09h52m08.2s    &29d14m10s    &1533(5)     &21.2(1.9)    &$-$21.98      &field\\
N3656            &IOpec    &11h23m38.4s    &53d50m31s    &2869(13)    &40(3)        &$-$23.74      &merg remn\\
N4150            &S0       &12h10m33.6s    &30d24m06s    &226(22)     &13.7(1.6)    &$-$21.70   &Coma Group\\
N4476            &SO$^{-}$ &12h29m59.1s    &12d20m55s    &1978(12)    &17.6(0.6)    &$-$21.76      &Virgo clust\\
N4526            &S0       &12h34m03.0s    &07d41m57s    &575(24)     &17.3(1.5)    &$-$24.71      &Virgo clust\\
N4459            &S0$^{+}$ &12h29m00s      &13d58m53s    &1210(16)    &16.1(0.4)    &$-$23.88      &Virgo clust\\
N5666            &?        &14h33m09.2s    &10d30m39s    &2221(6)     &31(2)        &$-$22.28     &field
\enddata 
\tablecomments{
Environments and classifications are taken from (1)Young 2002, (2) Young \etal\ 2008\\
and (3) RC3.  Distances are taken from (4) Young \etal\ 2009,\\ 
(5) NASA's Extragalactic Database (NED) and (6) Lyon Extragalactic Database (LEDA).\\  
Velocities are taken exclusively from NED; they refer to HI where available or to stellar\\
velocities.  K-Band magnitudes are taken from integrated magnitudes in the 2MASS catalog (Skrutskie \etal\ 2006).}
\end{deluxetable*}
 Testing whether photo-dissociation and star formation laws (empirical and theoretical) hold in early-type galaxies requires detailed information about the spatial distribution, column density, and kinematics of both the atomic and molecular gas components.  Recently it has been found that the most CO rich early-type galaxies are more likely to have high column density HI (Serra \etal\ 2012).  Therefore, in order to maximize our chances of detecting and resolving HI we begin this preliminary study using a sample of early-types already known to be CO-rich.  Our galaxy sample includes eleven of the fourteen galaxies recently mapped in $^{12}$CO 1$-$0 line with the Berkley-Illinois-Maryland Association (BIMA) millimeter interferometer at Hat Creek, CA by Young (2002, 2005) and Young \etal\ (2008).  The molecular gas in these eleven galaxies is located in very regular, symmetric rotating disks with diameters ranging from 0.7 to 12 kpc and they have H$_2$ masses in the range of 1.0$\times$10$^8$- 4.7$\times$10$^9$ M$_\odot$ (Young 2002, 2005; Young \etal\ 2008).

All of the galaxies in our sample have an r$^{1/4}$ profile or a classification as E, E/SO or S0 in several catalogs.  In some cases, these classifications are based on photographic evidence. In the case of NGC 5666, CCD imaging suggests that one of our sample galaxies may have a more complicated morphology (Donzelli \& Davoust 2003). 

Additionally, all of our early-type sample galaxies have estimates on the amount and location of star formation derived from high resolution radio continuum (5\arcsec), 24$\mu$ FIR, and UV maps which make them ideal objects for this preliminary study (Lucero \& Young 2007; Young \etal\ 2009).

The sample galaxies have a wide variety of properties and environments.  About half the galaxies are isolated, and the other half are located in loose groups or clusters (Virgo and Abell 569).  Some have very smooth, regular morphologies, whereas others are classified as ``peculiar'' because of dust lanes, stellar shells and ripples, etc. The targets' distances range up to 80 Mpc and optical luminosites are in the range $-$25.80$\leq$M$_K$$\leq$$-$21.7.  Thus, a study of these objects should give us a broad understanding of how the HI and CO are distributed in early-type galaxies.  

The sample galaxies discussed in this paper have a similar range in K-band stellar luminosity and M(HI)/L$_K$ as those in the \atlas\ survey.  However, our sample is heavily biased towards objects with H$_2$ masses above 10$^8$ M$_\odot$.  Additionally about half of or our galaxy sample reside at distances greater than 40 Mpc with a few a factor of two more distant.  Table \ref{hih2samp} contains ephemeris information for the galaxy sample.

One potential danger in making comparisons from single dish surveys is the large discrepancy of the fields of view between instruments used to observe HI and CO (e.g. 20\arcsec\ for CO from the IRAM 30 meter, versus 3\arcmin\ for HI from the Arecibo 300 meter).  However, for early-type galaxies at the distances of our sample, the evidence suggests there is little CO flux missing beyond the available fields of view.  For example, Welch \& Sage (2003) made multiple pointings on early-type galaxies with the 30m telescope but rarely found CO emission outside the central kpc.  Consistent with that result, a recent comparison of the CO fluxes from the IRAM 30 meter and the Combined Array for Millimeter Wave Astronomy (CARMA) for a subsample of the \atlas\ early-type galaxies shows that the CO is often slightly more extended than the IRAM 30 meter primary beam (Alatalo \etal\ 2012), but a single pointing with the 30m usually recovers at least 50\% of the total flux.  Interferometric surveys carried out with the Berkley Maryland Illinois Array (BIMA) and CARMA find that the CO in early-type galaxies rarely extends outside their respective primary beams (e.g. Young 2002; Young \etal\ 2005; Alatalo \etal\ 2012). \\
\section{Observations and Data Reduction}\label{obsre} 
\begin{deluxetable*}{lllccccccccc}[t!]
\tabletypesize{\scriptsize}
\tablecaption{\bfseries Observation and Image Parameters \label{HIparam}}
\tablehead{
\colhead{Galaxy}  & \colhead{Config}   & \colhead{Obs}   & \colhead{Program}   & \colhead{Flux}  & \colhead{Velocity}  & \colhead{}        & \colhead{}       & \colhead{Linear}      & \colhead{}         & \colhead{}        
                    & \colhead{}\\
\colhead{}          & \colhead{}         &\colhead{Dates}  & \colhead{ID}        & \colhead{Cal}   & \colhead{Range}     & \colhead{TOS$^a$}  & \colhead{Beam}  & \colhead{Resolution}  & \colhead{Channel}  & \colhead{Noise}    
                    &  \colhead{N$_{HI}$ Limit}\\
\colhead{}          & \colhead{}        &  \colhead{}     & \colhead{}          & \colhead{}      & \colhead{(km s$^{-1}$)}  & \colhead{(hrs)} & \colhead{(\arcsec)}   & \colhead{(kpc)}    & \colhead{(km s$^{-1}$)}   &  \colhead{(mJy beam$^{-1}$)} 
                    &\colhead{$(10^{19}$ cm$^{-2}$)}}
\startdata 
N83     &C      &2005 Aug    &AY159   &0137+331   &5900$-$6700    &6.6       &19.2$\times$17.3  &7.9$\times$7.1   &21                &0.3            &13\\
U1503   &C      &2002 Dec    &AY135   &0137+331   &4800$-$5400    &8.4       &13.8$\times$13.0  &4.8$\times$4.5   &21                &0.3            &23\\
NGC807  &D$^b$      &1985 Dec    &AD174   &3C48     &4100$-$5400      &16        &48.5$\times$44.2  &15.5$\times$14.1 &42  &0.5 &13\\
        &C      &2002 Jul    &AY135   &0137+331   &4100$-$5400    &9.0       &13.8$\times$13.1  &5.9$\times$5.5   &21             &0.2            &15\\
        &C\&D$^*$   &$-$    &$-$   &$-$   &4100$-$5400    &25        &32.9$\times$31.5  &10.5$\times$10.1 &42                &0.2            &5.3\\
N2320   &C      &2005 Sep    &AY159   &0137+331   &5200$-$6600    &7.5       &16.9$\times$15.8  &6.9$\times$6.4   &21                &0.3       &16\\
N3032$^c$  &C     &1992 Feb  &AL263   &0137+331 &1228$-$1873      &9.3       &18.9$\times$16.5  &1.9$\times$1.7 &10  &0.6 &13\\
N3656$^d$       &BCD   &1996/7 &AB819/791  &1328+307 &2365$-$3521 &28        &7.4$\times$7.2    &1.4$\times$1.4 &21  &0.16                       &42\\ 
N4476      &D  &2002 Jan    &AY128   &1331+305 &1637$-$2284  &4     &46.2$\times$42.6  &3.9$\times$3.5 &10.4  &1.0                                    &3.4\\ 
N4526       &D  &2005 Dec    &AY161   &1331+305   &0$-$1300       &1.1       &64.0$\times$50.5  &5.4$\times$4.2   &21              &0.7                    &3.0\\
N4459   &D      &2005 Dec    &AY161   &1331+305   &600$-$1800     &1.1       &45.0$\times$43.3  &3.5$\times$3.4   &21              &1.0                    &7.1\\
N4150   &D      &2005 Dec    &AY161   &1331+305   &$-$100$-$600   &1.1       &54.5$\times$52.0  &3.6$\times$3.5   &21              &1.0               &4.9\\
N5666$^e$  &AD    &1986 Feb  &AL111   &3C286    &1906$-$2533 &4.8   &55.0$\times$44.7  &8.3$\times$6.7 &21  &0.4                                           &2.3\\
           &C     &1986 Dec  &AL111   &3C286    &1906$-$2533 &13.3  &16.0$\times$14.5  &2.4$\times$2.2 &21  &0.5                                                     &36\\
           &C\&AD &$-$       &$-$     &$-$      &1906$-$2533 &18.1  &35.9$\times$31.7  &5.4$\times$4.8 &21  &0.7                                                               &6.1
\enddata
\tablecomments{
$^*$Combined C and D VLA configuration data.  $^{\star}$A array transition into D array.\\
$^a$TOS is the total time on source galaxy in hours.\\ 
HI column density limits in this table are derived from a conservative 6$\sigma$ of the rms noise in one channel.\\
{\bf References for archival data:} \\
$^b$Dressel 1987. \\
$^c$Observed by Ernest Sequist but not published.\\
$^d$Balcells \etal\ 2001\\
$^e$Lake \etal\ 1987}
\end{deluxetable*}
In order to test the current photo-dissociation and star formation models we require HI column density maps of the eleven early-type galaxies at angular resolutions comparable to what has already been obtained in CO.  These column density maps will allow us to calculate the surface density ratio $\Sigma_{H_2}/\Sigma_{HI}$  as a function of radius.  Therefore, new neutral hydrogen Very Large Array (VLA) observations were obtained for eight galaxies (NGC 83, UGC 1503, NGC 807, NGC 2320, NGC 3032, NGC 4526, NGC 4459, and NGC 4150).  Five of these galaxies (NGC 83, UGC 1503, NGC 807, NGC 2320, and NGC 3032) were observed in the C configuration (projected baselines 0-15 k$\lambda$) between 1992 February and 2005 August. The VLA's C configuration and its 15\arcsec\ beam is the best compromise between sensitivity and resolution for the purpose of comparisons between the atomic and molecular gas in our sample galaxies, and is a factor of three better than recent HI surveys with the WSRT (e.g. Serra \etal\ 2012).  NGC 4150, NGC 4459, and NGC 4526 were observed in the VLA D configuration (projected baselines 0-5 k$\lambda$) in 2005 December.  These three galaxies were observed in the lower resolution D array because they were observed with single dish telescopes (NGC 4150: Arecibo 305 meter, NGC 4459 and NGC 4526: Effelsberg 100 meter; Huchtmeier \& Richter 1986), but were not detected. The new VLA D configuration observations for these three galaxies are about a factor of 10 deeper than what was done at the single dish.  Recent Westerbork Synthesis Radio Telescope (WSRT) HI observations for NGC 4150 produce a clear detection (Morganti \etal\ 2006).  The WSRT data for NGC 4150 have a  similar resolution to that of our new D array data (70\arcsec$\times$35\arcsec), velocity resolution (16 km s$^{-1}$), and an rms noise of 0.6 mJy beam$^{-1}$.  VLA D configuration HI data for NGC 4476 were previously published by Lucero \etal\ (2005) and resulted in a non-detection.  We obtained and re-reduced this data using similar techniques described in  section 3.1 of Lucero \etal\ 2005 and were able to achieve a slightly better rms noise of 1 mJy beam$^{-1}$ in a primary beam corrected cube.  We derive a new upper limit to the HI mass of $<$6.7$\times10^{6}$ M$_{\odot}$. 
  
High-resolution ($\leq$ 15\arcsec) VLA HI maps already exist for NGC 3656 and NGC 5666. We have obtained a high resolution HI image of NGC 3656 from Jacqueline van Gorkom which is originally published by Balcells \etal\ (2001).  This image was made by combining VLA D, C, and B array data and yielded an HI mass of 2.0$\times10^{9}M_{\odot}$.  High resolution HI data for NGC 4476 were previously published by Lucero \etal\ (2005) and resulted in a non-detection and an upper limit to the HI mass of $<$6.7$\times10^{6}$ M$_{\odot}$.  We have re-reduced this data using similar techniques outlined in section 3.1 of Lucero \etal\ (2005) and were able to achieve a slightly lower rms noise of 1 mJy beam$^{-1}$ in a primary beam corrected cube.  Four hours of VLA C array data for NGC 5666 were published by Lake \etal\ (1987), but no HI flux or mass is quoted in that paper.  Additional time on NGC 5666 was obtained ($\sim$9 hrs in C array) but never published (Jacqueline van Gorkom in private communication).  Analysis of the full 13 hours of VLA C array data for NGC 5666 is presented in this paper.   

Table \ref{HIparam} gives specific dates, configurations, and time on-source for archive data as well as the new HI data.  All data calibration and image formation was done using standard calibration tasks in the Astronomical Image Processing System (AIPS) package (Greisen 2003).  Each galaxy was observed in one pointing centered roughly on the optical center of the galaxy. Phase drifts as a function of time are corrected by means of nearby point sources observed every 30 to 45 minutes. The absolute flux scale was set by observations of the sources 0137+331 or 1331+305 (whichever was closer to the galaxy in question).  These two sources are also used to correct variations in the gain as a function of frequency (bandpass calibration).  Comparisons of flux measurements on all the observed calibrators suggest that the absolute flux uncertainties are on the order of 10\%.  Variations in the bandpass are on the order of 1\%.  Initial imaging revealed which channel ranges were free of HI line emission. Continuum emission was subtracted directly from the raw uv data by making first-order fits to the line-free channels using the AIPS task UVLIN.  For the cases where no line emission is visible in any of the channels, the line width of the CO is used to determine which channels should be free of HI emission. The calibrated data are Fourier-transformed using several different uv data weighting schemes chosen to enhance the spatial resolution or the sensitivity to large-scale structures. Dirty images were cleaned down to a residual level of 1.0 times the rms noise fluctuations, and then primary beam corrected. Table \ref{HIparam} shows the velocity range covered and velocity resolution for each galaxy as well as the linear resolution (FWHM of the synthesized beam) and rms noise level in the final image cubes.  Integrated intensity and mean velocity maps are made using the AIPS task MOMNT, as follows.  The cleaned data cubes are smoothed in the velocity and spacial domains using Hanning and Gaussian functions, respectively.  The width of the Hanning function used was typically 3 channels.  The full width at half maximum (FWHM) of the Gaussian convolving kernal was typically 11 pixels.  Smoothed cubes are then clipped at a level of 1$\sigma$ and 2$\sigma$, respectively, defining a mask which is applied to the original cleaned cubes before integration over velocity.

The data reduction is straight forward in all cases except for the continuum subtraction for NGC 4459.  The D array data for NGC 4459 are dominated by errors due to the bright continuum source associated with M87 ($\sim1.6^o$ to the south of the phase center).  The following methods were used to try and remove these errors.  First we fit and subtracted the radio continuum using UVLIN on a cube who's phase center has been shifted to the position of M87.  This produced a cube with an rms of $\sim$ 0.7 mJy beam$^{-1}$, but with unstable baselines.  The task UVLIN is known to leave residuals at the position of strong continuum sources resulting in unstable baselines.  In order to minimize the residuals we next tried using the AIPS task UVSUB followed by UVLIN again shifting the phase center to the position on M87.  UVSUB subtracts from the visibility data a Fourier sum of the clean components estimated by deconvolution of the continuum image. The use of UVSUB produced no improvement.  We next subtracted the continuum in the image plane using the AIPS task IMLIN. IMLIN fits a low order polynomial to the line free part of the spectrum of each pixel in an image cube (Cornwell et al. 1992).  In this case the continuum model fit is derived from the line free channels using a first order polynomial.  Both UVLIN and IMLIN have similar problems dealing with point sources that are far from the phase center.  However, IMLIN may produce better results than UVLIN, because its errors scale with distance from the bright continuum source rather than from the phase center as they do when fitting with UVLIN.  Indeed IMLIN was much more successful at removing the sidelobes of M87, but produced an rms noise 1.4 times larger than that of the previous two methods ($\sim$1 mJy beam$^{-1}$).  HI upper limits in the following sections are derived from the IMLIN'd cube.

Lower resolution data for NGC 807 and NGC 5666 are reduced and combined with our new higher resolution data in order to maximize the sensitivity for the purpose of calculating as accurate a flux measurement as possible.  The lower resolution data for NGC 807 and NGC 5666 consists of 16 hours in the VLA D array (Dressel 1987), and 5 hours in the VLA AD array (Lake \etal\ 1987), respectively.  These two data sets were obtained from the VLA archive and reduced in a similar manner as described above through to continuum subtraction.  For NGC 807 the continuum-subtracted C array data is smoothed to the velocity resolution of the lower resolution data using the VLA task SPECR.  The lower resolution data is then shifted in frequency space using the AIPS task CVEL, which corrects topocentric to heliocentric velocities and resamples so that channel frequencies in the lower resolution data set correspond to the same channels as in the higher resolution data set.  Next the high and low resolution frequency corrected data sets are combined together using the AIPS task DBCON and then imaged using similar methods as those described above.  The two data sets are combined such that they contribute equally to the final cleaned image.  
\section{Results}
\subsection{HI Non-Detections}\label{res1}
\begin{figure}
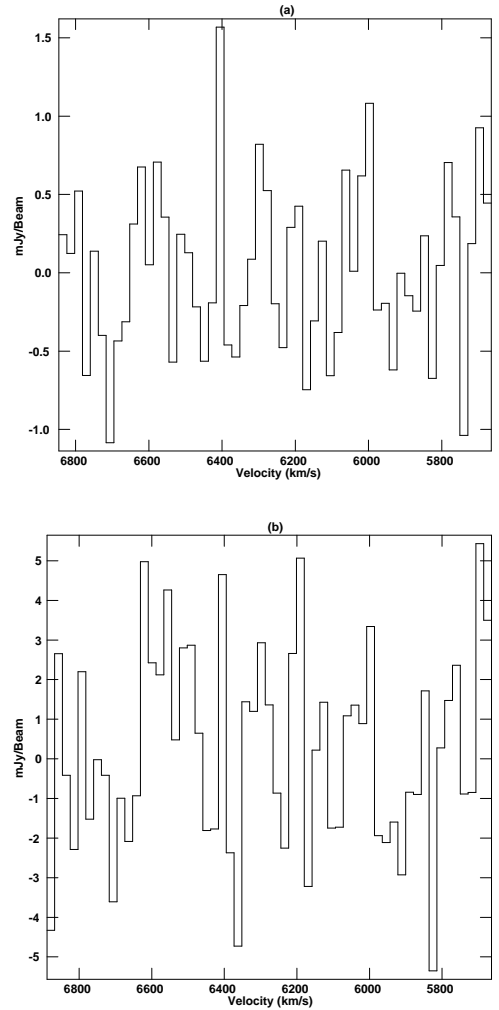

\begin{center}
\includegraphics*[scale=0.35]{f1a.eps}
\includegraphics*[scale=0.35]{f1b.eps}
\caption{\scriptsize NGC 83: (a) HI spectrum of a square region, 5\arcsec\ on a side, centered on the optical center of NGC 83. The spectrum was constructed by first using the CO image to define a rectangular mask region within which the CO emission is located.  The intensity was then integrated over the same spatial region for every channel, so the noise in the line free regions should be indicative of the noise on the line as well.  (b) HI spectrum of a square region, 1.5\arcmin\ on a side, centered on the optical center of NGC 83.  Integrated intensities quoted in the text are obtained from panel (a).  \label{n83ispec}}
\end{center}
\end{figure}
\begin{figure}
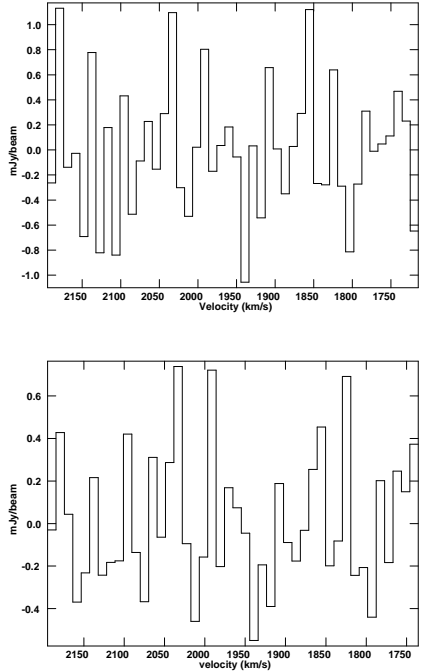

\begin{center}
\includegraphics*[scale=0.3]{f3t.eps}
\includegraphics*[scale=0.3]{f3b.eps}
\caption{\scriptsize NGC 4476: As in Figure \ref{n83ispec} but using box sizes of (t) 37\arcsec\ on a side, and (b)  1.1\arcmin\ on a side, centered on the optical center of NGC 4476.  \label{n4476ispec}}   
\end{center}
\end{figure}
\begin{figure}
\begin{center}
\includegraphics*[scale=0.35]{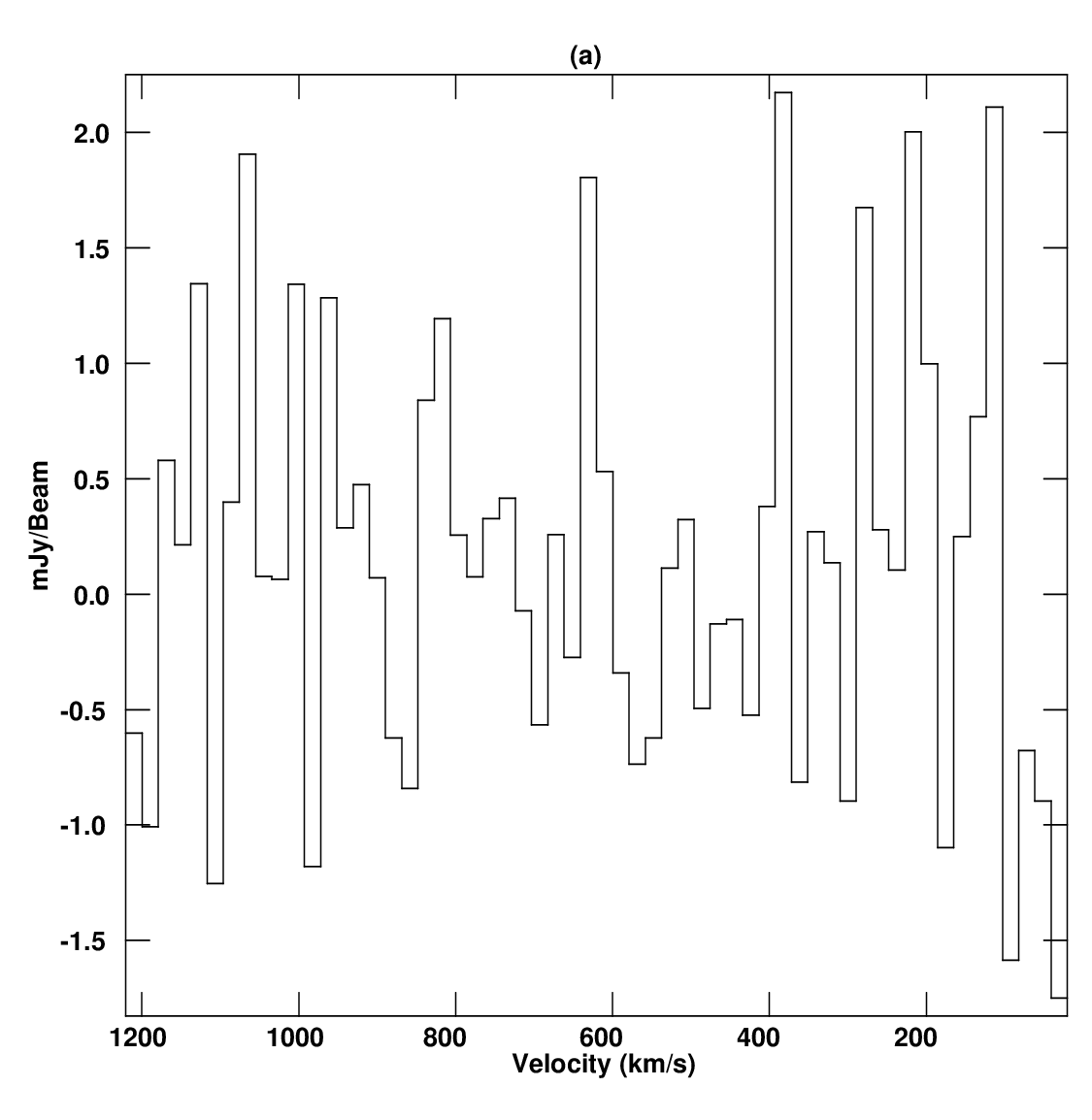}
\includegraphics*[scale=0.35]{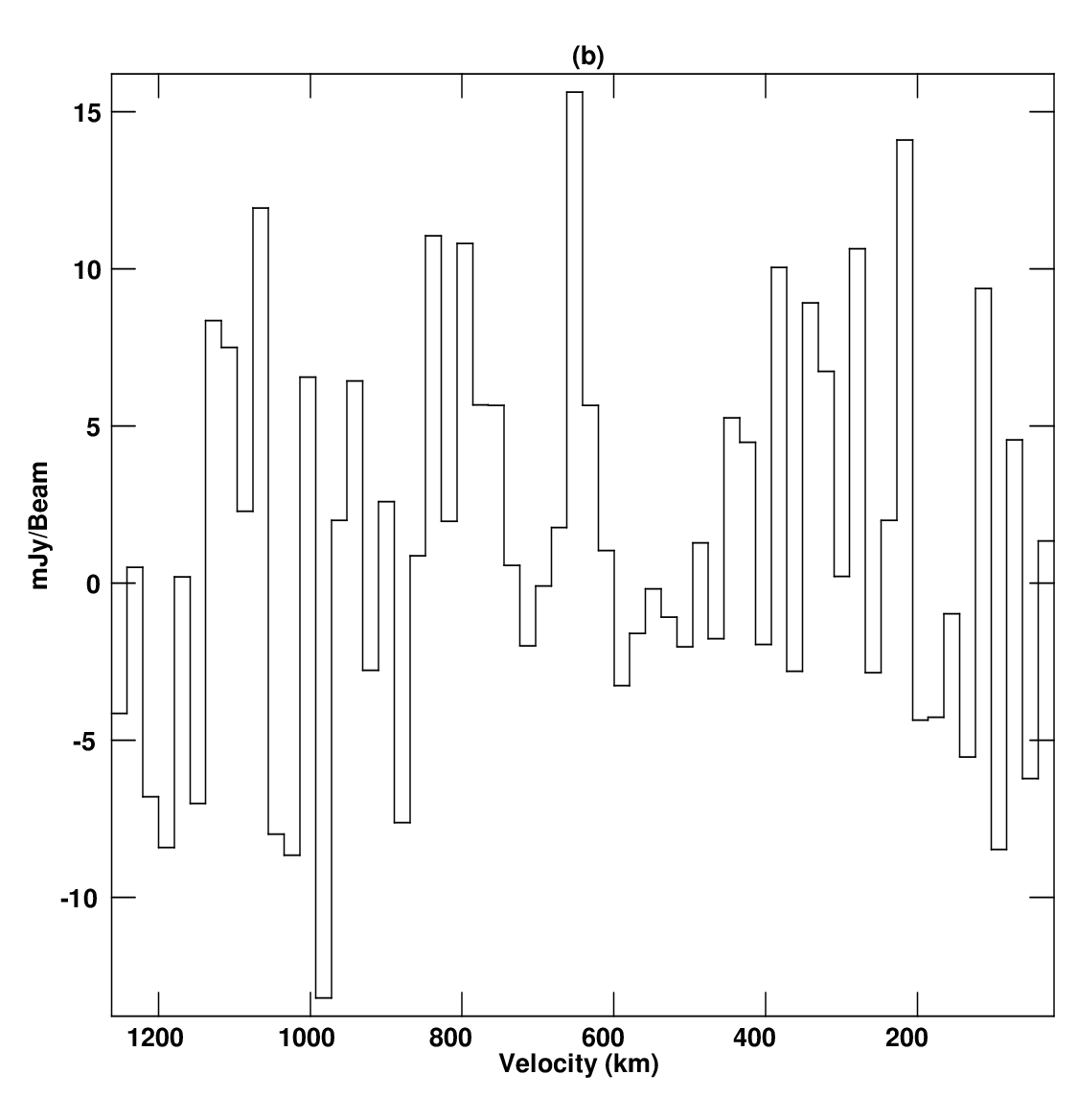}
\caption{\scriptsize NGC 4526: As in Figure \ref{n83ispec} but using box sizes of (a) 36\arcsec\ on a side and (b)  6.6\arcmin\ on a side, centered on the optical center of NGC 4526.  \label{n4526ispec}}   
\end{center}
\end{figure}
No HI emission or absorption is visible in the data cubes for NGC 83, NGC 4476, and NGC 4526.  Often the HI structures in early-type galaxies are much more extended than the CO emission, and in some cases a central HI hole may be present (Roberts \& Haynes 1994; van Driel \& van Woerden 1991; Serra \etal\ 2012). Therefore, two sets of primary beam corrected spectra were made for each of these galaxies.  For both spectra we assume that the HI covers the same velocity range as the CO.  The first spectrum is constructed by first using the CO maps to define a rectangular mask region within which "all" of the detected CO emission is located.  In the case of NGC 83 and NGC 4526 all of the detected CO emission fits within the central HI beam and so only a single pixel spectrum is extracted.  The intensity is then integrated over the same spatial region over the velocity range of the CO.  The second spectrum is made in a similar manner except emission is summed over a square box with a side length approximately equal to the diameter D$_{25}$ (de Vaucoleurs \etal\ 1991).  These larger boxes are typically 10 times the size of the CO diameter and 2.3 to 5.5 times the effective radius as defined in Cappellari \etal\ (2011), and they may reveal any extended, low surface brightness HI emission.  The primary beam corrected HI spectra are depicted in Figures \ref{n83ispec}, \ref{n4476ispec}, and \ref{n4526ispec}.  Spatial smoothing has not been employed, since the CO emission is smaller than the 15ÕÕ beam of the C configuration for NGC 83 and the 45ÕÕ beam of the D conÞguration for NGC 4526 and NGC 4476. 

Upper limits to the HI fluxes for NGC 83, NGC 4476, and NGC 4526 are determined from the unresolved/CO like spectra (see Table \ref{massnon}).  The uncertainty in the sum is calculated from the rms noise in the spectrum and the number of channels summed.  This estimate assumes that the channels are uncorrelated.  HI column densities and masses are obtained from three sigma upper limits of the uncertainty in the integrated intensity.  Numerically, this mass limit (a 3$\sigma$ limit on a sum over one beam and 20 to 30 channels) is within a factor of two of that proposed by Serra \etal\ (2012), which is a 3$\sigma$ limit on a sum over six beams and 40 km s$^{-1}$ (1 to 4 of our channels). Thus, the mass limit we quote here is appropriate either for an HI distribution which is spatially compact and follows the CO emission in velocity, or for a modestly extended but narrow linewidth "cloud" of the type observed around some nearby early-type galaxies by Serra \etal\ (2012).  HI column density and mass limits are obtained using "standard" formula, N$_{HI}=($1.1$\times$10$^{24}$ cm$^{-2}$)S$_{\nu,HI}$/($\theta_{max}\times\theta_{min}$) and M(HI)$=($2.36$\times$10$^5$ M$_\odot$)D$^2$S$_{\nu,HI}$, where S$_{\nu,HI}$ is the HI flux in units of Jy km s$^{-1}$, $\theta_{max}$ and $\theta_{min}$ are the major and minor axis of the synthesized beam in arcseconds, and D is the distance in Mpc.  All derived parameters for the HI non-detections can be found in Table \ref{massnon}.  No correction has been made for the presence of helium or effects due to inclination.
\subsection{NGC 4459: A Tentative Detection?}
\begin{figure}
\begin{center}
\includegraphics*[scale=0.45]{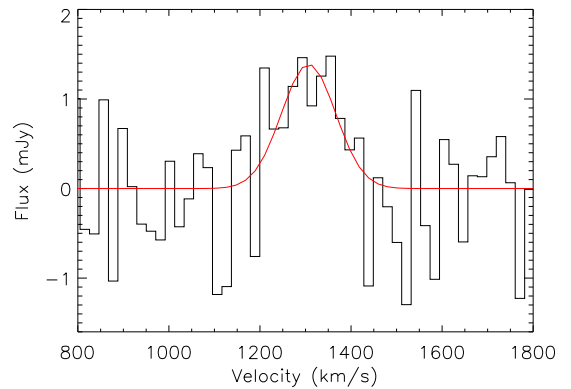}
\caption{\scriptsize NGC 4459: HI spectrum for the central pixel (15\arcsec\ on a side) centered on the optical center of NGC 4459. The CO in this galaxy fits inside the central pixel of the HI map.  The red line is a Gaussian fit to the entire usable spectrum.  The fit is centered on 1307 km s$^{-1}$, has an amplitude of 0.14 mJy, and a full width at half maximum, FWHM/2.355= 60.7 km s$^{-1}$.\label{n4459ispec}}
\end{center}
\end{figure}
For NGC 4459, spectra extracted from the central pixel (Fig. \ref{n4459ispec}) appear to show low level HI emission.  The IDL task GAUSSFIT is used to make a linear least square fit of a Gaussian to the spectrum in figure \ref{n4459ispec}.  The resulting fit produced a peak HI flux  of 0.14$\pm$0.04 mJy, a full width at half maximum of 143$\pm$45 km s$^{-1}$ centered on $1307\pm19$ km s$^{-1}$.  The HI column density and HI mass derived from the fitted parameters are $1.1\times10^{20}$ cm$^{-2}$ and $1.2\times10^6$ M$_\odot$, respectively.  The derived HI mass is consistent with a recent Westerbork Synthesis Telescope observation of similar resolution and slightly better rms noise (0.64 mJy beam$^{-1}$) which produces an upper limit to the HI mass of $<$7.41$\times$10$^6$ M$_\odot$ (Serra \etal\ 2012).  

We regard this as a tentative detection for several reasons.  First, The Gaussian fit is not in good agreement with the CO observations which give a CO line width of 400 km s$^{-1}$ centered on $1210\pm20$ km s$^{-1}$ (Young \etal\ 2005) or with estimates of the systemic stellar velocities $1232\pm40$ km s$^{-1}$ from Falco \etal\ (1999) and $1200\pm10$ km s$^{-1}$ from Emsellem \etal\ (2004).  Second, existing residual problems with the continuum subtraction produces spurious emission over large portions of the data cube (see section 3.1).  HI column density and mass upper limits from our VLA data are also obtained using the methods described in section \ref{res1} (see Table \ref{massnon}). 
\subsection{HI Detections: Fluxes and Line Widths}\label{res2}
\subsubsection{HI Absorption}\label{hiabs}
\begin{figure}
\begin{center}
\includegraphics*[scale=0.45]{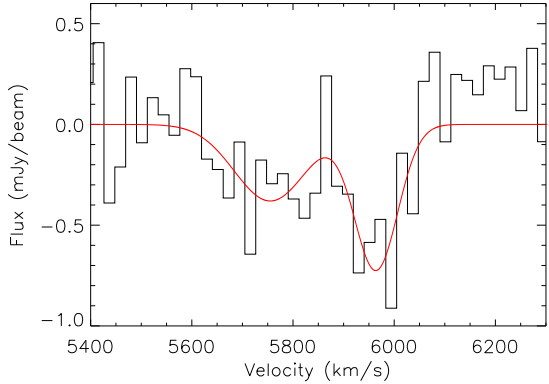}
\caption{\scriptsize NGC 2320: HI spectrum extracted from one pixel across the entire usable velocity range (1197 km s$^{-1}$) centered on NGC 2320.  The red line is a two component Gaussian fit to the entire usable spectrum.  The narrow component is centered on 5964 km/s, has an amplitude of -0.77 mJy beam$^{-1}$, and a full width at half maximum, FWHM/2.355=101 km s$^{-1}$.  The second component is centered on 5755 km s$^{-1}$, has an amplitude of -0.38 mJy beam$^{-1}$, and a full width at half maximum, FWHM/2.355=168 km s$^{-1}$. \label{n2320ispec}}
\end{center}
\end{figure}
We detect no HI in emission for NGC 2320.  There is HI emission present in the data cube for NGC 2320, but is associated with the spiral galaxy NGC 2321 (see appendix A).  Despite the fact that the data cube for NGC 2320 still contains some residual side lobes from a nearby bright continuum source, the continuum flux is consistent with that of the value quoted in NRAO VLA Sky Survey ($19.3\pm0.7$ mJy; Condon \etal\ 1998).  Figure \ref{n2320ispec} shows a spectrum extracted from the pixel containing the radio continuum peak.  This spectrum appears to show the presence of low level HI absorption.  The absorption is asymmetric with a peak of $\sim$ $-$$0.9\pm0.3$ mJy beam$^{-1}$ centered near 6000 km s$^{-1}$ and a weaker tail out to 5600 km s$^{-1}$.  The absorption peak is only three times the rms noise, and the broader portion of the absorption is at or below noise level.  Thus, we investigated the possibility that the absorption feature is just an instrumental effect by also inspecting continuum subtracted images of the phase and flux calibrators, but we find no similar central absorption features or other artifact associated with those cubes.  There is an HI absorption artifact associated with a spiral galaxy, NGC 2321, that is located 10\arcmin\ from the phase center (see appendix section \ref{aa}).  However, the channels which contain this artifact are offset by more than 100 km s$^{-1}$ from the absorption feature associated with NGC 2320 and so both features are probably not due to an interference spike.  We believe that the absorption artifact in NGC 2321 is probably caused either by problems with the continuum subtraction and/or imaging of a source that is far from the phase center; in other words, it is not related to the feature in NGC 2320.  It is also interesting to note here that the CO emission shows a small extension south west of the galaxy center, and the CO extension has similar velocities to the HI absorption peak (6032-6282 km s$^{-1}$; See Figure 12 of Young 2005).  This fact strengthens the possibility that at least the peak HI absorption is real.  We proceed now with the assumption that the entire absorption is real.

We fit a two component Gaussian to the usable portion of the absorption spectrum in Figure \ref{n2320ispec}.  The narrow component has a peak intensity of of $-$$0.72\pm0.17$ mJy beam$^{-1}$ and a line width of $101\pm31$ km s$^{-1}$ centered on $5964\pm13$ km s$^ {-1}$.  The broader component has a peak intensity of of $-$$0.38\pm0.14$ mJy beam$^{-1}$ and a line width of $168\pm80$ km s$^{-1}$ centered on $5755\pm31$ km s$^ {-1}$.  The total velocity width of the two component fit is 425 km s$^{-1}$ centered on 5862 km s$^{-1}$.  The center velocity of the two component fit is in close agreement with both the systemic velocity of the CO ($5886\pm20$ km s$^{-1}$; Young 2005) as well as measurements of the optical velocity ($5944\pm15$ km s$^{-1}$; Smith \etal\ 2000 and $5725\pm60$ km s$^{-1}$; de Vaucouleurs \etal\ 1991).

The optical depth calculated from the narrow and broad Gaussian fits are $0.043\pm0.013$ and $0.022\pm0.007$, respectively.  The estimated column density of the two components are $N_{HI}=8.4\pm2.7\times10^{20}$ cm$^{-2}$ and $N_{HI}=7.1\pm2.3\times10^{20}$ cm$^{-2}$, respectively.  If both components represent real absorption the total column density is $N_{HI_{tot}}=1.6\pm0.3\times10^{21}$ cm$^{-2}$.  Alternatively, if the absorption is in actuality contained in just one channel the optical depth is $0.053\pm0.010$, and the HI column density is $2.2\pm0.7\times10^{20}$ cm$^{-2}$ or 1.8$\pm$0.6 $M_\odot$ $pc^{-2}$.  The true HI column density probably falls somewhere between these three estimates.  If the HI is smoothly distributed inside one beam ($\sim$42.7 kpc$^2$), the total HI mass and H$_2$/HI ratios range from 0.8-5.5 $\times10^{8}M_\odot$ and 9 to 60, respectively. Both estimates of the HI mass and surface densities assume an HI harmonic mean spin temperature,  $<T_s>$, of $\sim$100 K (Dickey \& Lockman 1990) and are not corrected for the presence of helium.
\subsubsection{HI Emission}\label{hiem}
\begin{figure}
\includegraphics*[scale=0.4]{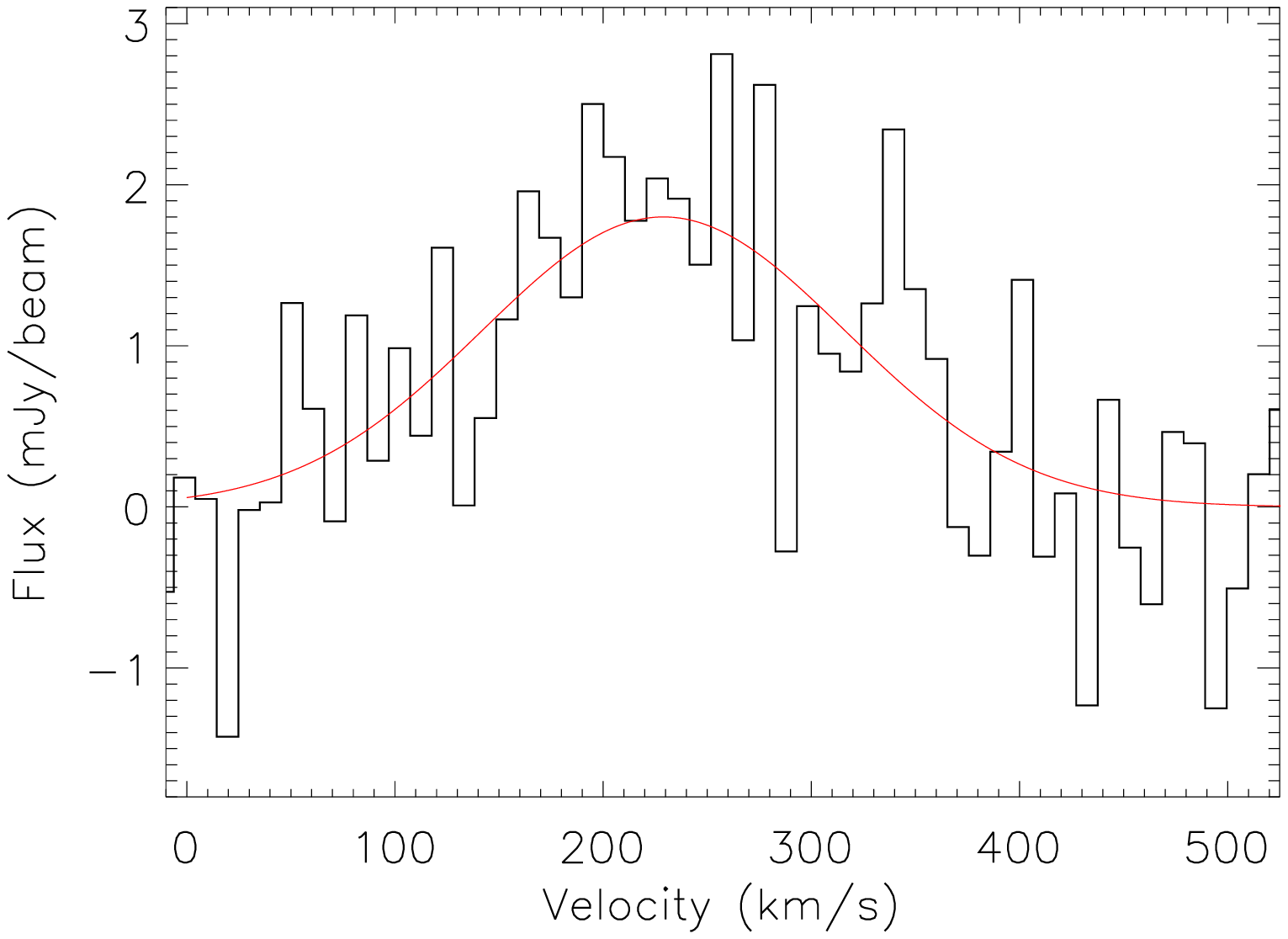}
\includegraphics*[scale=0.4]{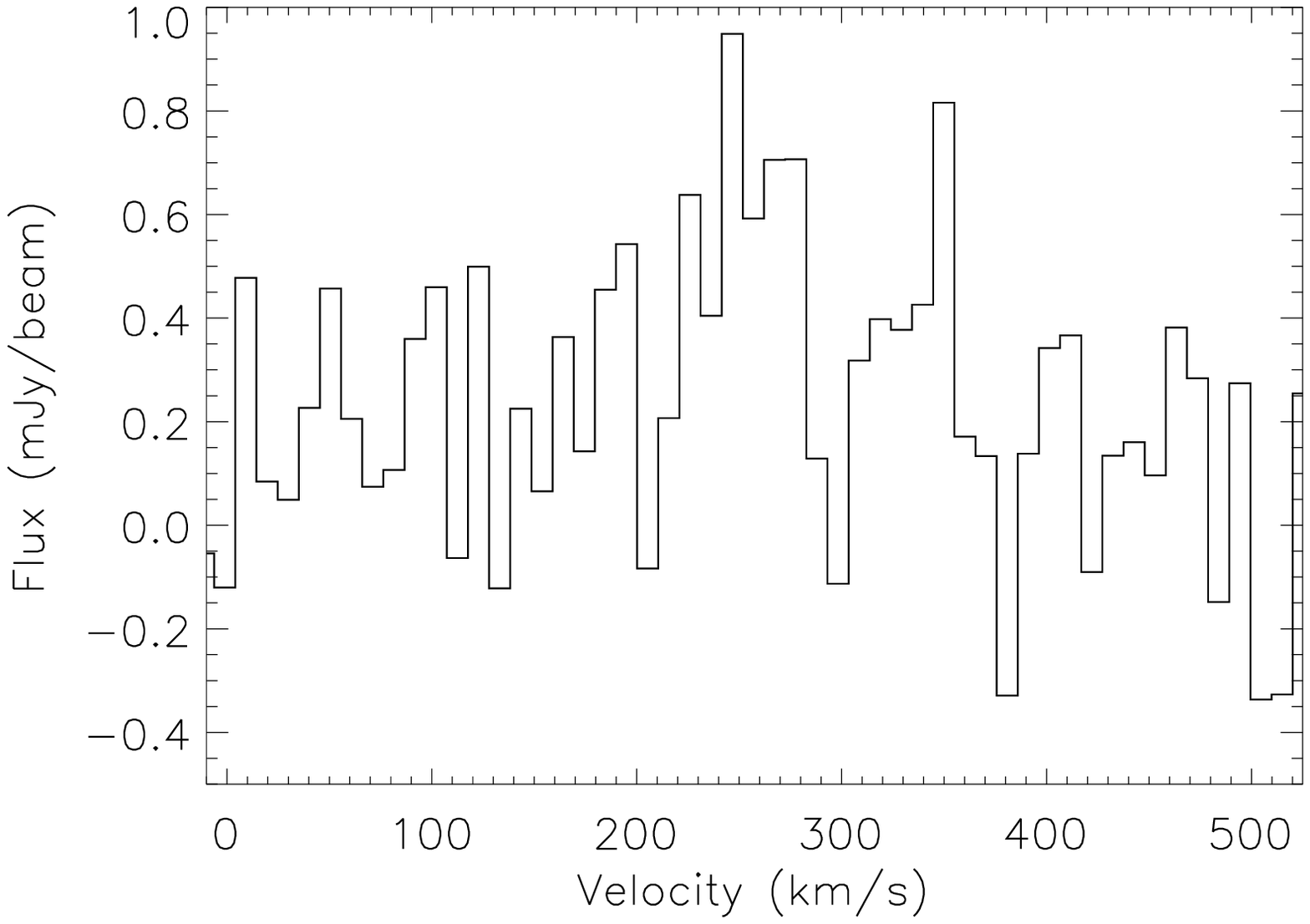}
\caption{\scriptsize NGC 4150: As in Figure \ref{n83ispec} but using box sizes of (a) 14\arcsec\ on a side and (b)  2.9\arcmin\ on a side, centered on the optical center of NGC 4150.  Integrated intensities quoted in the text are obtained from panel (a).  The red line in panel a is a Gaussian fit to the entire usable spectrum.  The fit is centered on 229 km s$^{-1}$, has an amplitude of 1.8 mJy beam$^{-1}$, and a full width at half maximum, FWHM/2.355=85.4 km s$^{-1}$.\label{n4150ispec}}
\end{figure}
For the galaxies with detected HI emission, Figures \ref{stuff6}-\ref{n5666con} show images of the integrated HI intensity, spectra, velocity fields, and individual channel maps.  Total fluxes in Table \ref{mass} are measured from the integrated intensity images in the top panels of Figures \ref{stuff6}-\ref{n5666con}.  The uncertainties in the HI fluxes are about 10$\%$, dominated mostly by the absolute calibration.  HI column densities and masses  are calculated using the fluxes listed in Table \ref{mass}, distances from Table \ref{hih2samp}, and synthesized beams from Table \ref{HIparam}.  No correction has been made for the presence of helium or inclination effects. 

A comparison of the new HI interferometric data with that of previous single dish observations yield the following:  Our HI flux for NGC 807 is 10.3 Jy km s$^{-1}$ and is within 10$\%$ of the HI flux measured by Huchtmeier \etal\ (1995), hereafter HSH95, using the Effelsberg 100m telescope (9.3\arcmin\ beam).  Our HI flux for NGC 3032 is 0.64 Jy km s$^{-1}$ which is smaller than 0.9 Jy km s$^{-1}$ measured with the Arecibo telescope and 0.96 Jy km s$^{-1}$ measured with the Westerbork at 45\arcsec\ resolution (Duprie \& Schneider 1996; Serra \etal\ 2012).  The Arecibo observations of NGC 3032 are three times less sensitive than our VLA C observations whereas the Westerbork observations are three times more sensitive.  If the much lower sensitivity Arecibo observations detected all of the flux, and the lower resolution but higher sensitivity Westerbork observations detect all of the flux, then the VLA C array observations must be resolving out $\sim$30$\%$ of the total HI flux.  This is surprising in that the detected HI emission is so compact  (i.e. there are probably no HI structures much larger than the primary beam).  At about 10\arcmin\ to the northeast of NGC 3032 there are three distinct HI sources.   A detailed presentation of the VLA HI observations for these three sources can be found in appendix A.  HSH95 also detect a larger HI flux from UGC 1503, 3.3 Jy km s$^{-1}$, than is found for the VLA C array image (1.8 Jy km s$^{-1}$), but an earlier HI detection (1.7 Jy km s$^{-1}$) by Haynes \& Giovanelli (1984) with the Arecibo telescope is consistent with the new observations.  It is unclear whether HI emission from UGC 1503 has been missed by the VLA C array.  The diameter of the HI emission in UGC 1503 is $\sim$1.5\arcmin\, so it is very unlikely that any emission could have been missed by the Arecibo observations. In the case of NGC 5666 the VLA C array observations recover a factor of 2 less flux than the VLA AD array (3.8 mJy km s$^{-1}$ versus 6.2 mJy km s$^{-1}$).  These two data sets have similar rms noises ($\sim$ 0.5 mJy beam$^{-1}$) and so it is likely that the C array observations have resolved out a significant amount of flux.  The HSH95 Effelsberg 100 meter single dish observations for NGC 5666 detect a smaller HI flux from NGC 5666, 4.5 Jy km s$^{-1}$,  than is found for the combined VLA AD and C array image (6.3 Jy km s$^{-1}$).  The rms noise of the HSH95 data, $\sim$1-2 mJy beam$^{-1}$, is about 2-4 times higher than the new observations. This is the most likely explanation for the smaller flux. 

We find weak and poorly resolved HI emission in the data cube for NGC 4150 and spectra extracted from the central pixel appear to show low level HI emission (Fig. \ref{n4150ispec}).  Fitting a 1-D Gaussian to the spectrum in Figure \ref{n4150ispec}a gives a full width at half maximum (FWHM) of 201$\pm$41 km s$^{-1}$ centered on 229$\pm$17 km s$^{-1}$ and a amplitude of 1.8$\pm$0.3 mJy beam$^{-1}$.  The center velocity of the emission is in close agreement with both the systemic velocity of CO (239$\pm$20 km s$^{-1}$; Young 2005) and measurements of the optical velocity (226$\pm$22 km s$^{-1}$; Fisher \etal\ 1995).  If the HI emission is unresolved the total flux is then $36\pm6$ mJy km s$^{-1}$ which gives a total mass of $1.6\pm0.3$$\times$10$^6$ M$_\odot$.  Similar WSRT observations yield an HI flux of 56 mJy km s$^{-1}$ (Morganti \etal\ 2006).  The WSRT observations have an angular resolution of 42.29\arcsec$\times$26.00\arcsec.  A total of  four twelve hour tracks produced an rms noise of 0.3 mJy beam$^{-1}$, three times more sensitive than the VLA observations.  The WSRT observations show that the HI in NGC 4150 is actually resolved (Morganti \etal\ 2006). 

Recent single dish Arecibo HI observations of NGC 4150 by Sage \& Welch (2006) produce spectra with similar line widths, but those authors quote an HI mass of 1.56$\times10^7$ M$_\odot$ which is more than an order of magnitude larger than those derived from the interferometric data.  One possible reason for the larger HI flux include is confusion with a nearby source. An analysis of the moment zero map (Fig. \ref{u4150map}) provided by Raffaella Morganti shows that roughly half of the detected HI is located inside the galaxy ($\sim$1.4$\times$10$^6$ M$_\odot$) while the other half is located outside of the galaxy ($\sim$1.6$\times$10$^6$ M$_\odot$) about 1\arcmin\ to the south at a position of RA. 12h 10m 29.79s and Dec. 30d 22m 45.96s.  The emission to the South does not appear to have any visible optical counterpart.  In fact, there are no visible counter parts (in projection or velocity of $\pm$ 500 km s$^{-1}$) within 40\arcmin\ of NGC 4150 listed in the NASA Extragalactic database (NED).  Therefore, it is unlikely that confusion with any additional field sources is occurring. A second possibility is confusion with high velocity galactic HI.  This possibility can be ruled out since the high velocity clouds along the line of sight toward the position of NGC 4150 have negative velocities (Kalberla \etal\ 2005).  A third possibility is that a significant amount of HI is in smooth, low density complexes that are only detected by the lower resolution instrument.  This is the most plausible explanation since HI disks around early-types often extend significantly outside of their optical components (Sadler \etal\ 2000; Morganti \etal\ 2006; Grossi \etal\ 2009).  
\subsection{HI Column Densities and the $\frac{M_{H_2}}{M_{HI}}$ Mass Ratio}
\begin{deluxetable*}{lcccclllcll}
\tabletypesize{\footnotesize}
\tablecolumns{10}
\tablecaption{\bfseries HI Non-detections: Flux, Gas Masses, and Ratio Limits \label{massnon}}
\tablehead{
\colhead{Galaxy}              &\colhead{$\Delta_{V_{CO}}^a$}         &\colhead{I$_\nu$$^b$}     &\colhead{S$_\nu^c$}              &\colhead{N$_{HI}^d$}     &\colhead{M$_{HI}^e$}  &\colhead{$log\left[\frac{M_{HI}}{L_{K}}\right]$} 
&\colhead{M$_{H_{2}}$}                         &\colhead{$\frac{M_{H_2}}{M_{HI}}$}                                                            &\colhead{H$_2$ ref}\\
\colhead{}              &\colhead{(km s$^{-1}$)}      &\colhead{(Jy km s$^{-1}$)}   &\colhead{(Jy km s$^{-1}$)}       &\colhead{(10$^{20}$ cm$^{-2}$)}     &\colhead{(10$^7$ M$_{\odot}$)}  
&\colhead{($\frac{M_{\odot}}{L_{K}}$)} 
&\colhead{(10$^8$ M$_{\odot}$)}                         &\colhead{}                                                            &\colhead{}
}
\startdata 
N83                      &6047-6464                           &0.043(0.030)                           &0.09                             &$<$3.0                                 &$<$15           &$<-$3.3           &19(3)         &$>$29     &1\\
N4459                  &1029-1392                            &0.034(0.092)                           &0.28                            &$<$1.6                               &$<$1.7            &$<-$3.6         &1.7(0.3)     &$>$21     &2\\
N4476                  &1870-2050                            &0.12(0.03)                             &0.09                            &$<$12                                  &$<$0.66          &$<-$3.2         &1.0(0.1)     &$>$15     &3\\
N4526                  &287-970                                &0.30(0.09)                               &0.27                           &$<$0.91                             &$<$1.9            &$<-$3.9         &6.4(1.5)     &$>$100   &2
\enddata
\tablecomments{
No correction is made for the presence of helium or inclination effects in either the HI or the H$_2$ masses.\\
H$_2$ mass references: (1) Young 2005; (2) Young \etal\ 2008; (3) Lucero \etal\ 2005.\\
$^a$ The observed width of the CO line.\\
$^b$HI Intensity integrated over the CO line width.\\
$^c$Upper limit of the HI flux for an unresolved source derived from three times the uncertainty in the integrated intensity.\\
$^d$Estimated upper limit to the peak HI column density derived from S$_{\nu}$.\\
$^e$Estimated upper limit to the HI Mass derived from S$_{\nu}$.}
\end{deluxetable*}
\begin{deluxetable*}{lccccccccll}
\tabletypesize{\scriptsize}
\tablecaption{\bfseries Detections: HI Fluxes, Gas Masses and Ratios \label{mass}}
\tablehead{
\colhead{Galaxy}  &\colhead{S$_\nu$}              &\colhead{N$_{HI}$}                         &\colhead{R$_{HI}$}  &\colhead{R$_{HI}$}  &\colhead{M$_{HI}$}                  & \colhead{log$\left[\frac{M_{HI}}{L_{K}}\right]$}  &\colhead{M$_{H_2}$}  &\colhead{$\frac{M_{H_2}}{M_{HI}}$} &\colhead{Class$^d$} &\colhead{ref}\\
\colhead{}              &\colhead{(Jy km s$^{-1}$)} &\colhead{(10$^{20}$ cm$^{-2}$)} &\colhead{(arcsec)}    &\colhead{(kpc)}          &\colhead{(10$^8$M$\odot$)} &\colhead{} &\colhead{(10$^8$M$\odot$)} &\colhead{}  &\colhead{} &\colhead{}}
\startdata 
U1503     &1.8(0.2)        &9.9               &47        &16(1)              &21(3)                               &$-$1.47                                &19(3)          &1     &D    &1\\
N807$^a$      &10.3(0.9)       &7.7          &223      &42(3)             &110(15)                            &$-$1.17                                &14(3)          &0.1     &u/D &1\\
N2320$^b$    &$-$        &2.2-16.0       &$-$      &$-$                              &1.2-7.4                 &$-$2.75                               &47(9)          &6-40 &d  &2\\
N3032    &0.64(0.06)    &13              &24        &2.5(0.3)  &0.68(0.13)                      &$-$2.27                               &4.9(1.1)      &7       &d  &3\\
N3656    &5.4(0.5)          &72               &41        &8.0(0.2)  &20(3)                               &$-$1.51                                &38(5)          &2      &u   &4\\
N4150    &0.036(0.007)   &0.49              &$-$      &$-$        &0.016(0.004)                  &$-$3.79                                 &0.58(1.1)    &36   &d &3\\
N4150$^c$   &0.032(0.010)         &0.29             &45        &3.0(0.3)  &0.014(0.003)                       &$-$3.85                                &0.58(1.1)    &41 &d       &5\\
N5666     &6.3(0.6)      &9.1              &96        &14(1)      &14(2)                               &$-$1.08                                &4.4(0.6)      &0.31   &D  &1
\enddata
\tablecomments{
The value quoted in the third column is the observed peak HI column density.  No correction is made for the presence of helium in either the HI or the H$_2$ masses.\\  
$^a$The HI radius quoted for the disk of NGC 807 is measured from the galaxy center out to the edge or the tidal arms.\\
$^b$NGC 2320 is detected in absorption.  The range in peak column density corresponds to the cases of narrow and broad absorption.\\
$^c$The HI flux and mass come from the WSRT observations.  The quoted HI flux accounts for only the HI associated\\
with the optical galaxy and excludes a patch of HI emission outside and $\sim$1\arcmin\ to the south of NGC 4150 (see section \ref{res2}).\\
$^d$Morphological HI classifications as defined by the \atlas\ classification scheme (Serra \etal\ 2012).\\ 
References: (1) Young 2002; (2) Young 2005; (3) Young \etal\ 2008; (4) Balcells \etal\ 2001; (5) Morganti \etal\ 2006}
\end{deluxetable*}
Our HI-detected sample galaxies have H$_2$/HI mass ratios ranging from 0.1 to 41 (see Table \ref{mass}).  This is a similar range in H$_2$/HI mass ratios found in both the single dish and interferometric volume limited surveys.  The single dish survey of Welch \& Sage (2006) find H$_2$/HI mass ratios in the range of 0.01-5.  The interferometric \atlas\ survey find H$_2$/HI mass ratios in the range of 0.01 to 36 (Serra \etal\ 2012).  The lower end of the range in our mass ratio is an order of magnitude larger than that of the volume limited samples.  This is probably due to the fact that our early-type sample is biased toward the most CO-rich objects.   All of our sample galaxies with H$_2$/HI mass ratios greater than 10 reside in group or cluster environments.  This is also consistent with what is found by the recent volume limited surveys.  We discuss the role of environment on the HI content of our sample galaxies in section \ref{envi}.

The interferometric observations for NGC 4150 give an M$_{H_{2}}$/M$_{HI}=41$.  However, this mass ratio may be too high since it is possible that the WSRT observations may have missed a significant amount of flux (see discussion in section \ref{hiem}).  If the HI mass is instead derived from the Arecibo flux (Sage \& Welch 2006) and the distance in Table 1, the mass ratio is much lower, M$_{H_{2}}$/M$_{HI}$$\sim2$. 
\section{HI Versus CO Morphology}\label{morph}
\begin{figure*}
\begin{center}
\epsscale{0.7}
\plotone{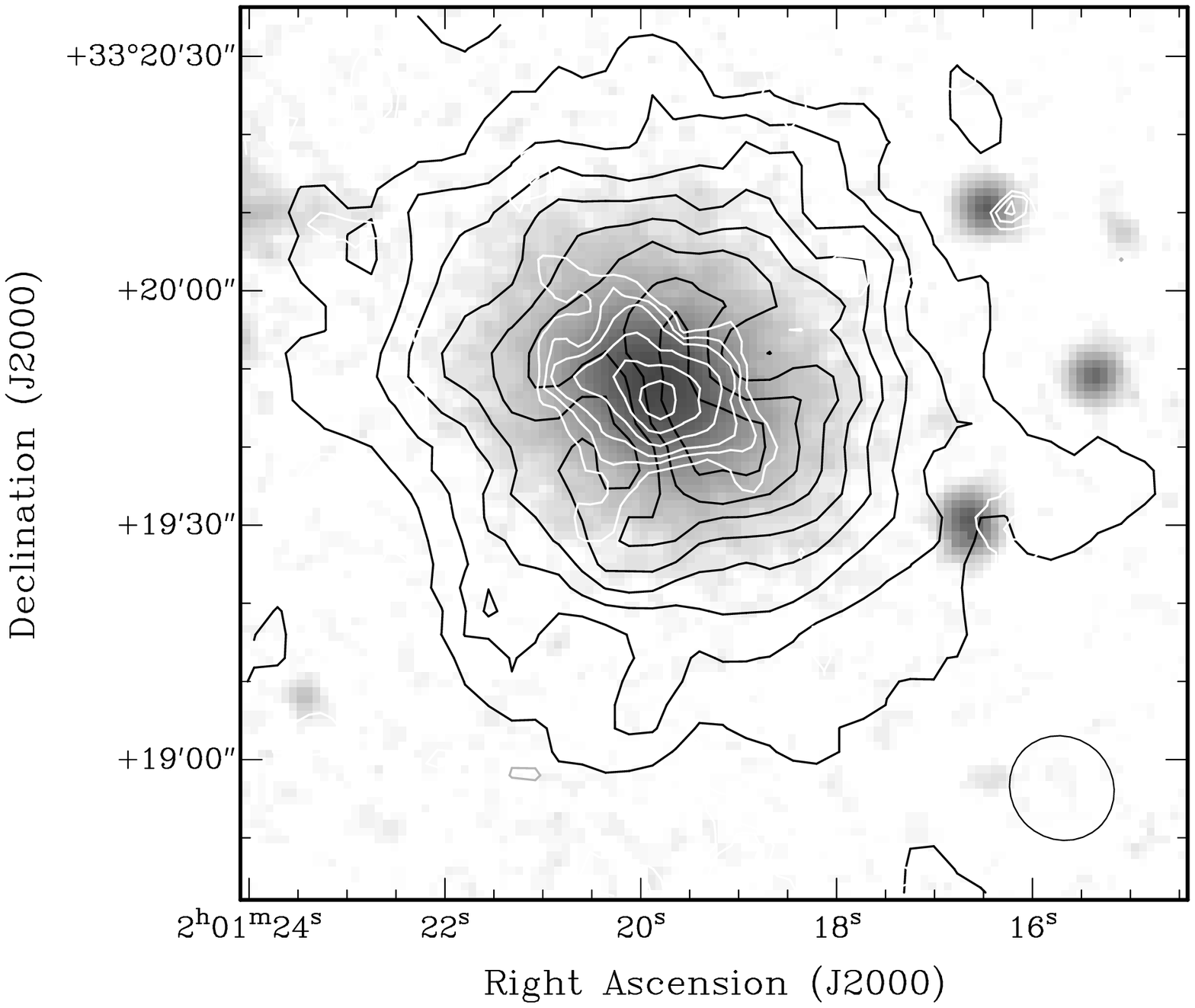}\\
\epsscale{1.0}
\plottwo{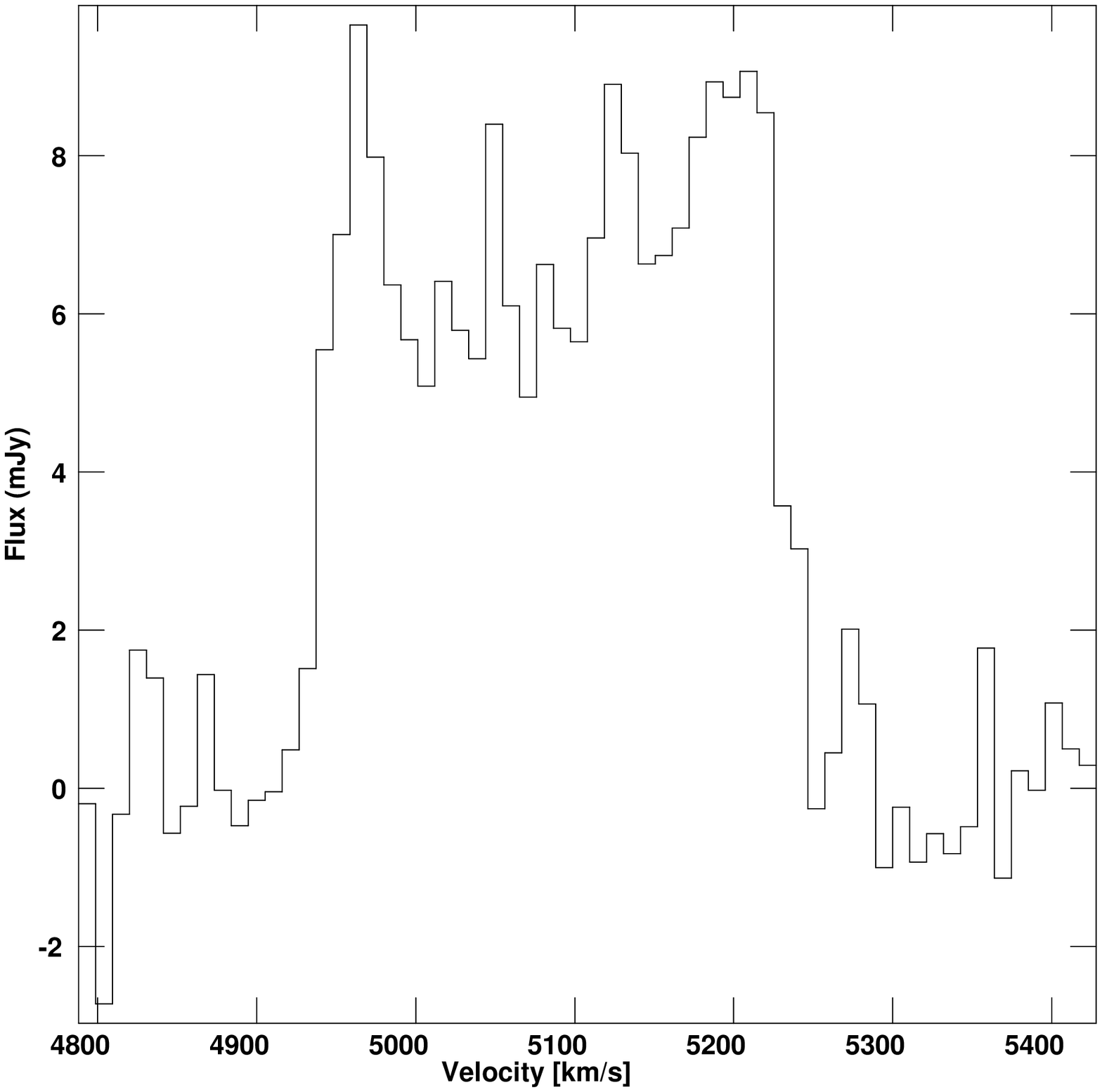}{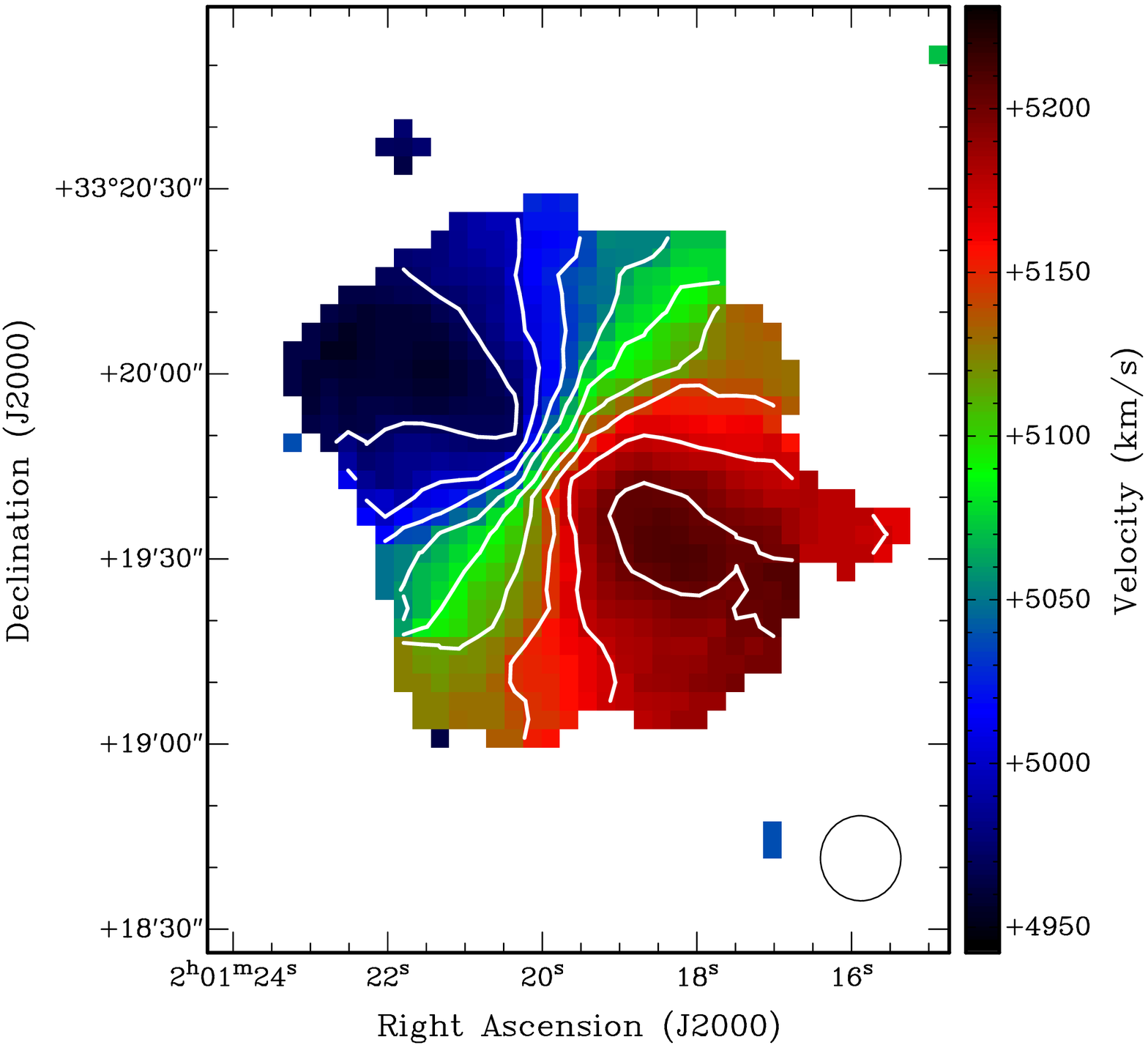}
\caption{\scriptsize  {\bf Top}:  UGC 1503: Solid black (positive) and grey (negative) contours show the HI integrated intensity in units of $-$10\%, $-$20\%, 10\%, 20\%, 30\%, 40\%, 50\%, 60\%, 70\%, 80\%, and 90\% of the peak (0.16 Jy beam$^{-1}$ km s$^{-1}$=9.9x10$^{20}$ cm$^2$). White contours show the CO integrated intensity in units of 10\%, 20\%, 30\%, 50\%, 70\%, and 90\% of the peak (6.3 Jy beam$^{-1}$ km s$^{-1}$$=$3.9$\times$10$^{21}$ cm$^{-2}$). The grey scale image is an SDSS2 R-band image.  {\bf Bottom left}: HI spectrum.   The spectrum was constructed by first using the integrated image (moment 0) to define a rectangular mask region within which the emission is located.  The intensity was then integrated over the same spacial region for every channel, so that the noise in the line free channels is indicative of the noise on the line as well.  {\bf Right bottom}:  Velocity field. The HI intensity-weighted mean velocity (moment 1) is shown in RGB color scale and in white contours from 4900 to 5200 km s$^{-1}$ in steps of 30 km s$^{-1}$.The black ellipse shows HI the beam size.  The CO resolution is 7.1\arcsec$\times$6.3\arcsec. \label{stuff6}}
\end{center}
\end{figure*}

\begin{figure*}
\begin{center}
\includegraphics*[scale=0.7]{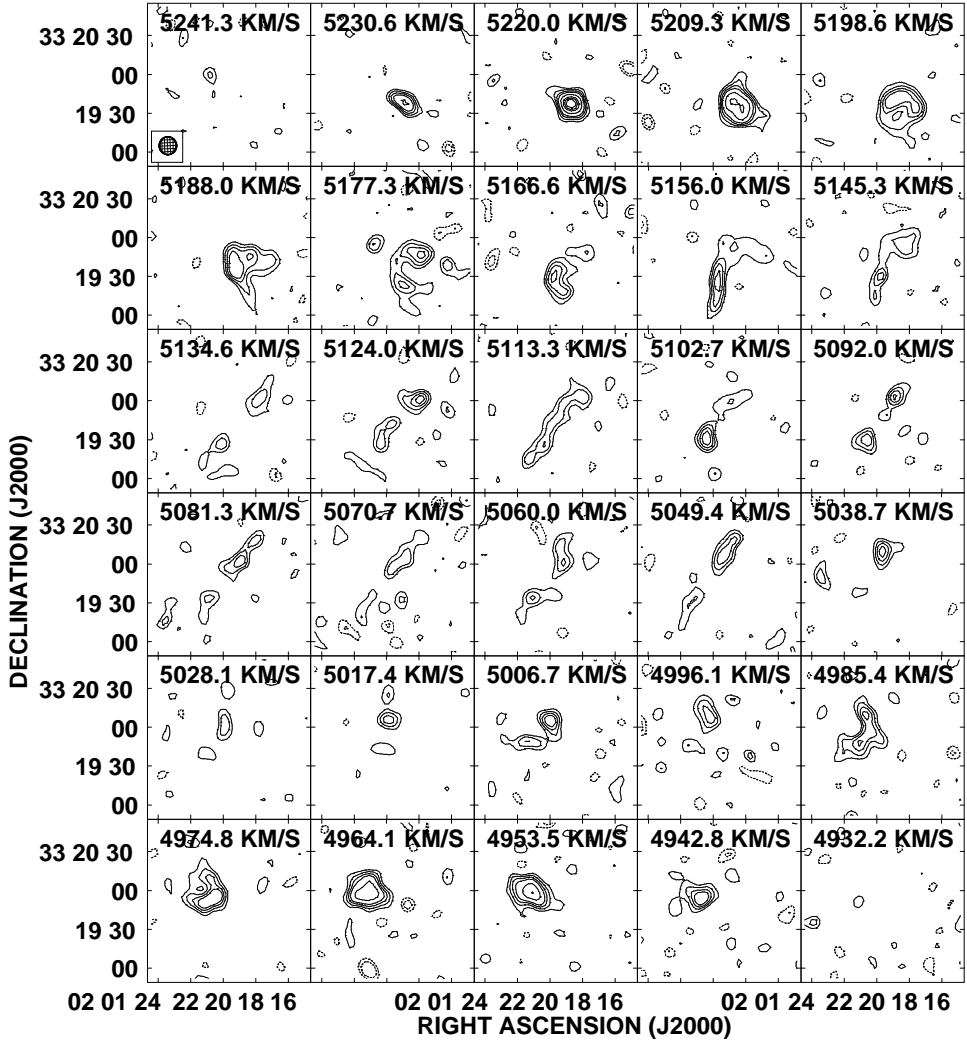}
\caption{\scriptsize UGC 1503. Individual channel maps showing HI emission.  Contour levels are -3, -2, 2, 3, 4, 5, 7, 9, 10, and 11 times 0.3 mJy beam$^{-1}$ $\sim$1$\sigma$. The velocity of each channel (in km s$^{-1}$) is indicated at the top of each panel and the beam size in the first panel in the bottom left corner. \label{u1503cont}}
\end{center}
\end{figure*}

\begin{figure*}
\begin{center}
\epsscale{0.6}
\plotone{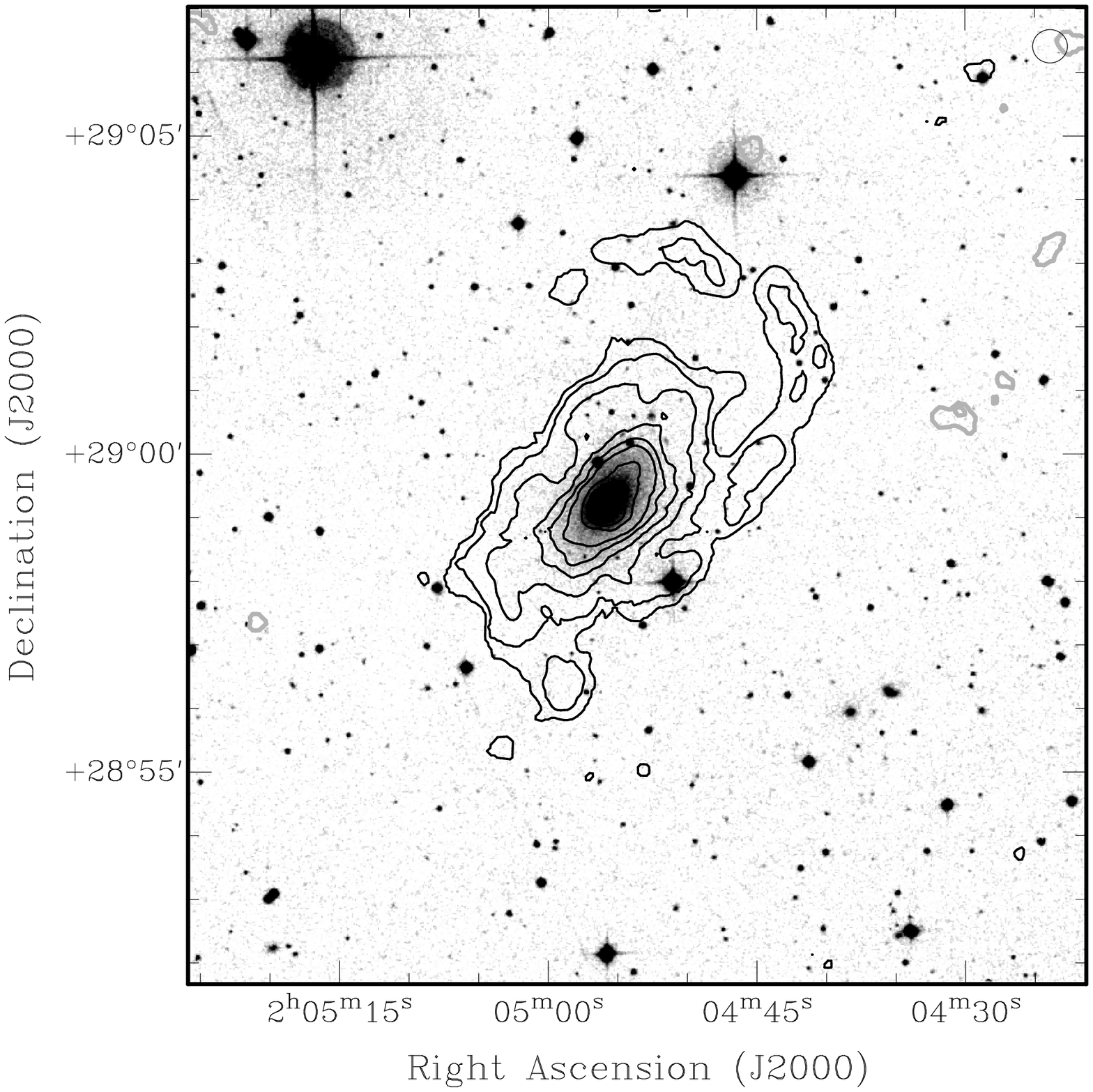}\\
\epsscale{0.7}
\plotone{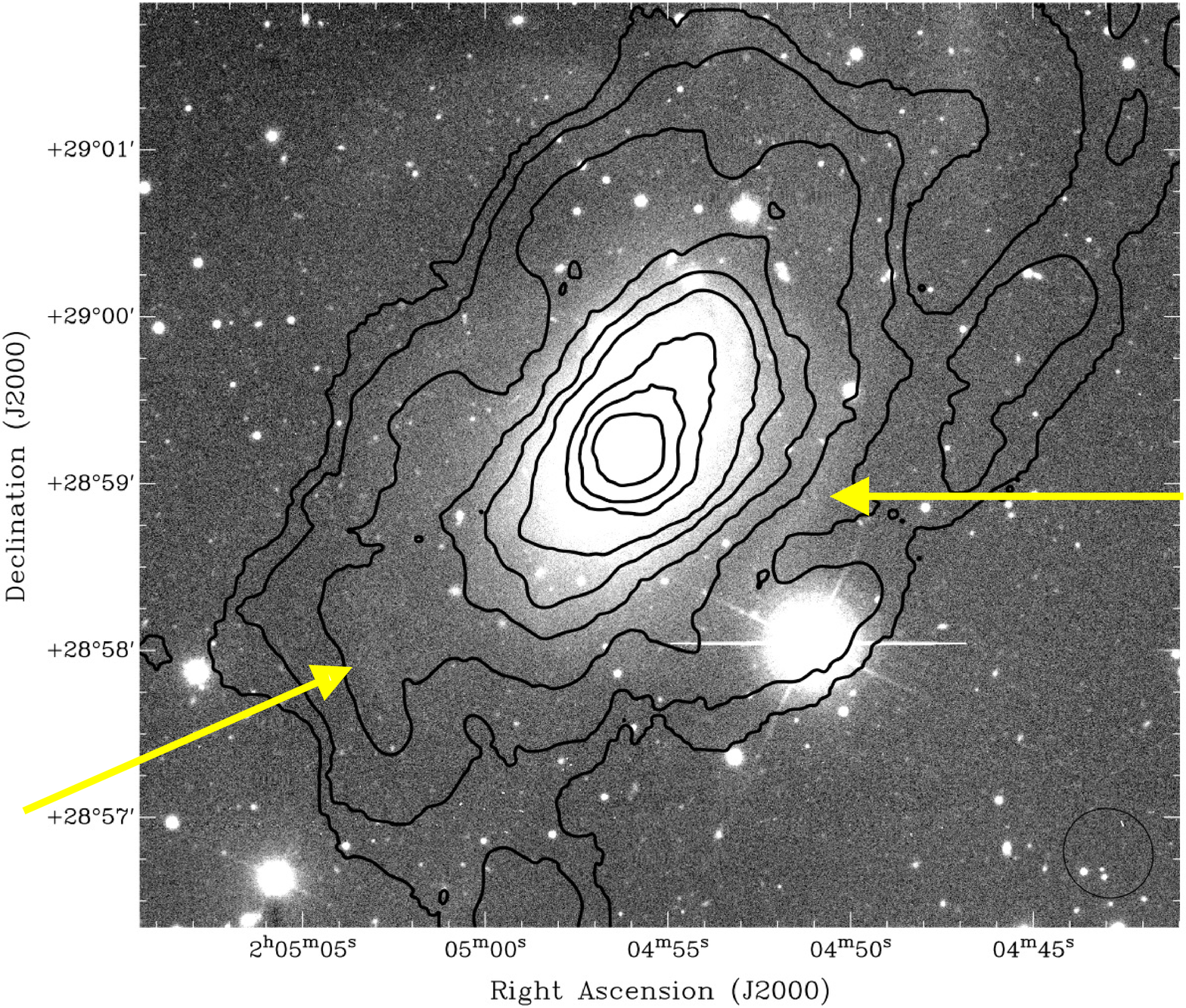}
\caption{\scriptsize  {\bf Top}:  NGC 807:  VLA C$+$D array HI Data.  Solid black (positive) and grey (negative) contours show the HI integrated intensity in units of -10\%, -5\%, 5\%, 10\%, 20\%, 30\%, 40\%, 50\%, 70\%, 80\%, and 90\% of the peak (0.72 Jy beam$^{-1}$ km s$^{-1}$$=$7.7$\times$10$^{20}$ cm$^{-2}$). The black ellipse shows the beam size.  The grey scale image is an SDSS2 R-band image. {\bf Bottom}:  HI contours overlaid on a WIYN 3.5 meter V band image.  The black ellipse shows the beam size.  Yellow arrows point to optical tidal features.  \label{stuff5}}
\end{center}
\end{figure*}

\begin{figure*}
\epsscale{0.8}
\plotone{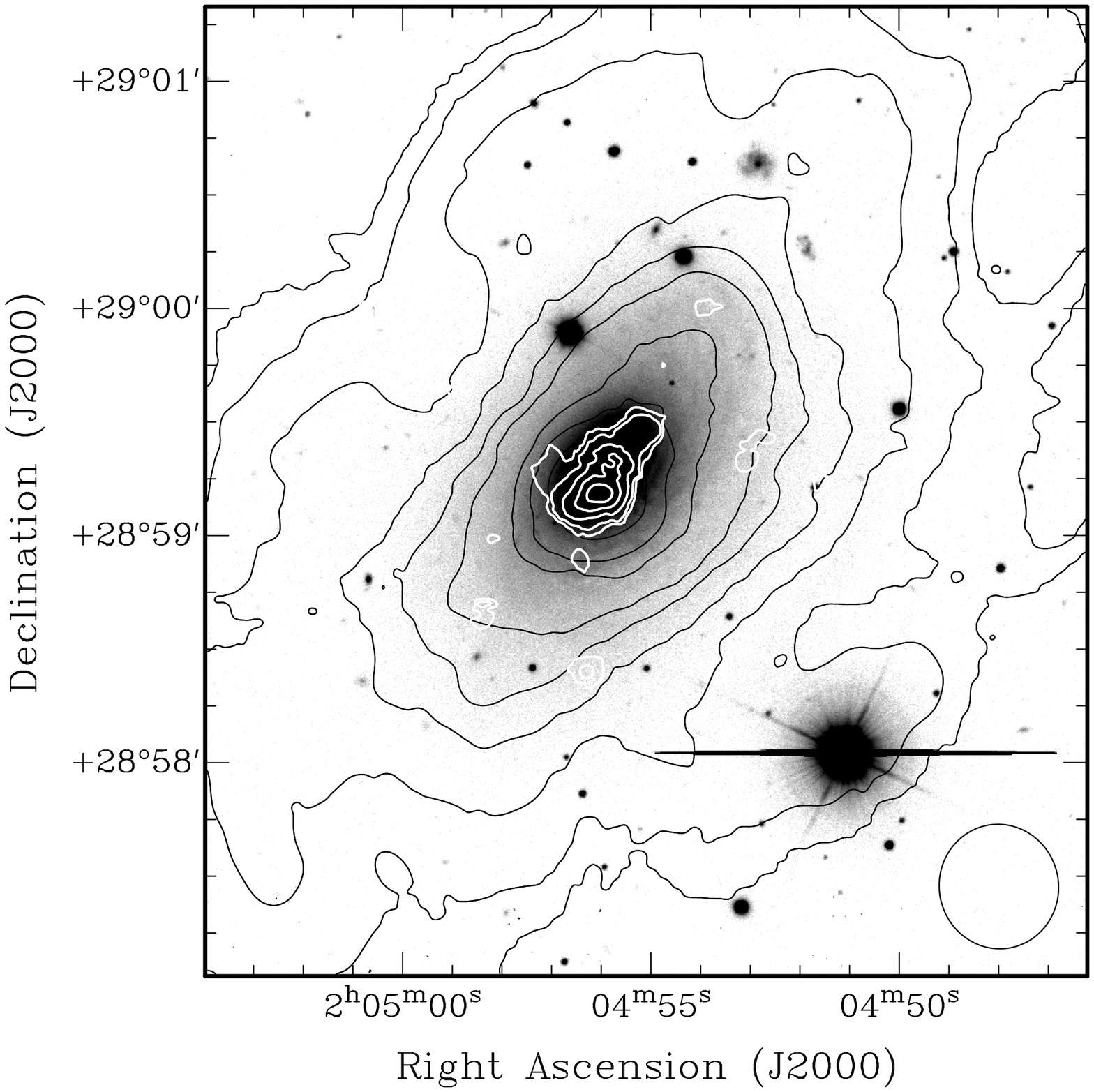}\\
\epsscale{1.0}
\plottwo{f10ba.eps}{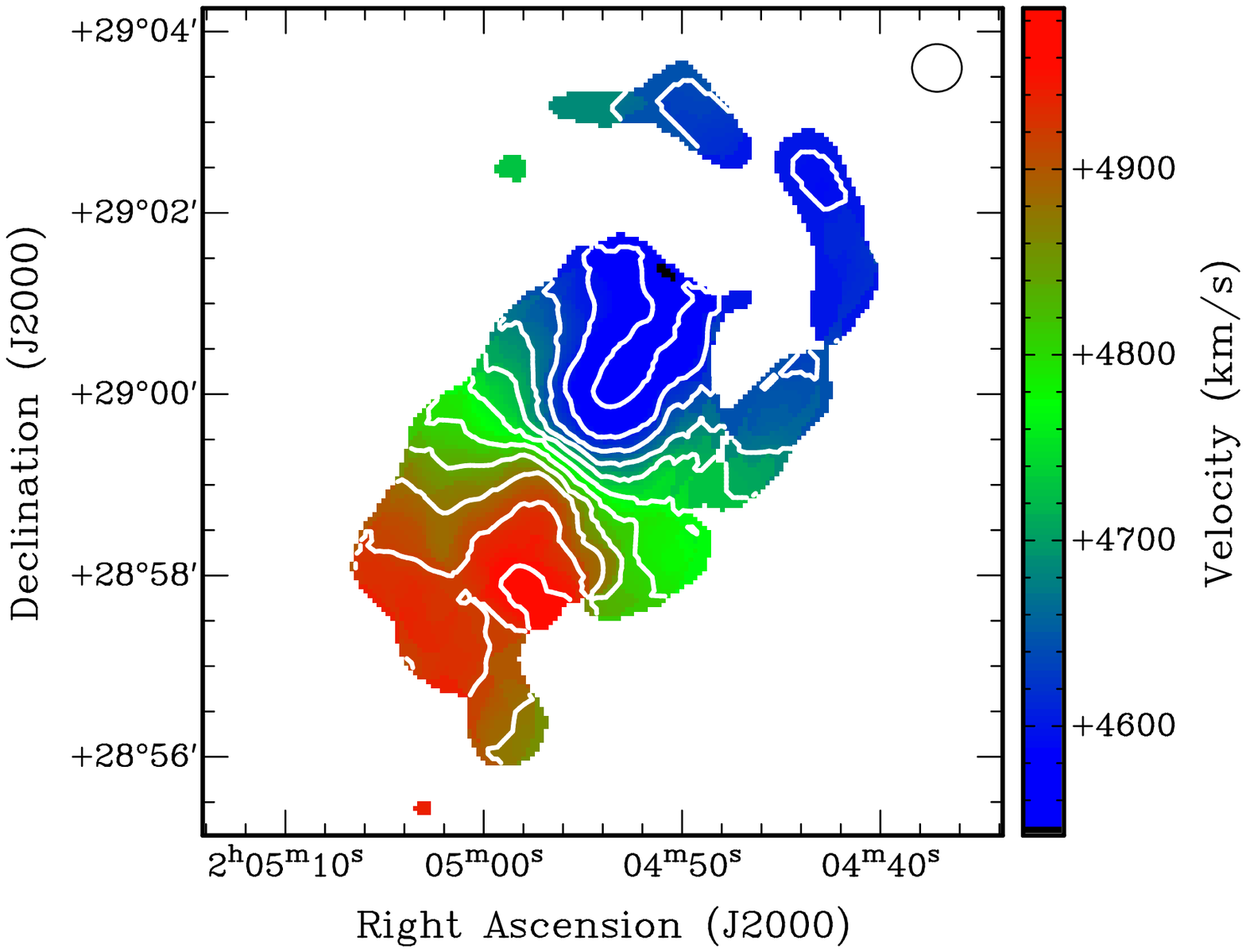}
\caption{\scriptsize  {\bf Top}:  NGC 807 zoomed in C+D array HI data overlaid with CO.  Solid black (positive) and grey (negative) contours show the HI integrated intensity in units of -10\%, -5\%, 5\%, 10\%, 20\%, 30\%, 40\%, 50\%, 70\%, 80\%, and 90\% of the peak (0.72 Jy beam$^{-1}$ km s$^{-1}$$=$7.7$\times$10$^{20}$ cm$^{-2}$). The black ellipse shows the beam size. white contours show the CO integrated intensity in units of 20\%, 30\%, 50\%, 70\%, and 90\% of the peak (7.6 Jy beam$^{-1}$ km s$^{-1}$$=$2.6$\times$10$^{21}$ cm$^{-2}$). The grey scale image is an WIYN 3.5 meter V band image. {\bf Left bottom}: HI spectrum. Constructed in a similar manner to UGC 1503.  {\bf Right bottom}: Velocity Field.  The HI intensity weighted mean velocity (moment 1) is shown in RGB color scale and in white contours from 4560 to 4960 in steps of 40 km s$^{-1}$. The black ellipse shows the HI beam size. The CO resolution is 7.0\arcsec$\times$6.3\arcsec. \label{stuff4}}
\end{figure*}

\begin{figure*}
\epsscale{1.0}
\plotone{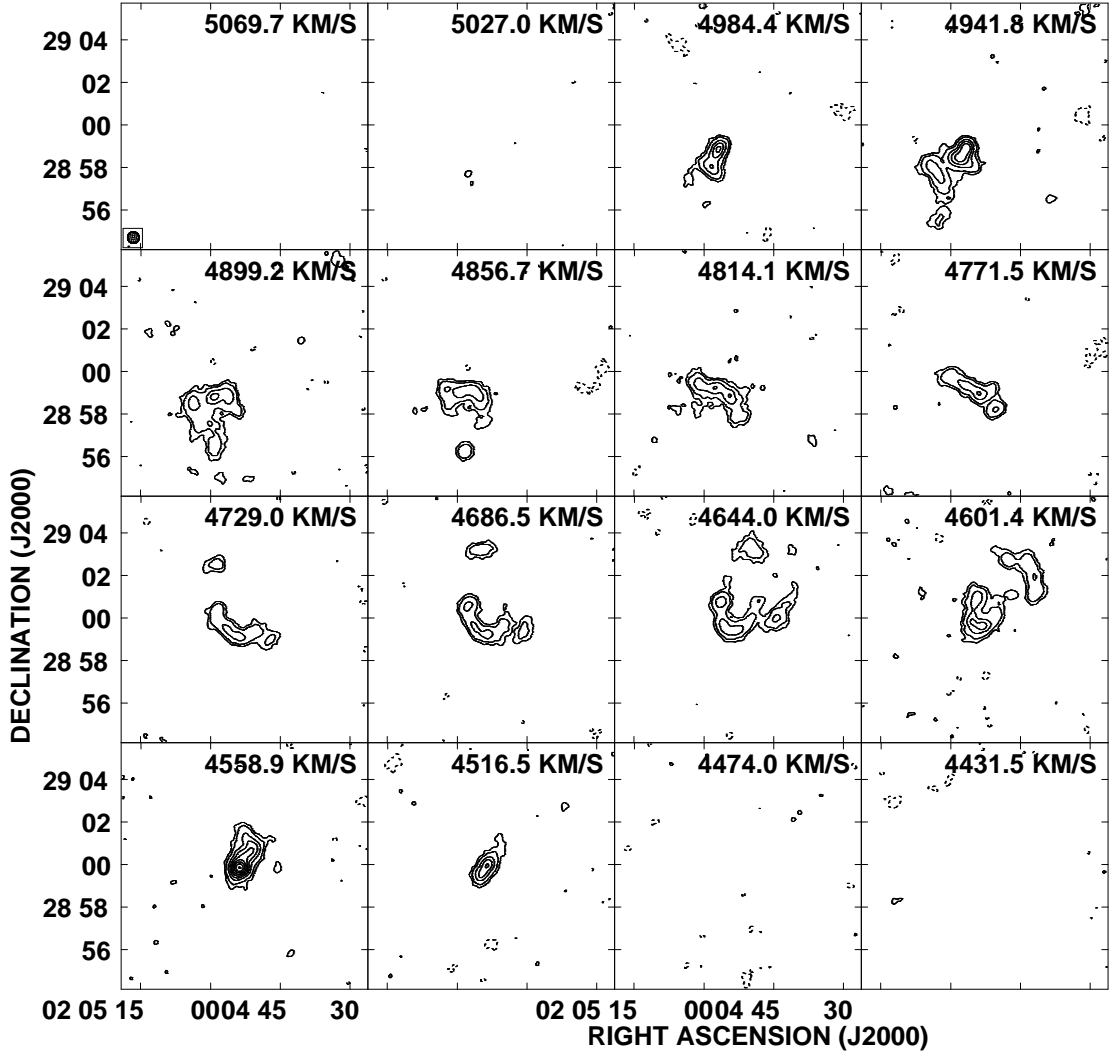}
\caption{\scriptsize NGC 807: Individual channel maps showing HI emission.  Contour levels are $-$5, $-$3, 3, 10, 15, 20, 25, 30, 35, and 40 times 0.2 mJy beam$^{-1}$ $\sim$1$\sigma$. The velocity of each channel (in km s$^{-1}$) is indicated at the top of each panel and the beam size in the first panel in the bottom left corner.\label{n807con}}
\end{figure*}

\begin{figure*}
\epsscale{0.7}
\plotone{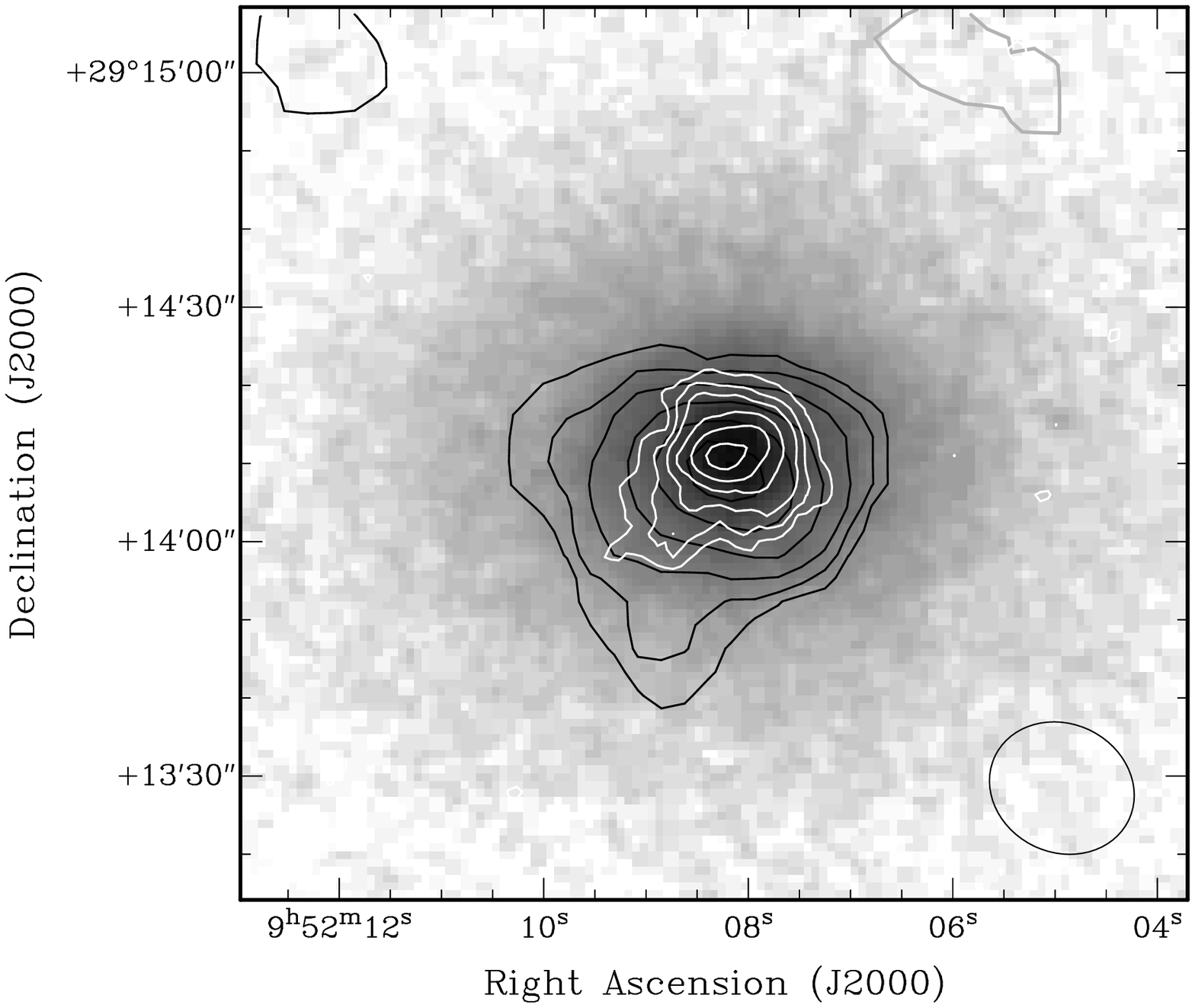}
\epsscale{1.0}
\plottwo{f12ba.eps}{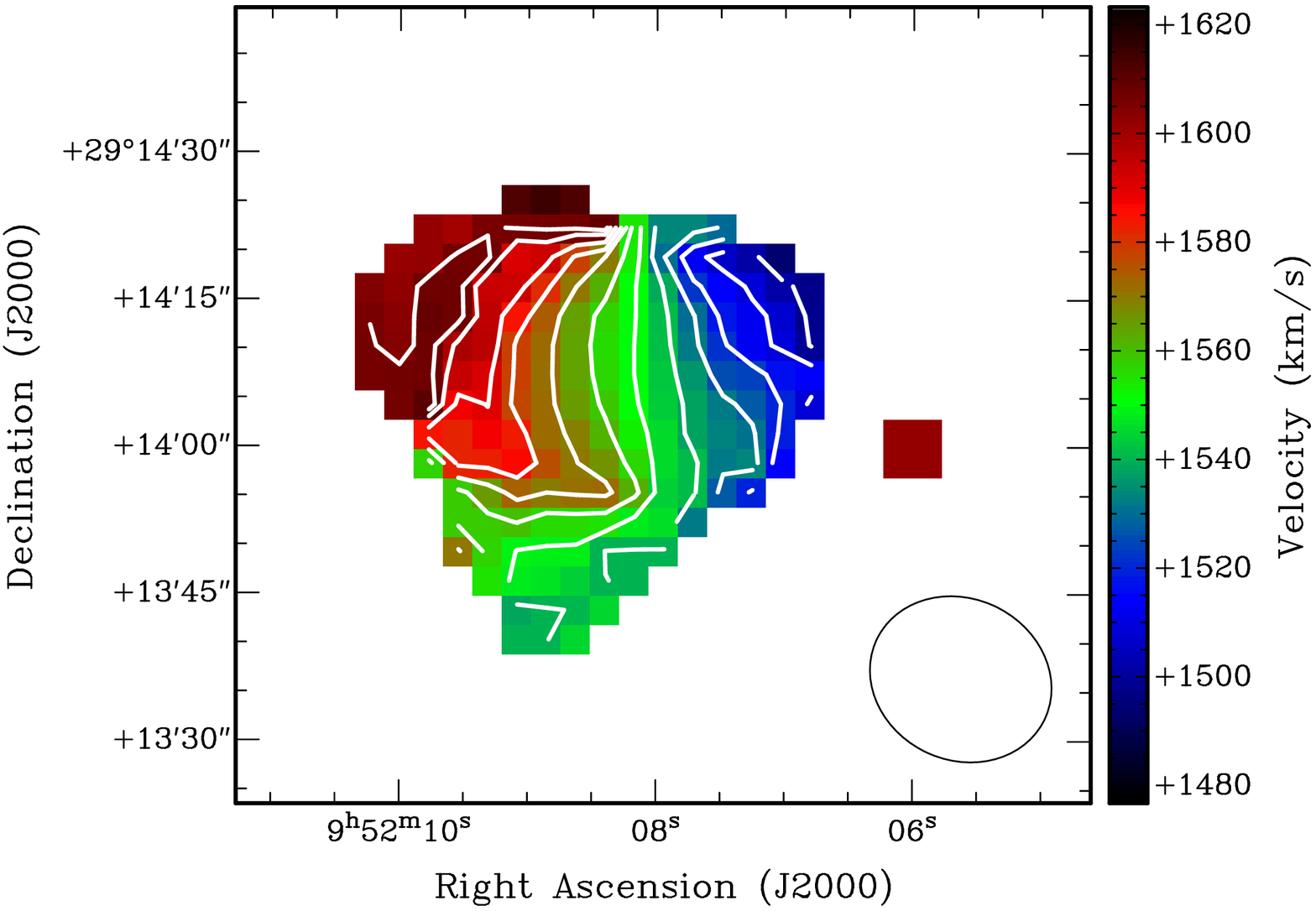}
\caption{\scriptsize  {\bf Top}:  NGC 3032:  Solid black (positive) and grey (negative) contours show the HI integrated intensity in units of -5\%, -15\%, 5\%, 15\%, 30\%, 50\%, 70\%, and 90\% of the peak (0.38 Jy  beam$^{-1}$ km s$^{-1}$=1.3$\times$10$^{21}$ cm$^2$). The ellipse shows the beam size. Green contours show the CO integrated intensity in units of $-$20\%, $-$10\%, 10\%, 20\%, 30\%, 50\%, 70\%, and 90\% of the peak (18.1 Jy beam$^{-1}$ km s$^{-1}$$=$1.6$\times$10$^{22}$ cm$^2$).  {\bf Left bottom}: HI spectrum. Constructed in a similar manner to UGC 1503.  {\bf Right bottom}: Velocity Field.  The HI intensity-weighted mean velocity (moment 1) is shown in RGB color scale and in white contours from 1500 to 1610 km s$^{-1}$ in steps of 10 km s$^{-1}$. The black ellipse shows the HI beam size. The CO resolution is 6.1\arcsec$\times$5.0\arcsec.  \label{stuff3}}
\end{figure*}

\begin{figure*}
\begin{center}
\includegraphics*[scale=0.7]{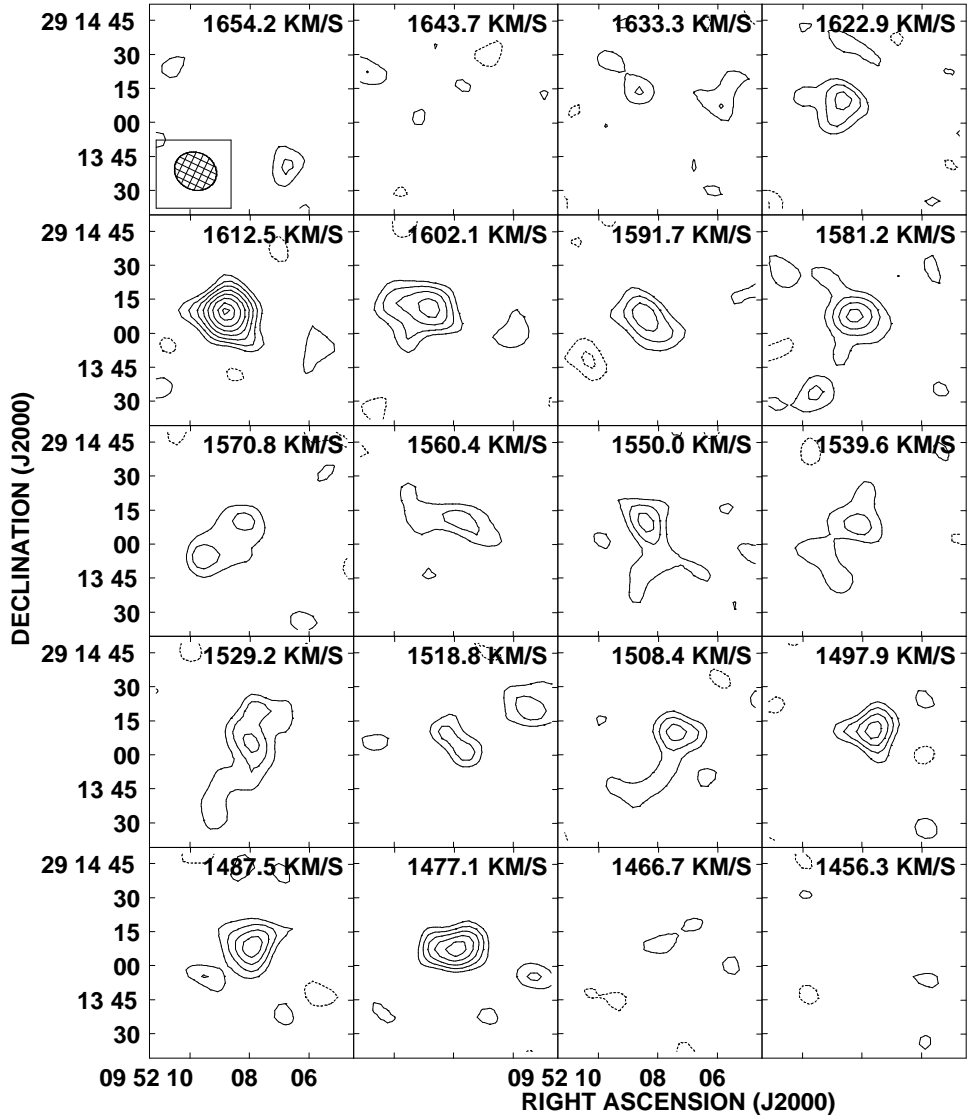}
\caption{\scriptsize NGC 3032. Individual channel maps showing HI emission.  Contour levels are -3, -2, 2, 3, 4, 5, 6, 7, 8, and 8.5 times 0.6 mJy beam$^{-1}$ $\sim$1$\sigma$. The velocity of each channel (in km s$^{-1}$) is indicated at the top of each panel and the beam size in the first panel in the bottom left corner.\label{n3032cont}}
\end{center}
\end{figure*}

\begin{figure*}
\begin{center}
\includegraphics*[scale=0.7]{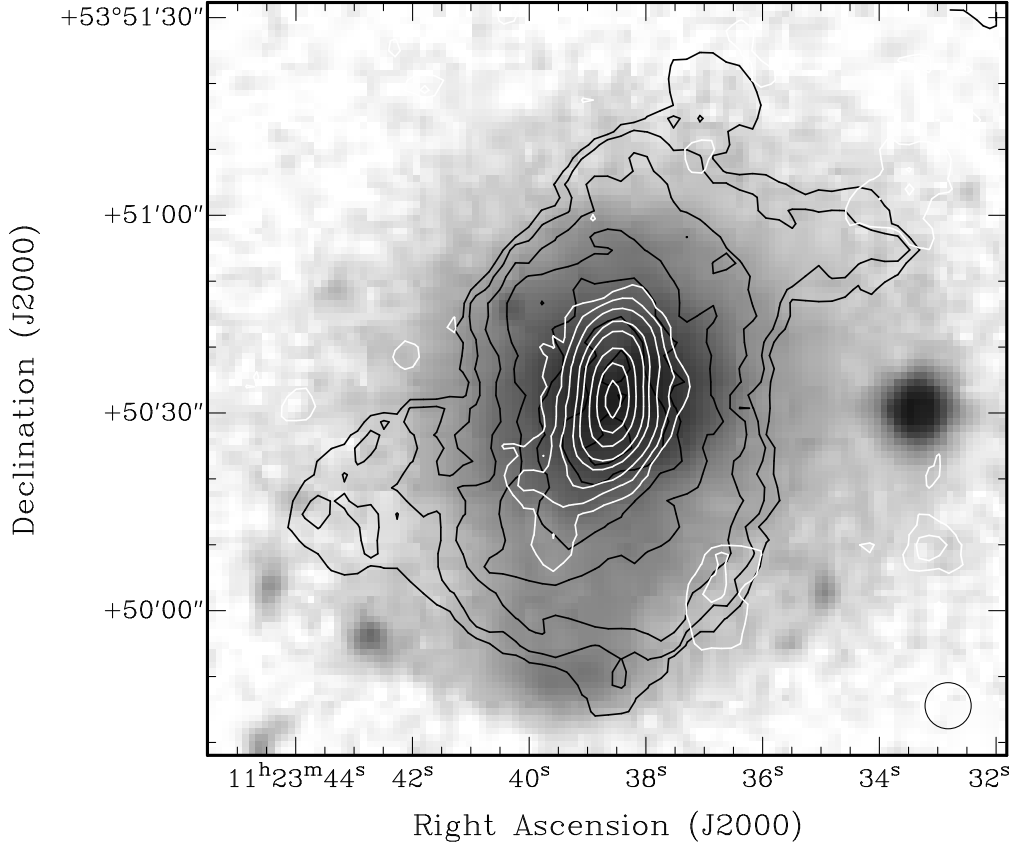}
\caption{\scriptsize NGC 3656. HI and CO overlaid on an SDSS2 R band image.  Solid black contours show the HI integrated intensity in units of 2$\%$, 6$\%$, 15$\%$, 30$\%$, 50$\%$, 70$\%$, and 90$\%$ of the peak (0.32 Jy beam$^{-1}$ km s$^{-1}=~$7.2$\times$10$^{21}$ cm$^{-2}$). The HI map is constructed from B, C, and D array VLA data (Balcells \etal\ 2001).  The circular ellipse shows the HI beam size.  Solid white contours show the CO integrated intensity in units 2$\%$, 5$\%$ 10$\%$, 20$\%$,30$\%$,50$\%$,70$\%$, and 90$\%$ of the peak (81.1 Jy beam$^{-1}$ km s$^{-1}=~$4.7$\times$10$^{22}$ cm$^{-2}$).  The CO resolution is 7.8\arcsec$\times$6.2\arcsec.  The HI spectra and velocity field are published in Balcells \etal\ 2001.  \label{n3656map}}
\end{center}
\end{figure*}

\begin{figure*}
\begin{center}
\epsscale{0.6}
\plotone{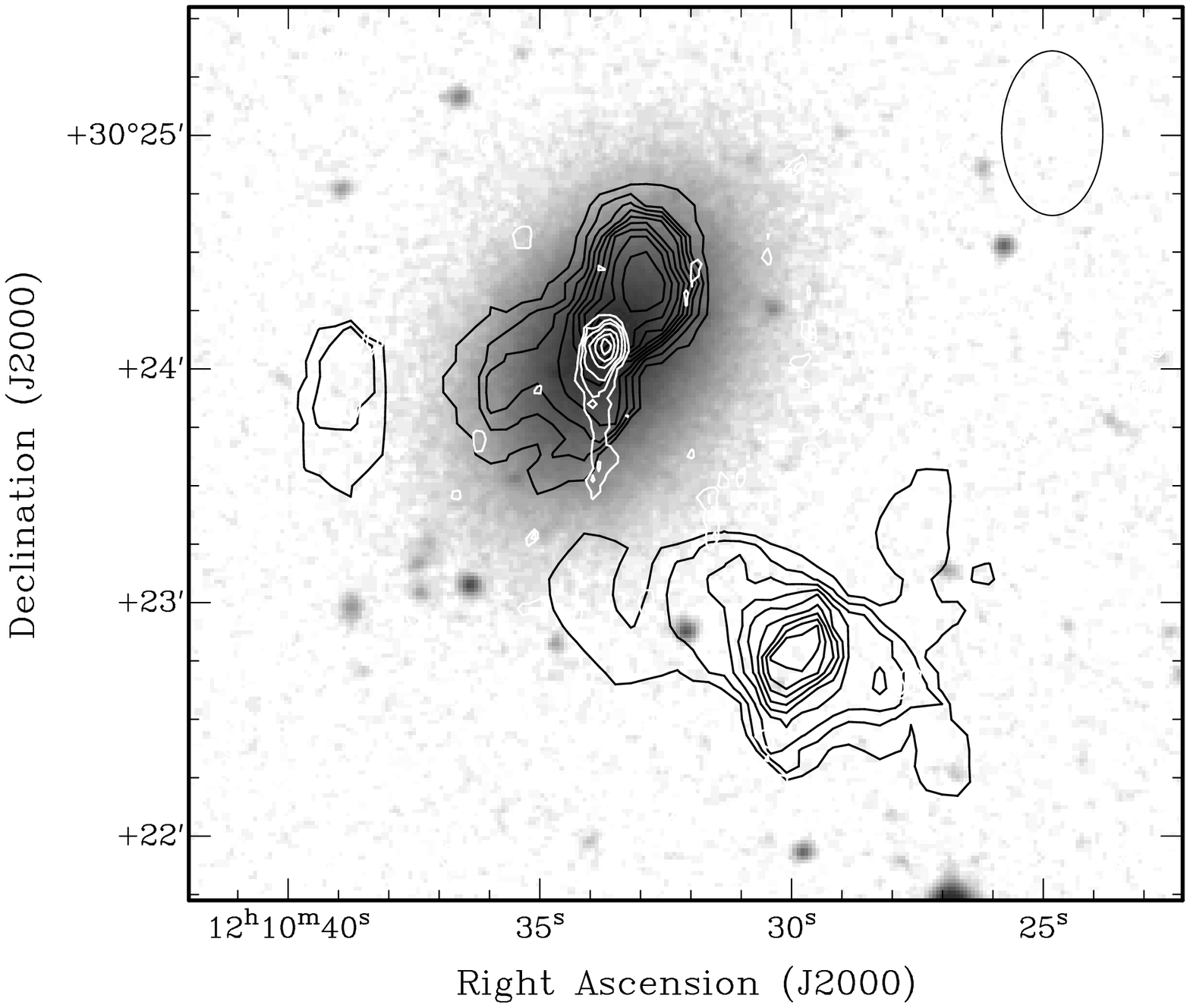}\\
\epsscale{0.9}
\plottwo{f15ba.eps}{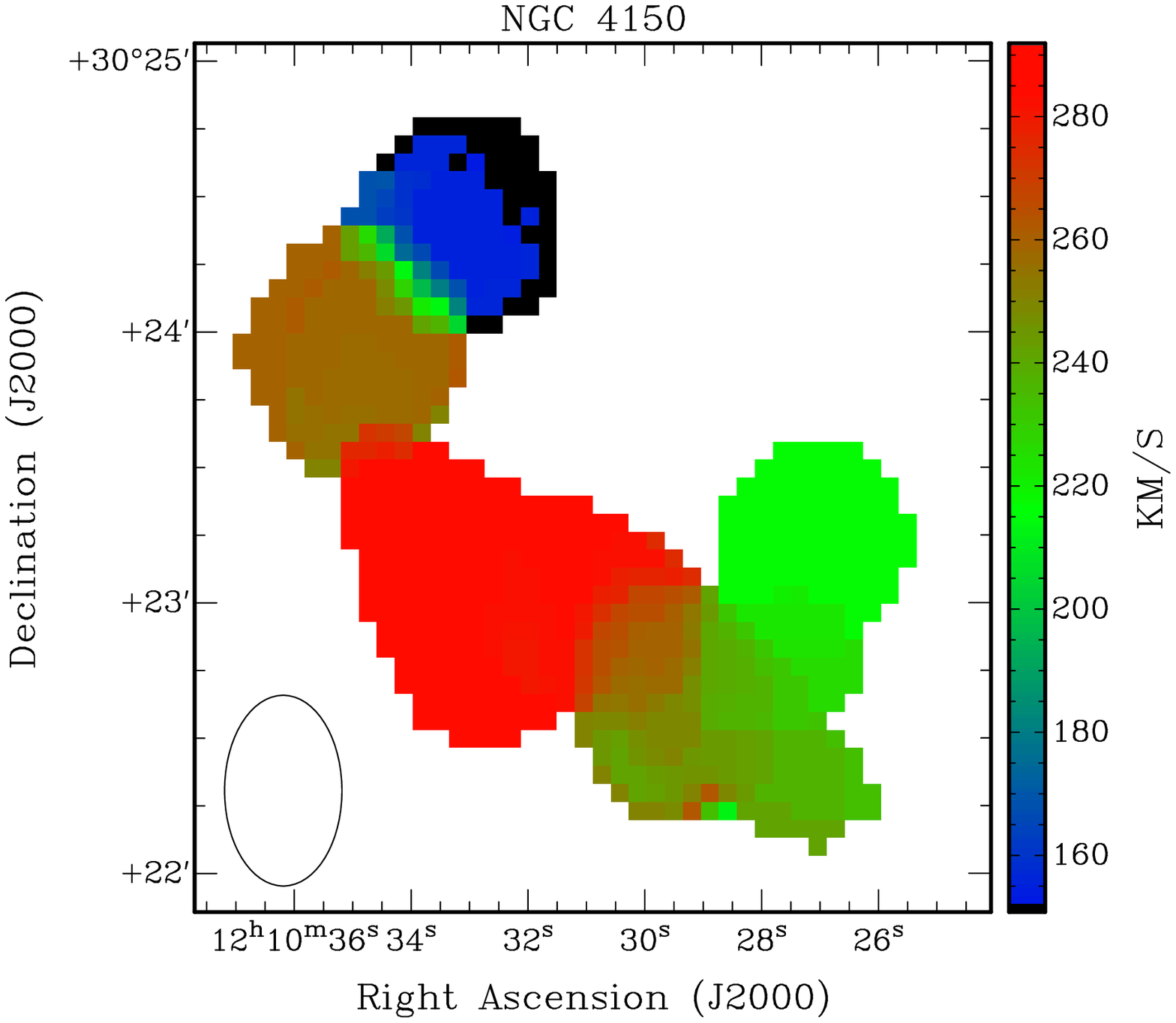}
\caption{\scriptsize {\bf Top}: NGC 4150. HI and CO. Solid white (positive) and yellow (negative) contours show the HI integrated intensity in units of 10\%, 20\%, 30\%, 40\%, 50\%, 60\%, 70\%, 80\%, and 90\% of the peak (28.9 mJy beam$^{-1}$ km s$^{-1}$=2.9x10$^{19}$ cm$^2$; Morganti \etal\ 2006).  Solid white contours show the CO integrated intensity in units of 10\%, 20\%, 30\%, 50\%, 70\%, and 90\% of the peak (17 Jy beam$^{-1}$ km s$^{-1}$$=$1.1$\times$10$^{22}$ cm$^2$). The grey scale image is an SDSS2 R-band image. {\bf Left bottom}: HI spectrum. Constructed in a similar manner to UGC 1503.  This spectra was carefully made so as to no include any of the HI emission from the unknown HI source to the south east.  {\bf Right bottom}: Velocity Field.  The HI intensity weighted mean velocity (moment 1) is shown in RGB color scale.  Note that we include the velocity structure of both NGC 4150 and the unknown HI source to the south east.  The black ellipse shows the HI beam size.  The CO resolution is 8.5\arcsec$\times$5.1\arcsec.   The WSRT HI data are formally described in Morganti et al (2006) and Oosterloo et al (2010), but the spectrum and the kinematic figures have not been published so we show them here with the permission of the original authors.\label{u4150map}}
\end{center}
\end{figure*}

\begin{figure*}
\begin{center}
\includegraphics*[scale=0.8]{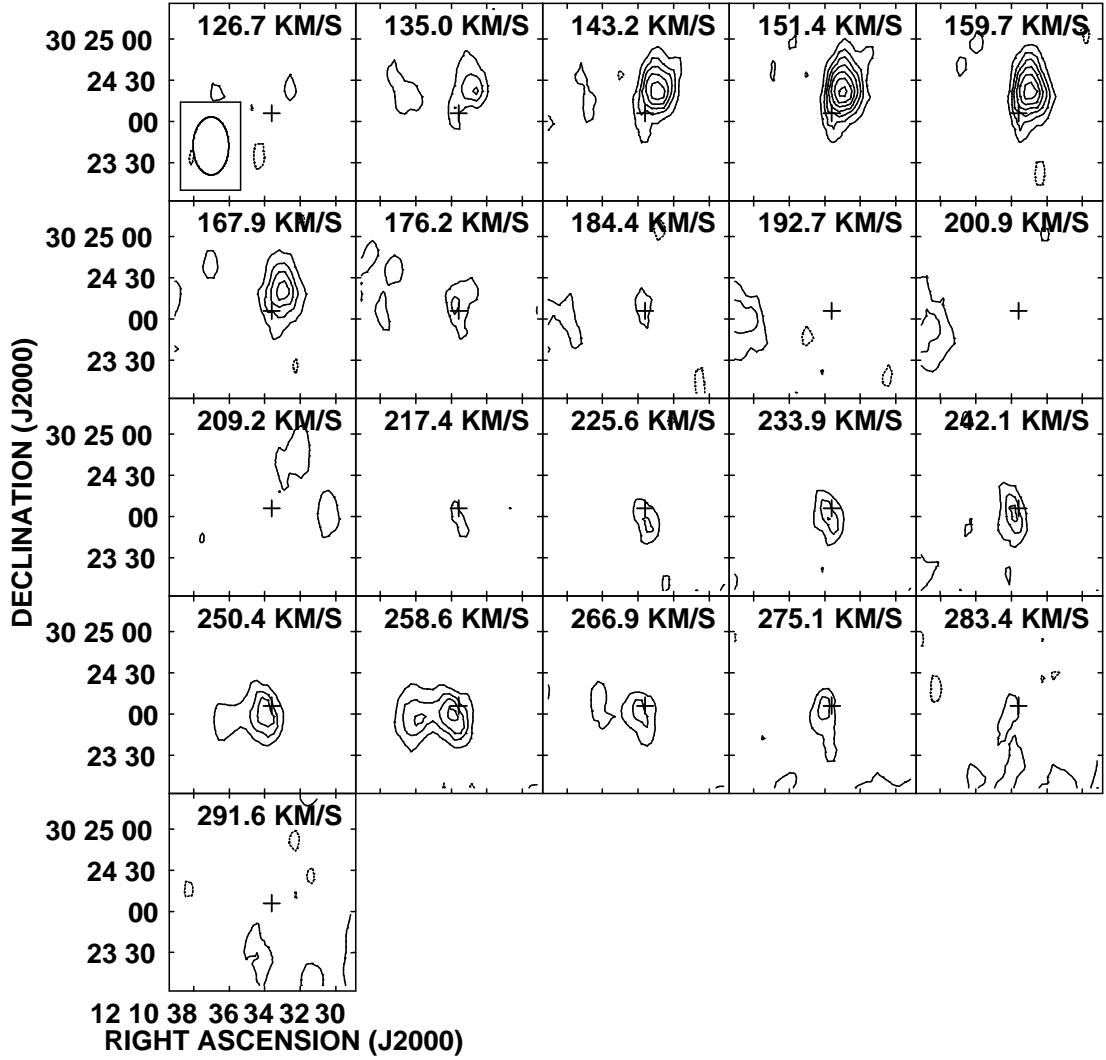}
\caption{\scriptsize NGC 4150. Individual channel maps showing HI emission from the WSRT data (Morganti \etal\ 2006).  Contour levels are -4, -2.5, 2.5, 4, 5, 6, 7, 8, 9, 10 and 11 times 0.11 mJy beam$^{-1}$ $\sim$1$\sigma$. The velocity of each channel (in km s$^{-1}$) is indicated at the top of each panel and the beam size in the first panel in the bottom left corner.  We note here that the HI emission emission for NGC 4150 is made up of two components that are separated by ~ 50 km s$^{-1}$ in velocity.  The WSRT HI data are formally described in Morganti et al (2006) and Oosterloo et al (2010), but channel maps have not been published so we show them here with the permission of the original authors. \label{n4150cont}}
\end{center}
\end{figure*}

\begin{figure*}
\epsscale{0.8}
\plotone{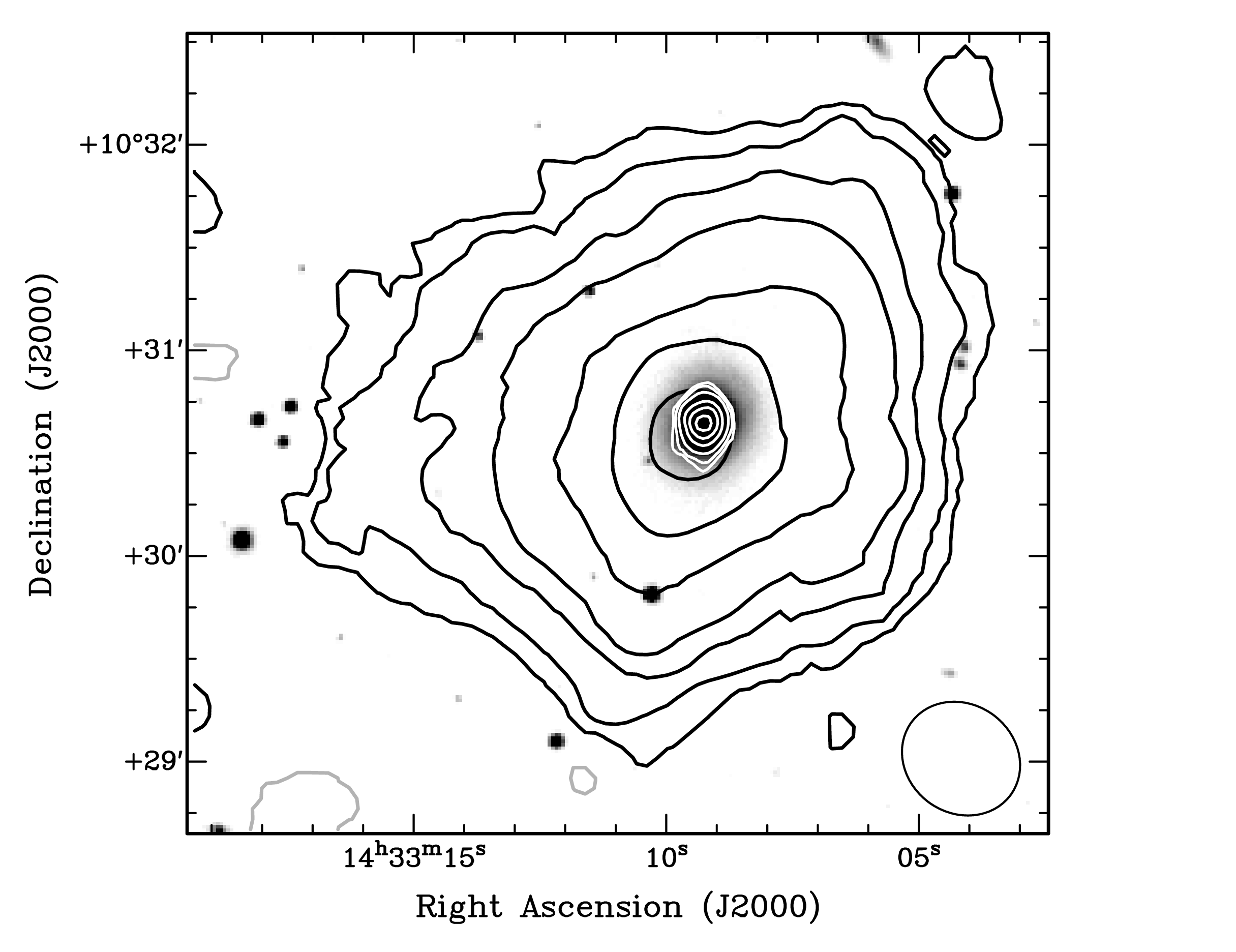}\\
\epsscale{0.9}
\plottwo{f17ba.eps}{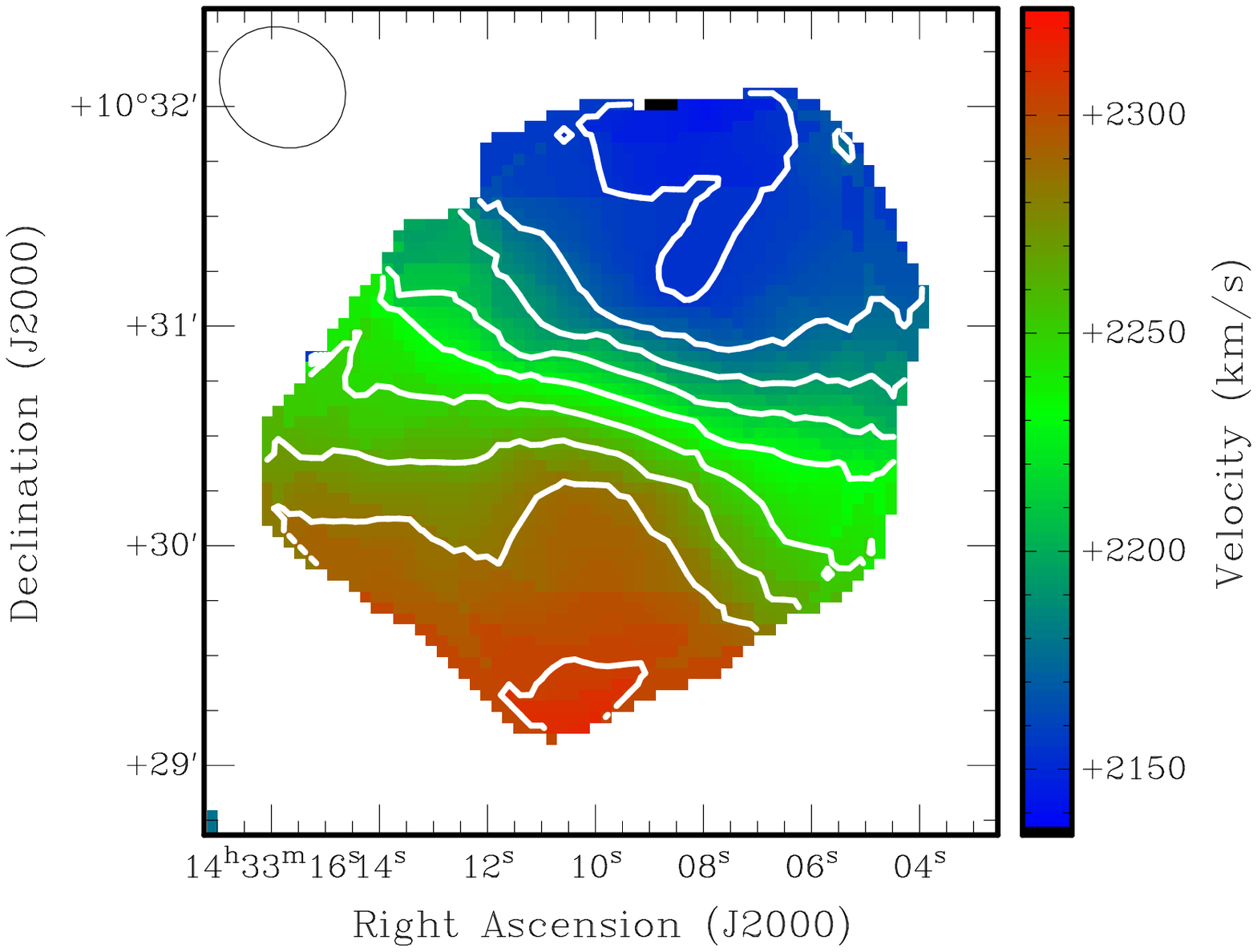}
\caption{\scriptsize NGC 5666. {\bf Top}: VLA C+D array HI overlaid with CO.  Solid black (positive) and solid grey (negative) contours show the HI integrated intensity in units of $-$10$\%$, $-$5$\%$, 5$\%$, 10$\%$, 20$\%$, 30$\%$, 50$\%$, 70$\%$, and 90$\%$ of the peak (0.94 Jy beam$^{-1}$ km s$^{-1}=~$9.1$\times$10$^{20}$ cm$^{-2}$). Solid white (positive) contours show the CO integrated intensity in units 15$\%$, 20$\%$,30$\%$,50$\%$,70$\%$, and 90$\%$ of the peak (21.3 Jy beam$^{-1}$ km s$^{-1}=~$7.5$\times$10$^{21}$ cm$^{-2}$). The grey scale image is an WYIN 3.5 meter V band image. {\bf Left bottom}: HI spectrum. Constructed in a similar manner to UGC 1503.  {\bf Right bottom}: Velocity Field.  The HI intensity weighted mean velocity (moment 1) is shown in RGB color scale and in white contours from 2136 to 2324 in steps of 20 km s$^{-1}$.  The black ellipse shows the HI beam size.  The CO resolution is 8.3\arcsec$\times$6.1\arcsec.  \label{stuff}}
\end{figure*}

\begin{figure*}
\epsscale{0.9}
\plotone{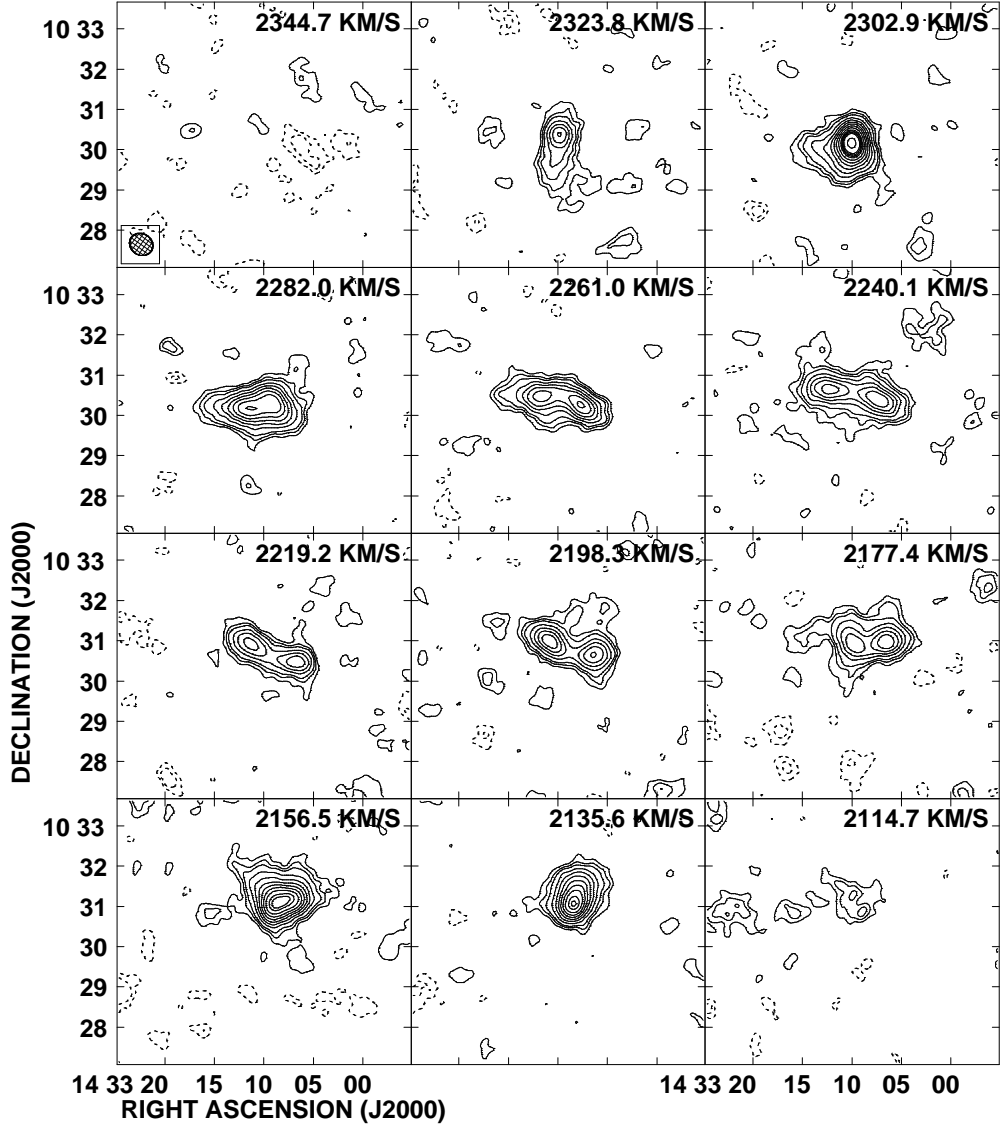}
\caption{\scriptsize NGC 5666: Individual channel maps showing HI emission.  Contour levels are $-$3, $-$2, 2, 3, 4, 6, 8, 10, 12, 14, 16, 18, 19, 20, 21, and 24 times 0.5 mJy beam$^{-1}$ $\sim$1$\sigma$. The velocity of each channel (in km s$^{-1}$) is indicated at the top of each panel and the beam size in the first panel in the bottom left corner.\label{n5666con}}
\end{figure*}

The top panels of Figures \ref{stuff6}, \ref{stuff4}, \ref{stuff3}, \ref{n3656map}, \ref{u4150map}, and \ref{stuff} show integrated HI maps overlaid with each galaxy's respective molecular gas map.  A wide-field image of the HI emission for NGC 807 is depicted in the top panel of Figure \ref{stuff5}.  Figures \ref{u1503cont}, \ref{n807con}, \ref{n3032cont}, \ref{n4150cont}, and \ref{n5666con} show the individual channel maps.  A detailed discussion of the HI morphology as compared to that of the CO is given for each galaxy below.  Table \ref{mass} gives morphological classifications of the HI according to the \atlas\ classification scheme (Serra \etal\ 2012).

{\bf UGC 1503}- This is a an isolated early-type galaxy.  The HI emission in this galaxy is detected over 288$\pm$10 km s$^{-1}$ centered on a systemic velocity of 5086$\pm$10 km s$^{-1}$.  The HI systemic velocity is in good agreement with stellar absorption line velocity measurements of 5062$\pm$43 km s$^{-1}$ from Falco \etal\ (1999) and 5098$\pm$25 km s$^{-1}$ from Wegner \etal\ (2003).  The HI systemic velocity is also in good agreement with the CO systemic velocity, 5080$\pm$15 km s$^{-1}$, after correction from the radio to the optical definition (Young \etal\ 2002).  The diameter of the HI emission is approximately 1.6\arcmin.  Like the CO emission, the HI in this galaxy is centrally peaked (0.16 Jy beam$^{-1}$ km s$^{-1}$$=9.9\times10^{20}$ cm$^{-2}$) and centered on the optical center of the galaxy.  An inspection of the integrated intensity map shows that HI is distributed into two concentric symmetric ring-like structures (Fig. \ref{stuff6}).  The central ring appears to be bisected by a bright bar like feature along the minor axis.  Circular depressions or holes are observed symmetrically oriented inside the inner ring and about the central bar like feature.  The highest surface densities are found in the bisecting bar like structure.  The inner ring like structure lies at a radius of 12\arcsec\ and has a fairly constant surface density of  $\sim$85 mJy beam$^{-1}$ km s$^{-1}$$=5.2\times10^{20}$ cm$^{-2}$.  The outer ring like structure is located at a radius of 24\arcsec\ and has a steadily declining surface density out to its edge.  Both ring like structures are about one beam width in diameter and so are unresolved.  It is interesting to note that the molecular gas does not appear to extend outside the inner edge of the inner ring.  

{\bf NGC 807}- The HI in this galaxy is detected over 468 km s$^{-1}$ centered on a systemic velocity of 4750$\pm$21 km s$^{-1}$.  The HI systemic velocity is in good agreement with stellar absorption line velocity measurements of 4747$\pm$41 km s$^{-1}$ from Falco \etal\ (2000) and 4721$\pm$30 km s$^{-1}$ from Wegner \etal\ (2003).  The HI systemic velocity is also in good agreement with the CO systemic velocity, 4728$\pm$25 km s$^{-1}$, after correction from the radio to the optical definition (Young \etal\ 2002).  The lower resolution map made from the combined AD and C array data gives the appearance of a smooth central HI disk with two tidal arms extending to the north west and south east along the major axis.  The higher resolution C array data shows that the HI distribution is actually very clumpy and complicated (see the top panel of Figure \ref{stuff5}).  The HI in this galaxy is distributed about equally between the inner disk and the outer tidal arms with $\sim5.5\times10^9 M_{\odot}$ in each component.  The diameter of the HI emission out to the edge of the tidal arms is approximately 7.4\arcmin.  The inner disk has a radius of $\sim$2.7\arcmin\ and appears to be distributed asymmetrically with about 60$\%$ of the HI located to the northwest.  The opposite is true of the tidal arms.  The bulk of the HI in the tidal arms, $\sim70\%$, is located in the arm to the southeast. The HI intensity is centrally peaked (0.72 Jy beam$^{-1}$ km s$^{-1}$$=7.7\times10^{20}$ cm$^{-2}$) and is roughly coincident with the CO intensity peak.  However, the intensity peaks are slightly offset ($\sim$ 5-9\arcsec) from the optical center of the galaxy. Interestingly, the CO distribution is also asymmetric, but the bulk of the CO emission ($\sim$70$\%$) is actually located to the southeast.  

{\bf NGC 2320}- NGC 2320 is a member of the Abell 569 cluster of galaxies.  The HI in this galaxy is detected only in absorption.  The absorption may be due to a small disk of HI which may not be large enough to be seen in emission due to the presence of the 17 mJy continuum source.  There is a small extension of CO emission south west of the galaxy center which also has roughly the same velocity as the HI absorption peak (6032-6282 km s$^{-1}$; See Figure 12 of Young 2005).  This indicates that the absorbing HI and the CO extension could be part of the same gas complex which is actively falling in toward the center of the galaxy. 

{\bf NGC 3032}- NGC 3032 is a low luminosity field lenticular with a small counter rotating stellar disk of radius of $\sim$2\arcsec\ at its center (McDermid \etal\ 2006a).  The HI in this galaxy is detected over 146 km s$^{-1}$ centered on a systemic velocity of 1550$\pm$5 km s$^{-1}$, so the HI systemic velocity is in good agreement with the CO systemic velocity (Young \etal\ 2009) as well as with stellar absorption line velocity measurements of 1555$\pm$41 km s$^{-1}$ from Falco \etal\ (2004) and 1559$\pm$10 km s$^{-1}$ from Emsellem \etal\ (2004).  The CO and HI line widths also match well.  The C array data gives the appearance of a smooth HI disk.  It is apparent from the individual channel maps that the HI structures are not well resolved.  The diameter of the HI emission is approximately 1\arcmin\ thus it is only slightly more extended than the CO emission.  There is an extension of the HI gas ($\sim 6.8\times10^6 M_{\odot}$, roughly 10$\%$ of the total mass) outside the main disk to the south. This small amount of HI is oriented in roughly the same direction as a similar extension of the CO.  The HI intensity is centrally peaked (0.38 Jy beam$^{-1}$ km s$^{-1}$$=1.3\times10^{21}$ cm$^{-2}$) and roughly coincident with the optical and CO emission peaks. 

{\bf NGC 3656}- The HI data published by Balcells \etal\ (2001) shows that the HI in NGC 3656 is located in a nearly edge-on, warped minor-axis gaseous disk 7 kpc in diameter.  The rest is situated outside the galaxy in what appear to be two tidal tails or perhaps a disrupted outer HI disk or ring.  Balcells \etal\ (2001) note that the velocity structure of the HI distribution is asymmetric, indicating non-circular motions, and they suggest that this is due to recent or ongoing accretion.  Their derived position angle of the inner HI disk is $170^o$ and closely matches the position angle derived from the Young's (2002) CO data.  Balcells \etal\ (2001) HI systemic velocity of 2850$\pm$11 km s$^{-1}$ is in good agreement with the CO systemic velocity of 2840$\pm$15 km s$^{-1}$ (Young 2002) as well as with a stellar absorption line velocity measurement of 2890$\pm$11 km s$^{-1}$ from Rothberg \& Joseph (2006).  The CO and HI line widths also match well.  The HI intensity is centrally peaked (0.32 Jy beam$^{-1}$ km s$^{-1}$$=6.6\times10^{21}$ cm$^{-2}$) and roughly coincident with the optical and CO emission peaks.

{\bf NGC 4150}-  NGC 4150 is a small nearby lenticular galaxy located on the outskirts of the Virgo Cluster at a projected distance of 18 degrees from M87 (Karachentsev \etal\ 2003).  Morganti \etal\ (2006) quote an WSRT HI mass of 2.5$\times$10$^6$ M$_\odot$, and describe the morphology of the HI inside the galaxy as that of a rotating disk with a diameter of about 1\arcmin.  A close inspection of the integrated intensity map reveals that the HI distribution in the disk is highly asymmetric.  The bulk of the HI ($\sim 60\%$) is located in the Northwest portion of the stellar disk.  The HI peaks at the center (30.4 mJy beam$^{-1}$ km s$^{-1}$$=$3.04$\times$$10^{19}$ cm$^{-2}$) and again at a radius of 17\arcsec\ along the major axis to the Northwest.  There is a small unresolved clump of HI at the center of the disk that appears to be spatially coincident with the CO disk.  The HI gas in the southern part of the disk has column densities up to 5 times lower than those in the northern and central components. The low surface density of the gas and the asymmetric distribution of the HI imply that the gas is currently being accreted from an outside source.  Single dish observations show that the HI and CO have matching line width (Welch \& Sage 2003).  

{\bf NGC 5666}- This is an isolated, low-luminosity field early-type galaxy.  The HI emission in this galaxy is detected over 210$\pm$11 km s$^{-1}$ centered on a systemic velocity of 2219$\pm$10 km s$^{-1}$.  The HI systemic velocity is in good agreement with stellar absorption line velocity measurements of 2222$\pm$40 km s$^{-1}$ from Falco \etal\ (1999) and 2224$\pm$8 km s$^{-1}$ from Wegner \etal\ (2003) and the CO systemic velocity of 2225$\pm$15 km s$^{-1}$ (Young 2002).  The CO and HI line widths also match well.  The lower resolution map made from the combined AD and C array data gives the appearance of a smooth HI disk.  The diameter of the HI emission is approximately 3.2\arcmin.  There is a slight asymmetry in the HI distribution in that the spatial extent of the HI is larger on the eastern side of the galaxy (i.e. the contours of the integrated intensity are father apart in the east).  Like the CO emission, the HI in this galaxy is centrally peaked (0.94 Jy beam$^{-1}$ km s$^{-1}$$=9.1\times$10$^{20}$ cm$^{-2}$) and centered on the optical center of the galaxy. An inspection of the higher resolution C array data shows that the HI distribution is actually clumpy, with the overall appearance of a ring with a bright clump/disk of HI at its center.   The HI ring begins at a radius of $\sim45$\arcsec.  

The HI detected sample galaxies fit well into the \atlas\ HI morphological classification scheme.  The HI detected sample galaxies can be divided up into three categories: (D) Very extended regularly rotating HI disks (UGC 1503 and NGC 5666).  (d) small regularly rotating HI disks that don't extend outside the optical body of the galaxy (NGC 2320, NGC 3032, and NGC 4150), and (u) unsettled HI morphologies (NGC 807 and NGC 3656).  In the case of NGC 2320 we don't actually know the true extent of the HI emission, but we think it likely to be distributed like the CO, in a compact disk within the optical body of the galaxy.  We prefer to classify NGC 807 as an intermediate between the D and u \atlas\ subclasses, because the kinematics of the HI are surprisingly regular despite obvious tidal features.  We have no galaxies in our sample where the HI is found to be in small scattered external clouds.  This could be due to the fact that the galaxies are biased toward the most CO rich objects.  

None of the HI detected galaxies exhibit central HI holes.  In all cases except for NGC 807 the HI and CO emission is centrally peaked.  In the case of NGC 807 the HI and CO emission peaks are offset from the optical center but still coincide with each other.  All of our sample HI detected galaxies have peak HI column densities at or above 10$^{21}$ cm$^{-2}$ except for NGC 4150 which peaks at 2.9$\times$10$^{19}$ cm$^{-2}$.  This is consistent with the peak HI column densities derived from the recent single dish surveys.  Serra \etal\ (2012) note that the early-type galaxies in their sample with HI column densities of 10$^{21}$ cm$^{-2}$ all have the largest amounts of molecular gas, complex and prominent dust distributions, and strong evidence of star formation.  The same is true for our HI detected sample galaxies except for NGC 2320 where there is no evidence of star formation (Young \etal\ 2007, Crocker \etal\ 2011).  Serra \etal\ (2012) find that 50\% of the very extended HI disks (class D) differ in orientation and kinematics from their stellar components, whereas the small HI disks (class d) are tightly coupled to their host galaxy.  Our sample only exhibits the later trend.  We see no misalignments between the HI major axis and the stellar isophotal major axis in any of our sample D galaxies.  This could be an additional result of selection bias toward the CO richest objects or simply to the small sample size.   
\section{HI Versus CO Kinematics}\label{kin}
\begin{deluxetable*}{lcccccccc}
\tabletypesize{\scriptsize}
\tablecaption{\bfseries HI Kinematic Position Angles and Dynamical Masses \label{test}}
\tablehead{
\colhead{Galaxy}  &\colhead{PA} &\colhead{i}          &\colhead{V$\sin(i)$}      &\colhead{V}              &\colhead{R$_{HI}$}   &\colhead{M$_{dyn}$}           &$\frac{M(HI)}{M_{dyn}}$    &$t_{orb}$\\
\colhead{}     &\colhead{($^\circ$)}   &\colhead{($^\circ$)}  &\colhead{(km s$^{-1}$)}    &\colhead{(km s$^{-1}$)}   &\colhead{(kpc)}     &\colhead{10$^{10}$M$_\odot$}   &\colhead{}                &\colhead{($10^8$ yr)}}
\startdata
U1503  &$-$125        &51(4)         &144(5)                &185(12)              &16(1)          &13                                   &0.016                                     &5.3\\
N807     &146       &65(6)         &234(10)              &258(17)              &42(2)          &65                                    &0.016                                  &10\\
N3032   &90       &44(2)         &73(5)                  &105(8)                 &2.5(0.3)     &0.78                                 &0.010                                   &1.8\\
N3656    &191      &74-90         &213(11)              &218(8)                 &8.0(0.2)      &8.9                                   &0.022                                  &2.3\\
N5666    &162      &27(9)         &105(11)             &231(75)               &14.5(0.9)   &18                                   &0.010                                   &3.9\\
\enddata
\tablecomments{Position angles are derived using the AIPS task GAL.}
\end{deluxetable*}
\begin{figure*}
\epsscale{1.0}
\plotone{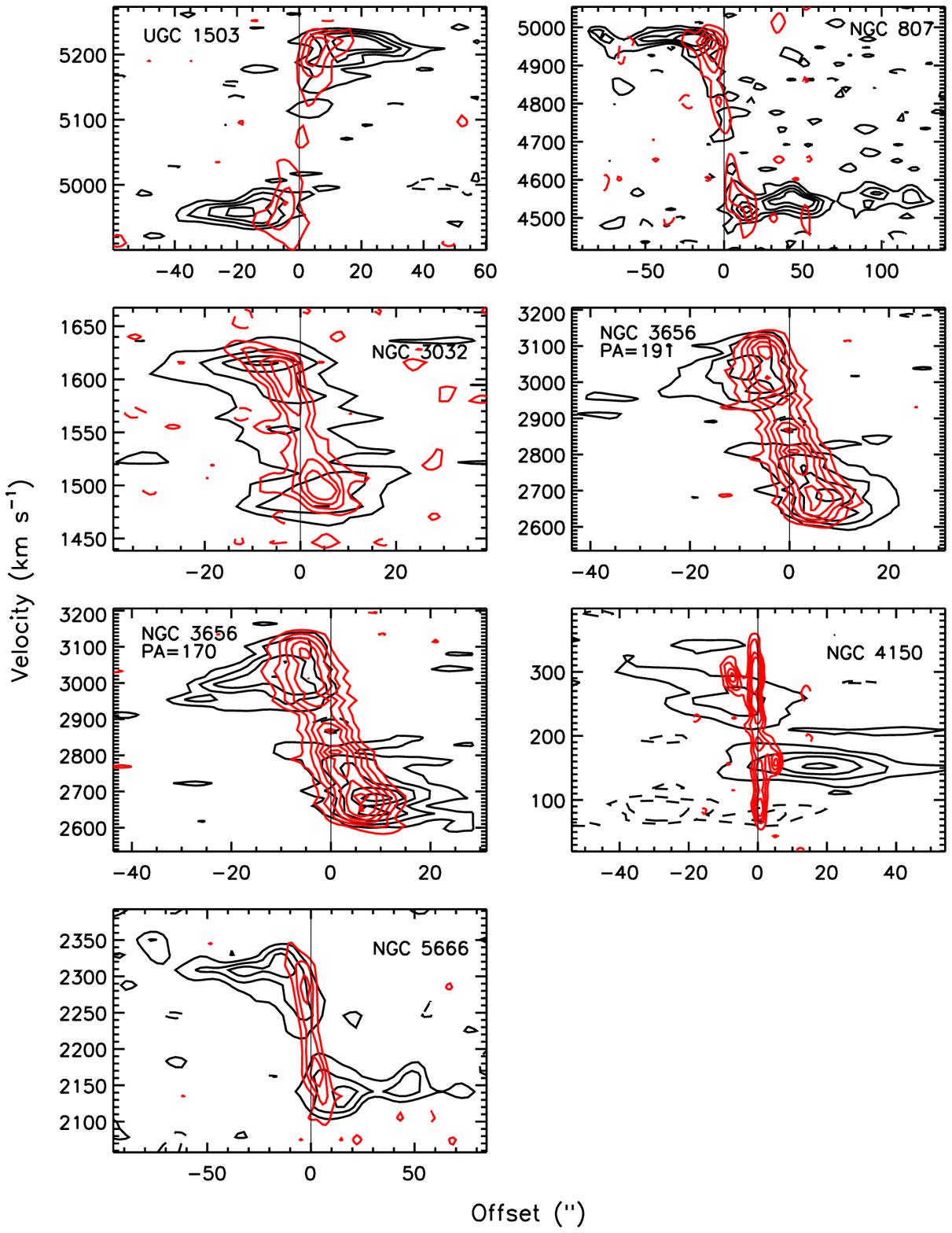}
\caption{\scriptsize Position velocity curves. CO (red contours) and HI (black contours).  The data cubes are sliced through the galaxy nucleus at the kinematic position angles derived from the GAL fits to the HI velocity maps (see Table \ref{test} and section \ref{kin}).  The derived HI kinematic position angles are within a couple of degrees of the those derived from the CO data.  The HI contour levels are all ($-$2, 2, 4, 6, ...) times the rms noise from Table \ref{HIparam}.  The CO contour levels are all ($-$2, 2, 4, 6, ...) times the rms noise in Tables 2, 3, and 3 of Young (2002), Young (2005), and Young \etal\ (2007), respectively.  In the case of NGC 3656  the contours are multiplied  by an additional factor of 1.3.  \label{posvel}}
\end{figure*}
Global kinematic position angles, inclinations, and systemic velocities are derived by fitting exponentially rising model rotation curves to the HI velocity fields depicted in the bottom right of Figures \ref{stuff6}, \ref{stuff4}, \ref{stuff3}, \ref{u4150map}, and \ref{stuff} using the National Radio Astronomy Observatory's AIPS task GAL.  In the case of NGC 3656 we produced a HI velocity field by obtaining and re-reducing the VLA B and C array data using similar techniques as in Balcells \etal\ (2001); the data are mapped with a Briggs robust parameter of 1.0 giving a beam size of 11.6\arcsec$\times$10.2\arcsec\ and an rms noise level 0.18 mJy beam$^{-1}$ in a 21 km s$^{-1}$ channel. In the case of NGC 3032 the spatial resolution is sufficiently poor that the exponential fits are not robust.  A solid body fit to their velocity fields proved more stable for NGC 3032.  Estimates of the position angle of the kinematic major axis, inclination angle, and systemic velocities derived from these fits are consistent with values derived from the dust and CO maps of these five galaxies (Young \etal\ 2008; Young \etal\ 2009).  

The GAL fits are weighted by the integrated intensity, so the fitted position angle is actually the kinematic axis at small radii. In order to get a good idea of how the position angle changes with radius, constant velocity rotation profiles were fitted to the velocity fields using concentric annuli.  Annuli of 30\arcsec\ are used for NGC 807 and annuli 10\arcsec\ are used for the galaxies with smaller HI diameters.  The constant velocity method also produces fitted position angles which agree with those derived from the CO to within a few degrees.  The HI position angles of the major axis for all the galaxies discussed in this section are constant with increasing radii (within the margin of error), and never vary more than $\sim$5-10 degrees.  There is a hint of a possible warp suggested by the slight presence of an integral sign shape in the zero velocity curve of the HI velocity field for NGC 3032 (Bottom left panel of Fig. \ref{stuff3}).  This integral sign feature may also be present in the CO velocity field of NGC 3032 (Young \etal\ 2008).  However, the spacial and velocity resolution are poor for both HI and CO data sets.  

In the case of NGC 4150 not only is the spacial resolution poor, but the distribution of the velocities is also highly non-uniform (see Figure \ref{u4150map}c).  The AIPS task GAL assumes a circularly rotating disk and so may not give correct results for the HI in NGC 4150.  For this galaxy we employed the kinemetry techniques of Krajnovic \etal\ (2006) to measure the position angle as a function of radius.  The measured position angles for NGC 4150 derived from this technique inside 12\arcsec\ are consistent with the stellar morphological and kinematic positions angles as well as the kinematic position angle derived from CO measurements. 

Position-velocity plots for UGC 1503, NGC 807, NGC 3656, NGC 4150, and NGC 5666 are constructed using the kinematic centers and position angles derived from the model fits described above (see Figure \ref{posvel}).  Both the HI (black contours) and CO (red contours) data cubes are sliced through the galaxy nucleus at the kinematic position angle listed in Table \ref{test}.  The position-velocity diagrams for UGC 1503, NGC 807 ,and NGC 3032 show steeply rising, approximately solid body rotation regions in their centers followed by a flattening at the edges of the HI distribution.  In the case of NGC 5666 the position-velocity plot represents the superposition of two separate components, an inner HI disk (radius of $<$25\arcsec) and an outer HI ring (radius of $>$25\arcsec).  The HI in the inner ring does not appear to extend past the region of solid body rotation while the material in the outer ring has only a flat constant velocity component.  It is interesting to note that the material in the ring appears to be rotating 20 km s$^{-1}$ or 15$\%$ slower than the material at the edge of the inner HI disk.  The appearance of slower rotation of the ring could be due to a central mass concentration as is observed for the Milky Way or perhaps to a change in the orientation of the velocity field.  However, the GAL constant velocity fits for NGC 5666 show that the position angle in the vicinity of the ring does not change more than a few degrees from that of the disk component.  Thus, it is not clear whether these two components are kinematically distinct. 

In the case of NGC 3656 we use two position angles to construct the position-velocity plots in Figure \ref{posvel}. The angle of 191 degrees is derived from the CO map, weighted towards the inner part of the CO, and there is some evidence of a warp as the outer regions tend towards position angles less than 180 degrees.  Balcells \etal\ (2001) use 170 degrees and that position angle also cuts through a prominent HI shell in the southern part of the galaxy.  The resulting position-velocity overlays produce obvious signatures of non-circular gas motions in the outer portions of the HI disk.  As in the other cases the CO does not appear to extend past the solid body portion of the rotation curve and appears to be for the most part kinematically settled.  The HI also appears to be kinematically settled within the CO radius.  

The position-velocity overlays show that the kinematics of the CO and the HI match for UGC 1503, NGC 807, NGC 3032, NGC 3656, and NGC 5666  which strongly suggests a common origin for the two gas phases in these five galaxies.  The position-velocity plot of NGC 4150 shows that the HI and CO share the same sense of rotation, but have very different line widths.  This is likely due to the fact that much of the HI in this galaxy is currently unsettled due to a recent interaction (section \ref{morph}).  We do not have an HI velocity map for NGC 2320, however its gas morphology as discussed in section \ref{morph} also implies a common origin between both gas phases.  

The dynamical mass interior to the HI disk's outer edge can be calculated using the observed gas velocities: 
\begin{equation}
M_{dyn}=(2.33\times10^5 M_{\odot})V^2R
\end{equation}
where R is the radius of the outer edge in kpc and V is the observed velocity in km s$^{-1}$, corrected for inclination.  This is assuming that the gas disks are intrinsically circular, and that the gas itself moves along circular orbits.  The implied dynamical masses (Table \ref{test}) range from a few $\times10^{9}M_{\odot}$ to a few $\times$$10^{11}M_{\odot}$ interior to the edge of the HI disk, and the observed masses of atomic gas are a few percent of these dynamical masses.  Table \ref{test} also gives the orbital time for gas at the edges of the HI disks.   
\section{Discussion}\label{disc}
\subsection{The Effect of Environment on the HI content}\label{envi}
As in other recent HI early-type surveys we find that the most HI-deficient galaxies either reside in cluster/group environments, show indications of recent interactions or both.  The most HI rich galaxies reside in the field, but several of these galaxies also show indications of recent interactions.  

The deficiency of atomic gas in clusters galaxies is thought to be due to mechanisms such as (1) tidal interactions, (2) ram pressure (Gunn \& Gott 1972), (3) turbulent/viscous stripping (Nulsen 1982), and (4) fast evaporation of the atomic gas due to the presence of a hot intracluster medium (Grossi \etal\ 2009).  Grossi \etal\ (2009) suggest that these mechanisms would also aid in limiting the number of HI rich dwarfs in clusters, thus limiting the available reservoirs of gas from which other larger galaxies can feed.  In the three HI-undetected lenticulars (NGC 4459, NGC 4476, NGC 4526) there is no indication in the CO or optical data of a recent tidal interaction (Young 2002, Young \etal\ 2008).  Therefore, we conclude that the HI in these galaxies has most likely been removed by one or a combination of the last three mechanisms mentioned above.  Additionally, current photodissociation models imply that a sudden increase in pressure will increase the gas surface density such that large portions of a galaxy could spontaneously become molecular without converting the diffuse gas to self-gravitating clouds (e.g. Elmegreen 1993).  This is perhaps a plausible explanation for why no HI is detected over the entirety of the CO disks of these three galaxies.  We will explore this possibility in more depth in paper II (Lucero \& Young in prep). 

NGC 4150 is detected in HI despite being a Coma group member.  The presence of HI in this galaxy can be explained by the fact that it resides near the edge of the cluster where the ICM densities are probably lower (i.e. ram pressure and evaporation are much less effective), and the population of HI-rich dwarfs is probably higher.  Indeed it appears likely that NGC 4150 is accreting gas from a nearby low surface brightness dwarf galaxy.  The HI surface densities of both NGC 4150 and the unknown HI source peak at $\sim$0.2 M$_{\odot}$ pc$^{2}$ (corrected for helium and inclination) which is more than 25 times smaller than the lowest peak surface density of the HI detected field galaxies.  This implies that the stripping/evaporation mechanisms are either still affecting the gas content at a reduced efficiency or only existed for a shorter period of time compared to those cases where the HI has been completely stripped. 

Many of the same mechanisms discussed above could also be at work in group environments, and could also explain the HI deficiency  in NGC 83 and NGC 2320.  Both galaxies are located relatively nearby to strong sources of X-ray emission.  NGC 83 is a member of the NGC 83 Group.  This group is very near (in velocity and spatially) to the galaxy cluster PCC S49-147.  A GIS X-ray mosaic image shows that the NGC 83 group could be embedded in the diffuse X-ray intracluster medium associated with PCC S49-147 (Nakazawa \etal\ 2007a).  NGC 2320 is about 2$^o$ or 2.9 Mpc in projection from NGC 2329 which lies at the center of the Abell 569 cluster.  The Abell 569 cluster was recently observed with the Einstein satellite (Burns \etal\ 1994).  The X-ray emission in this cluster is relatively compact, is quite bright, and is centered on NGC 2329.   

NGC 83 and NGC 2320 are the most luminous galaxies in the present sample.  Grossi \etal\ (2009) suggest that the presence of a hot X-ray ISM in luminous early-type galaxies may prevent an atomic phase from forming altogether or cause the HI (whether it originates from mass loss or accretion) to evaporate.  Unfortunately, neither NGC 2320 and NGC 83 have been searched deeply for hot gas, and no evidence of a hot ISM has been reported for either galaxy in the literature outside of what has been discussed above.  The blue luminosities of NGC 83 and NGC 2320 are well in the range in which O'sullivan \etal\ (2001) find copious amounts of hot gas.  The lower luminosity sample galaxies live in the lower range where the X-ray luminosity is dominated by emission from X-ray binary systems rather than hot gas (see Figure 9 of O'sullivan \etal\ 2001).

Interestingly, the CO properties of the HI-deficient galaxies do not differ that much from their field counterparts.  The only difference worth mentioning is that the cluster/group galaxies have larger peak H$_{2}$ surface densities than both the group and field galaxies.  One might expect the density of the molecular gas in the center of cluster/group galaxies to be increased due to ICM/ram pressures (Nakanishi \etal\ 2006).  We will explore the effects of external pressures on the gas surface density in depth in paper II (Lucero \& Young in prep).  Of course the larger peak H$_{2}$ surface densities could be a resolution effect as all of the cluster galaxies are 2 to 5 times nearer than the other galaxies in the sample.
\subsection{The Origin of the Cold Gas in Early-Type Galaxies}\label{ori}
We find an overwhelming amount of evidence that suggests that both the neutral hydrogen and molecular gas in our early-type galaxy sample share a common origin be it external or internal.  The kinematics of the HI and CO gas phases in the sample galaxies are remarkably similar.  The HI and CO line widths, intensity peaks, and kinematic position angles match in all cases. We find that if the HI is in a relaxed disk, so is its CO.  Alternatively, if the HI appears disturbed, so does its CO.   Thus, our results are consistent with the \atlas\ volume limited survey which also find that the two gas phases always exhibit the same kinematics.  Other early-type galaxies with differing CO and HI linewidths (e.g. Welch \& Sage 2003; Sage \& Welch 2006; Sage \etal\ 2007; Welch \etal\ 2010) are probably cases in which the HI is either still falling in or being affected by tidal or ram pressure stripping.  The existing interferometric data shows that the HI is often extended over tens of kiloparsecs while the CO is confined to the inner 1-4 kpc.  The kinematics of the HI and CO always match where the two phases overlap.  Outside the CO radius the HI often exhibits a range of other morphologies and kinematics. 
  
As in the previous HI surveys, the total gas content (HI$+$H$_{2}$), HI content, and  H$_{2}$ content only weakly correlate with galaxy luminosity.  This is an indication that the cold gas may have an external origin.  Indeed, a large number (45$\%$) of the sample galaxies show evidence that their cold gas has been obtained or is currently being accreted from outside sources.  These include NGC 2320, NGC 3032, NGC 3656, NGC 4150, and NGC 4476.  The most obvious cases are NGC 3032 and NGC 4476.  The cold gas systems (HI and CO) in these two galaxies are counter-rotating with the bulk of the stars, and therefore an internal origin can be ruled out (Young \etal\ 2008; Crocker \etal\ 2010).  Additionally, small extensions of HI and CO in NGC 3032 and NGC 4150 oriented in roughly the same direction imply that this gas is actively falling onto these galaxies.  NGC 2320 contains an absorbing HI complex and an extension of CO emission which have similar velocities redshifted with respect to the systemic velocity indicating gas is actively falling in toward the galaxy.  The radio continuum emission in NGC 2320 is thought to be powered by an AGN.  Perhaps the cold gas in NGC 2320 is helping to fuel the central AGN activity (Young 2005). Tidal features (optical and HI) and the non-circular motions implied by the velocity structure of the HI in NGC 3656 also suggest that the cold gas in this galaxy has been acquired during a recent interaction (Balcells \etal\ 2001).    

The origin of the cold gas in NGC 83, UGC 1503, NGC 807, NGC 4459, NGC 4526, and NGC 5666 is less clear.  The kinematics of the HI and or CO disks in the field early-types UGC 1503 and NGC 5666 suggest that the gas/dust disks in these systems are well settled into dynamic equilibrium.  The orbital timescales at the edge of the HI disks in UGC 1503 and NGC 5666 are 0.5 Gyr and 0.4 Gyr respectively.  If the cold gas in these two galaxies was acquired from an external source it must have occurred several orbital timescales ago.  There is no overwhelming evidence for an outside origin of the molecular gas for the cluster/group galaxies NGC 83, NGC 4459, and NGC 4526 except perhaps slight asymmetries in the CO distributions of NGC 4459 and NGC 4526 (Young \etal\ 2008).  The generally symmetric appearance of the optical images of these three galaxies also suggests that it has been well over a Gyr since any mergers, if any, have occurred.  One then might conclude that their cold gas originated solely from mass loss.  However, Davis \etal\ (2010) suggest that the tell tale signatures of an external origin for the cold gas may get erased by environmental processes such as ram pressure or by galaxy wide processes in the most massive objects such as NGC 83.  If true, prograde gas cannot completely rule out an external origin of the cold gas in these particular galaxies.  The HI in NGC 807 appears to be in a relaxed disk out to a radius of $\sim$1.4\arcmin, and then outside this radius the HI appears to be tidally disrupted.  A deep optical image taken by the WYIN telescope also shows the presence of faint stellar tidal arms oriented in a similar manner to that of the HI tidal features (bottom panel of Fig \ref{stuff5}).  The kinematics of the HI and CO in NGC 807 are surprisingly regular.  The uniformity of the kinematics probably mean that the cold gas was already distributed in a relaxed regular rotating structure like those observed in NGC 5666 and UGC 1503 before the interaction occurred.  This of course does not rule out the possibility that the cold gas in NGC 807 was acquired a few orbital timescales before the most recent interaction.  A better way to distinguish between external and internal gas origins would be to look for significant differences in angular momentum between the cold gas and the stars.  Unfortunately, we lack the necessary stellar kinematic information to carry out this type of analysis for these particular galaxies.  
\section{Conclusions}\label{concl}
In this paper we present an analysis of new VLA and archive HI observations of 11 CO rich early-type galaxies as well as a preliminary comparison between the HI and CO morphologies and kinematics.  The new HI observations have 2 to 16 times better resolution than that of the recent volume limited HI surveys of early-type galaxies.  In summary:

Six of the eleven sample galaxies (UGC 1503, NGC 807, NGC 2320, NGC 3032, NGC 3656 and NGC 4150) are detected in HI; the other five are not, even though they have strong CO detections.  The HI detected galaxies have a wide range of HI masses (1.4$\times$10$^{6}$ to 1.1$\times$10$^{10}$ M$_{\odot}$).  There does not appear to be any correlation between the HI mass or morphology (E versus S0). 

HI absorption is detected for NGC 2320.  The peak of this absorption is red shifted with respect to the optical systemic velocity by 60 km s$^{-1}$  and may be associated with a small extension of CO emission which is at a similar offset velocity as the HI.  We suggest that the absorbing HI and the CO extension are probably part of the same gas structure which is actively falling toward the center of the galaxy.

The HI in UGC 1503, NGC 3032, NGC 4150, and NGC 5666 is distributed in disk-like structures in regular rotation with diameters of a few to 16 kpc.  The HI emission for these galaxies peaks at the disk center ($5-9\times10^{20}$ cm$^{-2}$).  There are no kpc sized HI holes like those observed in early-type spiral galaxies.  NGC 807 and NGC 3656 have a significant amount of HI located in tidal arms far from the main disk.  The cold gas kinematics of NGC 807 are surprisingly regular despite the fact that the gas has recently been disturbed.  

The relatively high resolution of the HI maps presented in this work enable more detailed comparisons of CO and HI kinematics than have previously been possible for early-type galaxies.  In the regions where both phases are detected, they show identical kinematics within our measurement errors.  In the three cases where HI traces a clear turnover in the circular rotation speed, the molecular disks extend out to the turnover (so they would be useful in a Tully-Fisher analysis) but not far beyond that point (so the molecular gas is strongly concentrated in the solid-body part of the rotation curve).

A little over half of the early-type sample galaxies show evidence that their cold gas has been obtained or is currently being accreted from outside sources.  We find no evidence in the present early-type galaxy sample to support the suggestion that the HI and CO have different origins.  There is also no indication that the origin of the cold gas is dependent on galaxy morphology (E versus S0), stellar luminosity or environment.

As in the recent volume limited surveys of early-type galaxies, there is a clear correlation with environment and HI.  All of the HI non-detected galaxies reside in group or cluster environments.  The galaxy with the smallest detected HI mass (NGC 4150) resides at the edge of the Virgo Cluster where ICM densities are expected to be low.   

 Overall, the results of these observations are consistent with the recent HI surveys which also find a mix of HI morphologies and masses, low HI peak surface densities, and a paucity of HI in early-type galaxies which reside in high density environments.  

The fact that all of the isolated field galaxies are detected in HI is consistent with the assumption that once an early-type has obtained some cold gas, it is much easier to hold on to it in low density environments.  The very regular kinematics and highly extended HI disks in UGC 1503 and NGC 5666 imply that their gas systems are relatively old, at least a few orbital time scales (on the order of a few Gyrs).  NGC 807, NGC 3032, and NGC 3656 show indications that their cold gas has recently been acquired or disturbed.  This is good evidence to support the proposition that there is still a significant amount of cold gas in field environments on which early-type galaxies can feed.

The conclusions presented in this work are based on a relatively small sample of galaxies biased toward early-type galaxies known to contain substantial amounts of molecular gas. Clearly, we must extend our analysis to a much larger and more homogenous sample of early-type galaxies.  
\acknowledgments
We thank the reviewer for his/her thorough review and highly appreciate the comments and suggestions, which significantly contributed to improving the quality of the publication.  We are grateful to Jacqueline van Gorkom for providing the HI map and of NGC 3656 as well as Raffaella Morganti and Tom Oosterloo for providing the WSRT HI map and image cube for NGC 4150.  This work is based on observations collected with the Very Large Array operated by the National Radio Astronomy Observatory.  The National Radio Astronomy Observatory is a facility of the National Science Foundation operated under cooperative agreement by Associated Universities, Inc.  This research has also made use of the NASA/IPAC Extragalactic database (NED) which is operated by the Jet Propulsion Laboratory, California Institute of Technology, under contract with the National Aeronautics and Space Administration.  This work is partially supported by NSF grants AST-0507432 and AST-1109803.
{\it Facilities:} \facility{VLA}.
\appendix
\section{HI Detection of sources in the vicinity of NGC 2320 and NGC 3032}
In this section we report new VLA observations of several HI sources detected in the fields of NGC 2320 and NGC 3032.  HI images, spectra, velocity fields, and individual channel maps can be found in Figures \ref{stuff2}-\ref{stuff11}.  The properties of the HI detections for these sources can be found in Table \ref{field1}.  A description of the data reduction can be found in section \ref{obsre}.  All of these field sources are outside the field of view the BIMA CO observations of Young (2002, 2005) and Young \etal\ (2008).
\subsection{NGC 2321}\label{aa}
\begin{figure*}
\epsscale{0.7}
\plotone{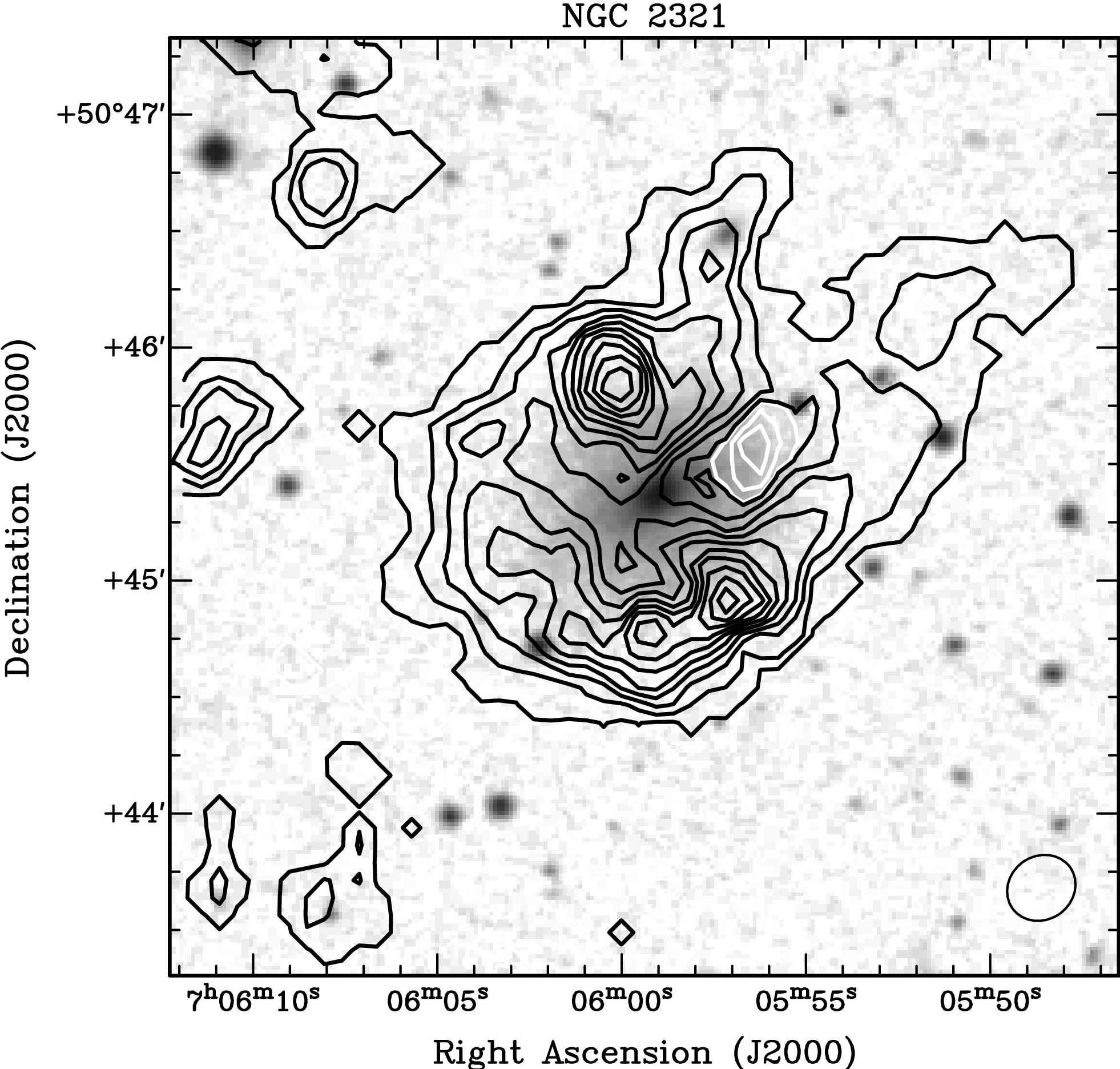}
\epsscale{1.0}
\plottwo{f20ba.eps}{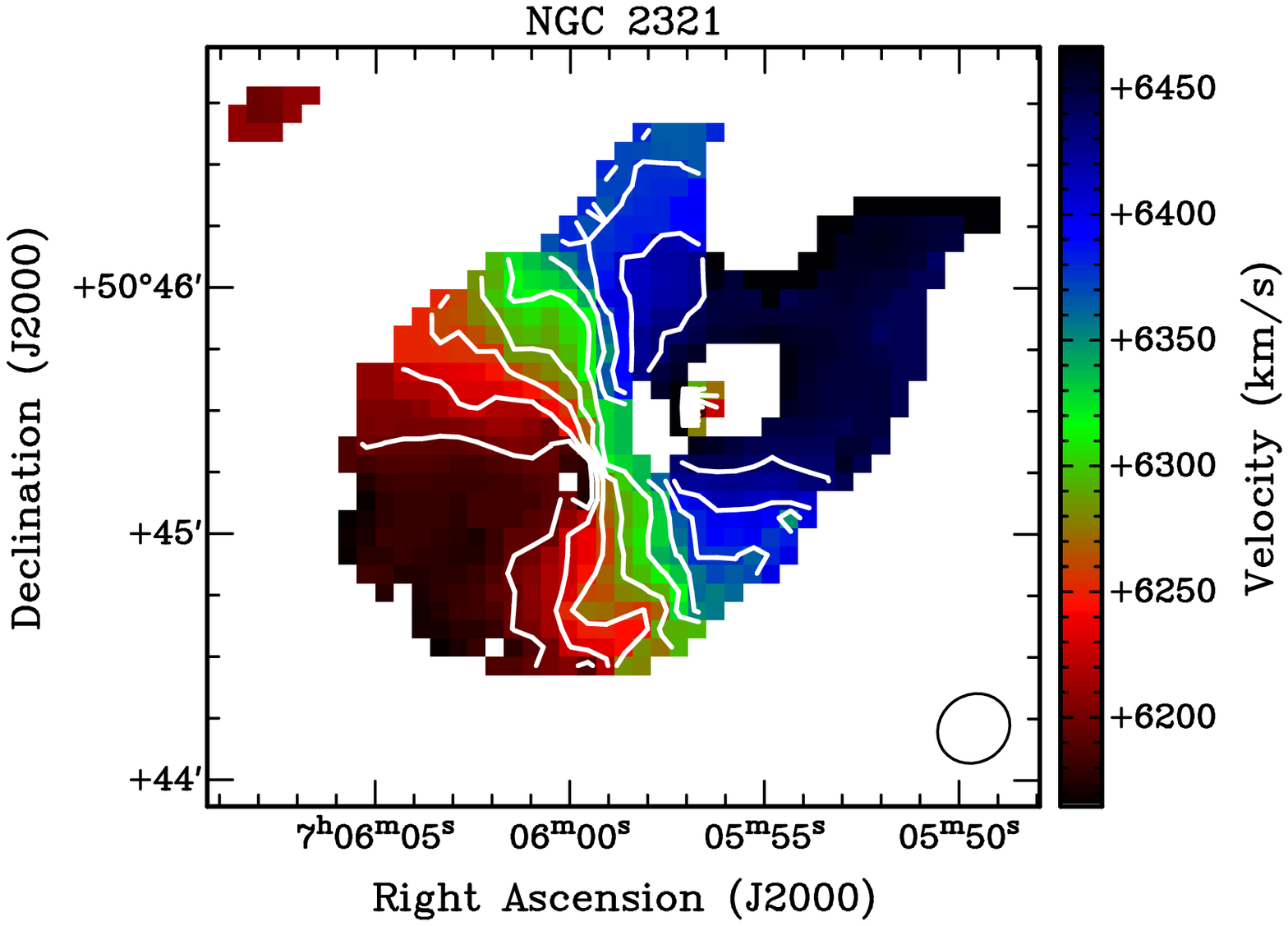}
\caption{\scriptsize  {\bf Top}: Solid black (positive) and solid white (negative) contours show the HI integrated intensity in units of $-20\%$, $-10\%$, $-$1$\%$, 1$\%$, 10$\%$, 20$\%$, 30$\%$, 40$\%$, 50$\%$, 60$\%$, 70$\%$, 80$\%$ and 90$\%$ of the peak (0.94 Jy beam$^{-1}$ km s$^{-1}=~$9.1$\times$10$^{20}$ cm$^{-2}$).  The grey scale image is an SDSS2 red band image. {\bf Left bottom}: HI spectrum. Constructed in a similar manner to UGC 1503.  {\bf Right bottom}: Velocity Field.  The HI intensity weighted mean velocity (moment 1) is shown in RGB color scale and in white contours from 6150 to 6450 km s$^{-1}$ in steps of 21 km s$^{-1}$.  The black ellipse shows the beam size. \label{stuff2}}
\end{figure*}
\begin{figure*}
\epsscale{0.75}
\plotone{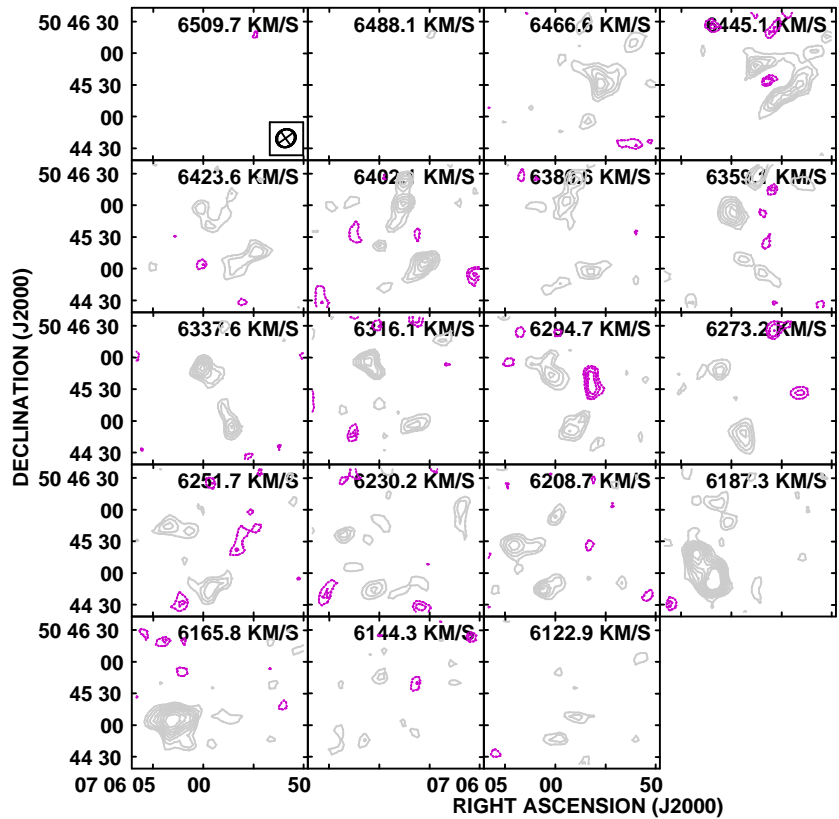}
\caption{\scriptsize NGC 2321: Individual channel maps showing HI emission.  Contour levels are -5, -4, -3, 3, 4, 5, 6, 7, 8, 9, and 10 times 0.3 mJy beam$^{-1}$ $\sim$1$\sigma$.  Negative contours are purple.  The velocity of each channel (in km s$^{-1}$) is indicated at the top of each panel and the beam size in the first panel in the bottom left corner. \label{n2321cont}}
\end{figure*}
NGC 2321 is classified as a barred spiral galaxy in the RC3 catalogue.  NGC 2321 is also a likely member of the Abell 596 cluster as its systemic velocity (6318 km s$^{-1}$) is well within the velocity dispersion of that cluster ($\sim$300 km s$^{-1}$).  NGC 2321 is about 10' from NGC 2320 and about 8' from the cluster center in projection.

Figures \ref{stuff2} and \ref{n2321cont} show the integrated intensity, the HI spectrum, velocity map, and the individual channel maps.  The total HI flux is 2.6$\pm$0.3 Jy km s$^{-1}$ which translates to an HI mass of 4.1$\times$10$^{9}$ M$_{\odot}$.  The HI emission is detected over 301$\pm$11 km s$^{-1}$ centered on a systemic velocity of 6316$\pm$11 km s$^{-1}$ which gives a circular velocity of V$_{circ}$$\sin{i}=$151 km s$^{-1}$.  This flux is consistent with single dish HI observations made with the meridian transit Nancay radiotelescope (Theureau \etal\ 1998) which quote an HI flux of 2.3$\pm$1.0 Jy km s$^{-1}$.

The channels maps show the characteristic butterfly pattern, and so the HI is probably distributed in a disk in semi-regular rotation.  The HI is not centrally peaked, but peaks inside the disk.  In the northeastern part of the disk there is a marginally resolved region of what appears to be HI in absorption (peak $-$68 Jy beam$^{-1}$ km s$^{-1}$ at RA 7h 05m 56.357s Dec 50d 45m 35.30s).  However, there is no evidence of any continuum source at the position of the absorption.  There is a 2.9 mJy source listed in the NVSS about 1.4\arcmin\ to the northeast (Condon \etal\ 1998) which is clearly visible in a radio continuum maps made from the new VLA data, but can't be associated with the absorption.  An inspection of the channel maps shows that there is significant negative emission in only two channels, and these channels are quite far apart in velocity space.  We conclude that the negative emission is a 5 sigma artifact due to problems with imaging a source that is located significantly far from the pointing center and continuum subtraction (similar to those of NGC 4459, see section 2.3).  We attempted to use various other continuum subtraction techniques, but are unable to produce a data cube without the artifact.  It is surprising that all of the flux is recovered, despite the artifact.   
\begin{deluxetable*}{lccccccc}
\tabletypesize{\scriptsize}
\tablecaption{\bfseries HI properties of Galaxies in the Field of NGC 2320 and NGC 3032 \label{field1}}
\tablehead{
\colhead{Galaxy}                  &\colhead{morph}          &\colhead{RA}             &\colhead{DEC}          &\colhead{V$_{helio}$}      &D           &\colhead{S$_{\nu,HI}$}                &M(HI)\\
\colhead{}                            &\colhead{}                   &\colhead{(J2000)}       &\colhead{(J2000)}      &\colhead{(km s$^{-1}$)}    &(Mpc)     &\colhead{(Jy km s$^{-1}$)}          &\colhead{($\times$10$^8$M${_\odot}$)}}
\startdata
NGC 2321                           &SBa                           &07h05m59.0s             &50d45m22s              & 6318(13)                         &87         &2.3(0.2)                                     &41\\
KUG 0950$+$295                &Irr                              &09h52m57.3s             &29d18m38s              &1628(42)                           &22         &0.94(0.10)                                 &1.1\\
DS96 0949$+$2935$^a$      &dwarf                          &09h52m43.8s             &29d20m53s              &1456(16)                           &20         &1.2(0.1)                                     &1.1\\
Unknown$^b$                     &?                                &09h52m52.2s             &29d21m08s              &1602(10)                            &22         &0.2(0.02)                                   &0.23\\
\enddata
\tablecomments{Morphological types, positions and systemic velocities are taken from NED if available.  The distance is calculated from the systemic velocity and H$_{o}=$73 km s$^{-1}$ Mpc$^{-1}$. \\
No correction is made for the presence of helium or inclination effects in the HI masses.\\
$^a$ Systemic velocity measured from the HI line observations with the Arecibo 300 meter telescope (Duprie \& Schneider 1996)\\
$^b$ Position and systemic velocity measured from the HI line from the VLA C array observations presented in this paper.}
\end{deluxetable*}
\subsection{KUG 0950+295 and DS96 0949+2935}
\begin{figure*}
\epsscale{0.50}
\plotone{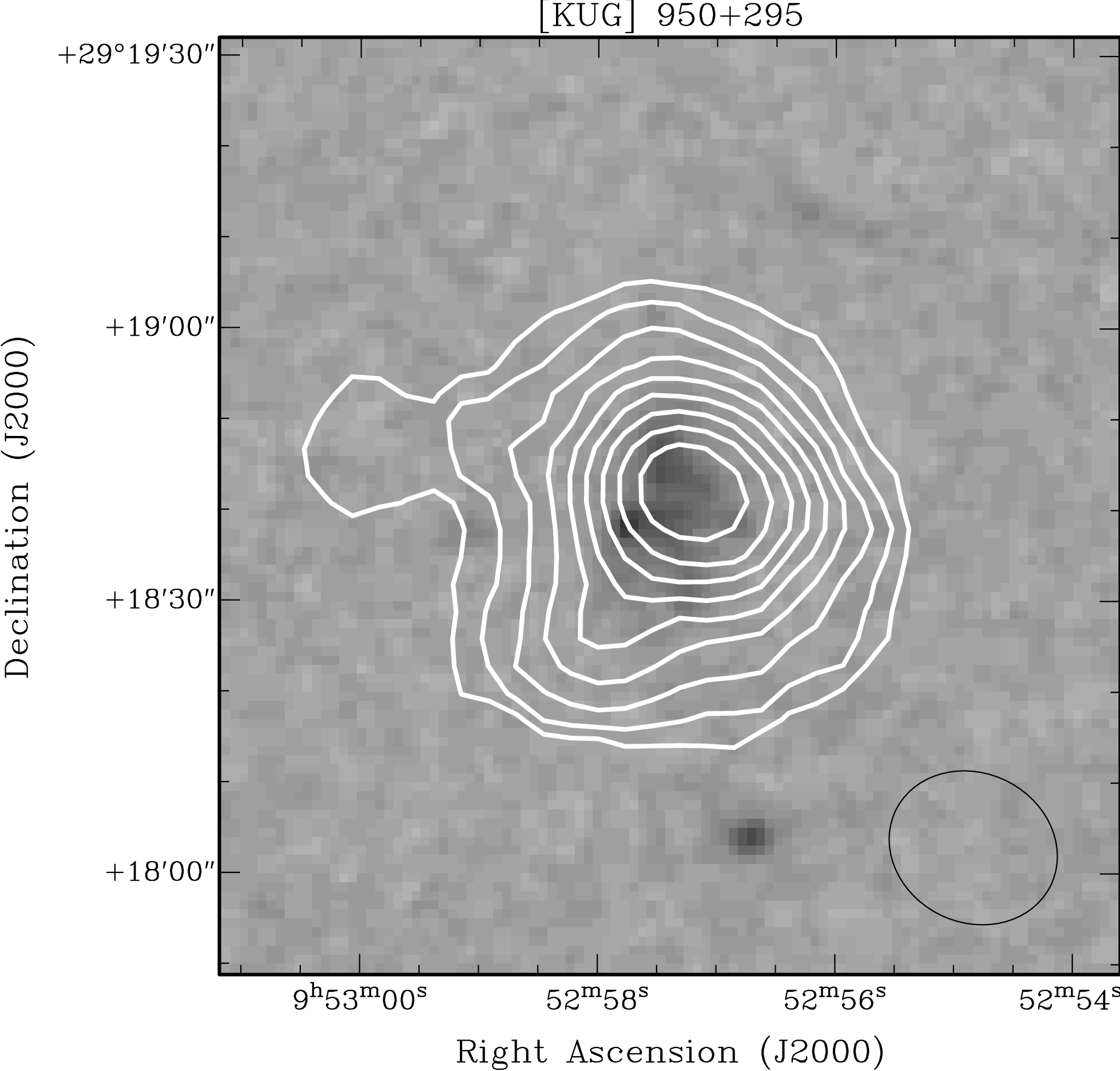}\\
\epsscale{0.9}
\plottwo{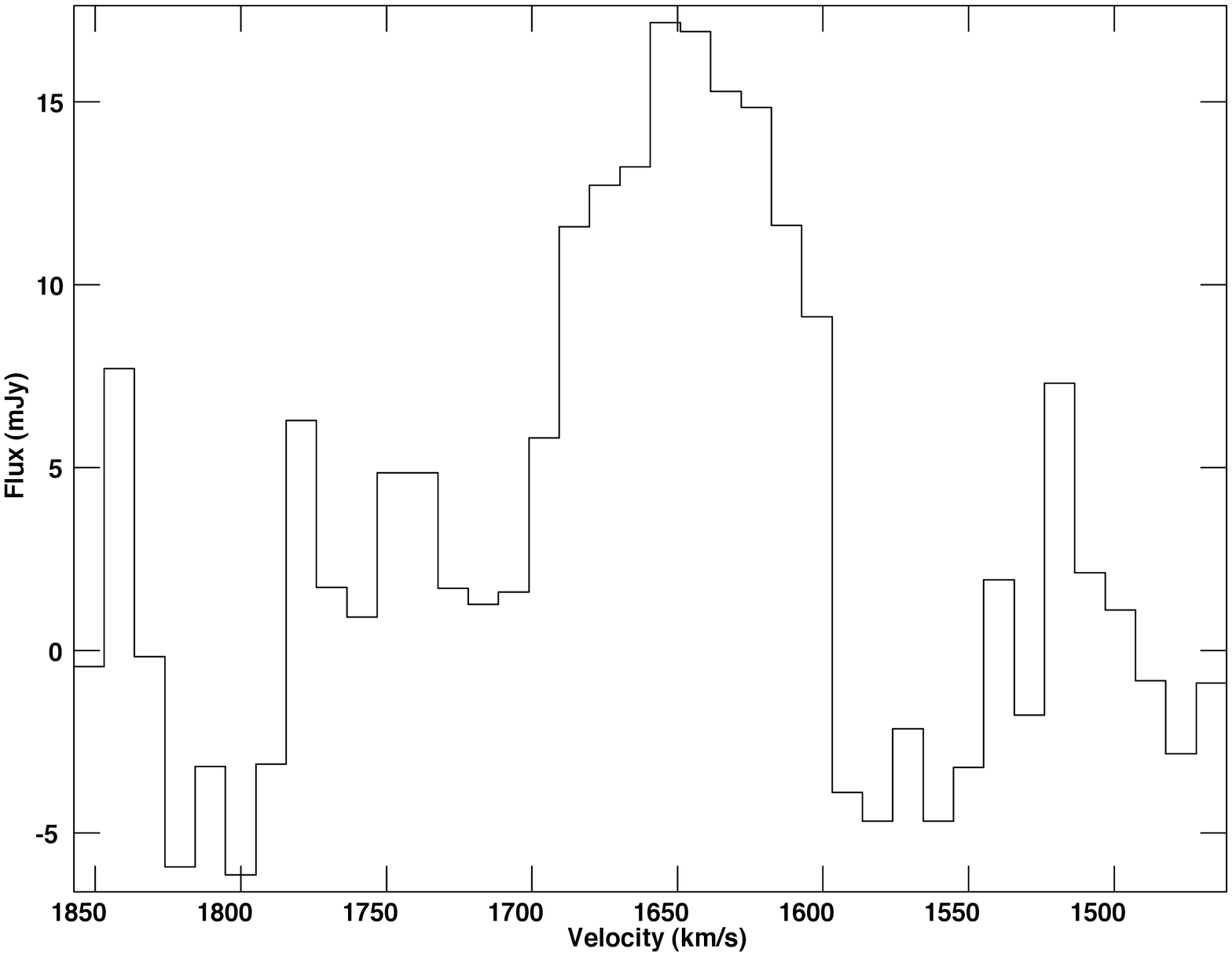}{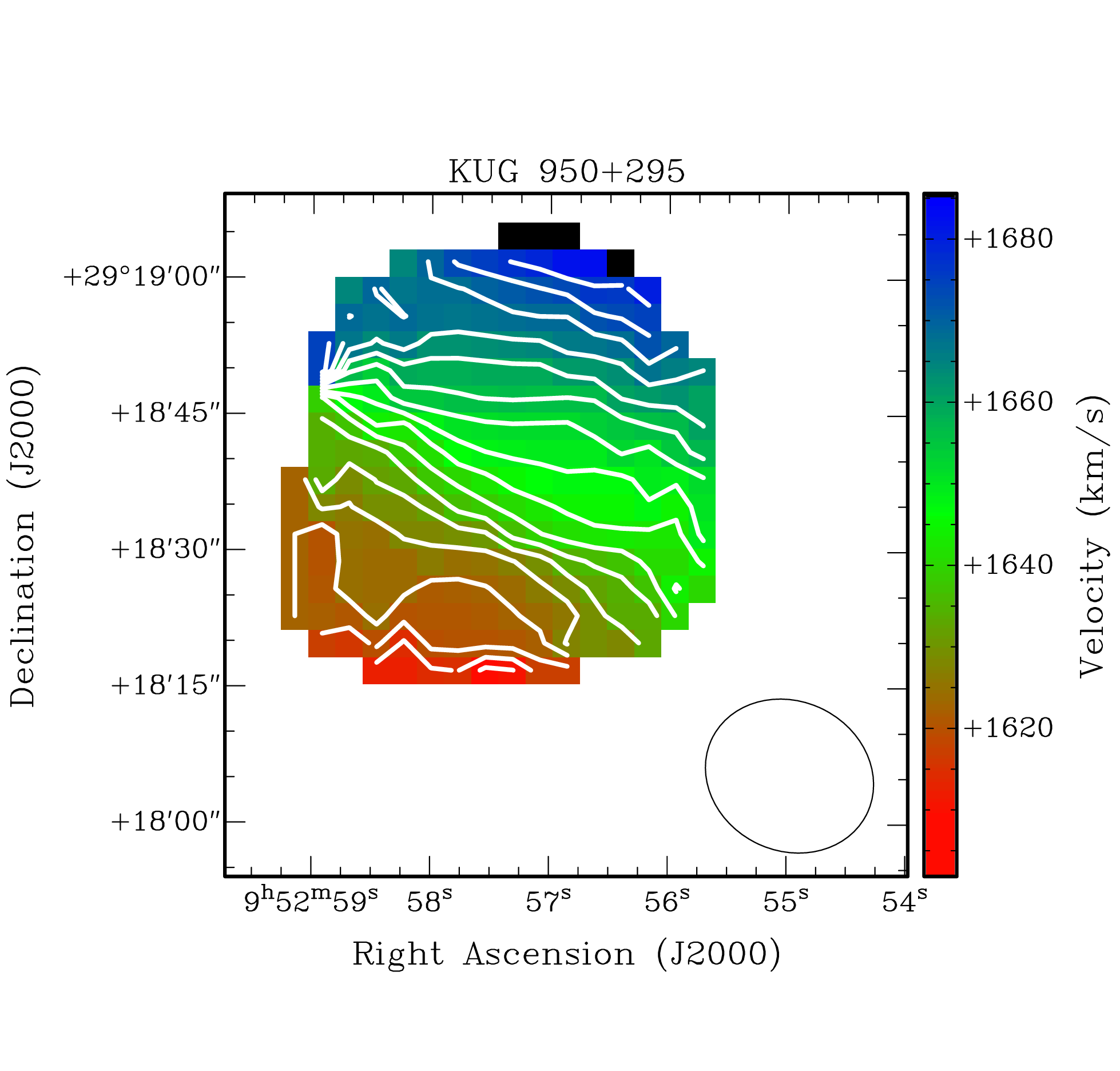}
\caption{\scriptsize  {\bf Top}: Solid black (positive) and solid grey (negative) contours show the HI integrated intensity in units of $-20\%$, $-10\%$, $-$1$\%$, 1$\%$, 10$\%$, 20$\%$, 30$\%$, 40$\%$, 50$\%$, 60$\%$, 70$\%$, 80$\%$ and 90$\%$ of the peak (0.33 Jy beam$^{-1}$ km s$^{-1}=~$1.1$\times$10$^{21}$ cm$^{-2}$).  The grey scale image is an SDSS2 red band image. {\bf Left bottom}: HI spectrum. Constructed in a similar manner to UGC 1503.{\bf Right bottom}: Velocity Field.  The HI intensity weighted mean velocity (moment 1) is shown in RGB color scale and in white contours from 6150 to 6450 km s$^{-1}$ in steps of 21 km s$^{-1}$.  The black ellipse shows the beam size. \label{stuff7}}
\end{figure*}
\begin{figure*}
\epsscale{0.90}
\plotone{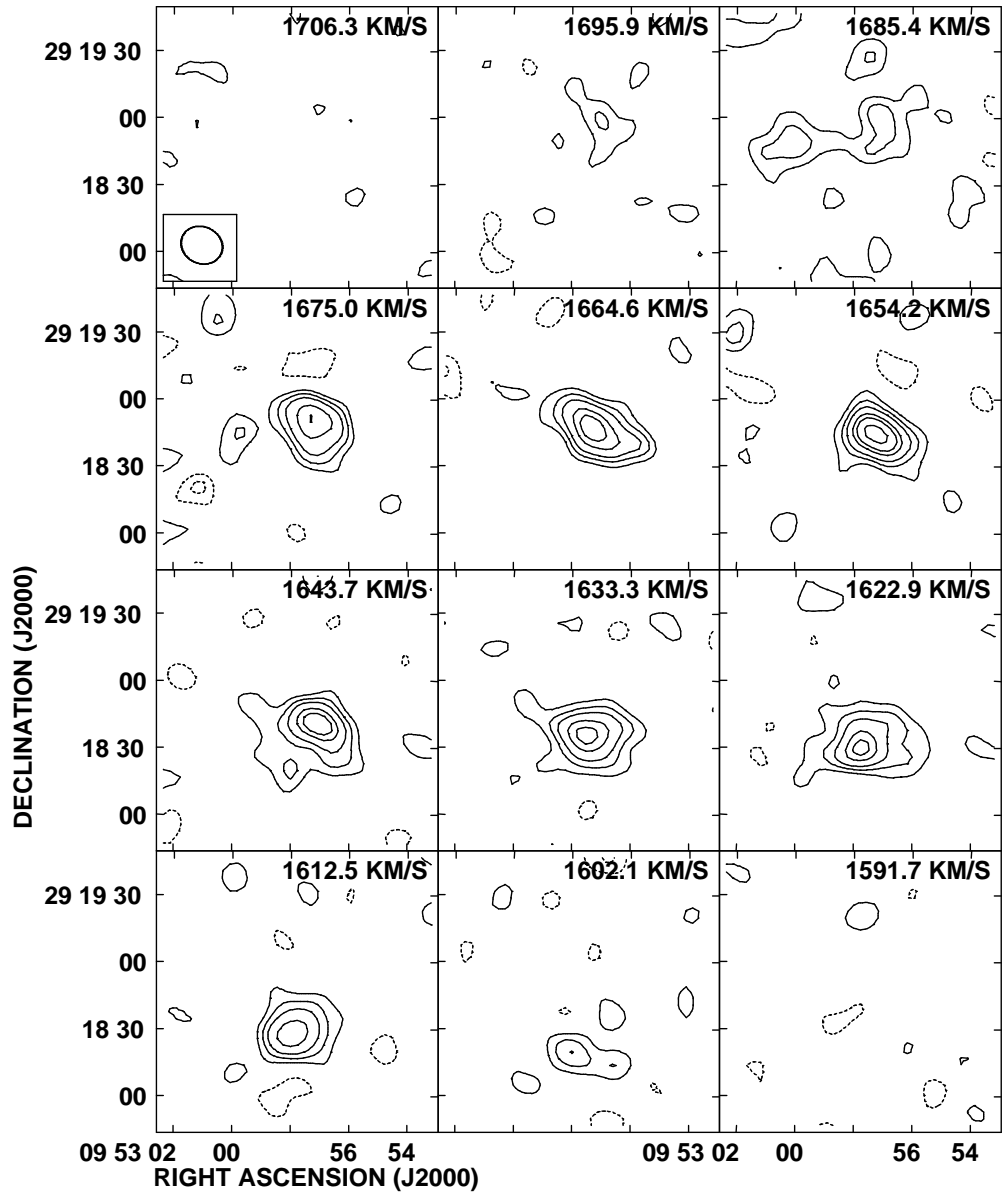}
\caption{\scriptsize   [KUG] 950$+$295: Individual channel maps showing HI emission.  Contour levels are $-$5, $-$3, 3, 5, 12, and 15 times 0.6 mJy beam$^{-1}$ $\sim$1$\sigma$.  Negative contours are grey.  The velocity of each channel (in km s$^{-1}$) is indicated at the top of each panel and the beam size in the first panel in the bottom left corner.\label{stuff8}}
\end{figure*}
\begin{figure*}
\epsscale{0.50}
\plotone{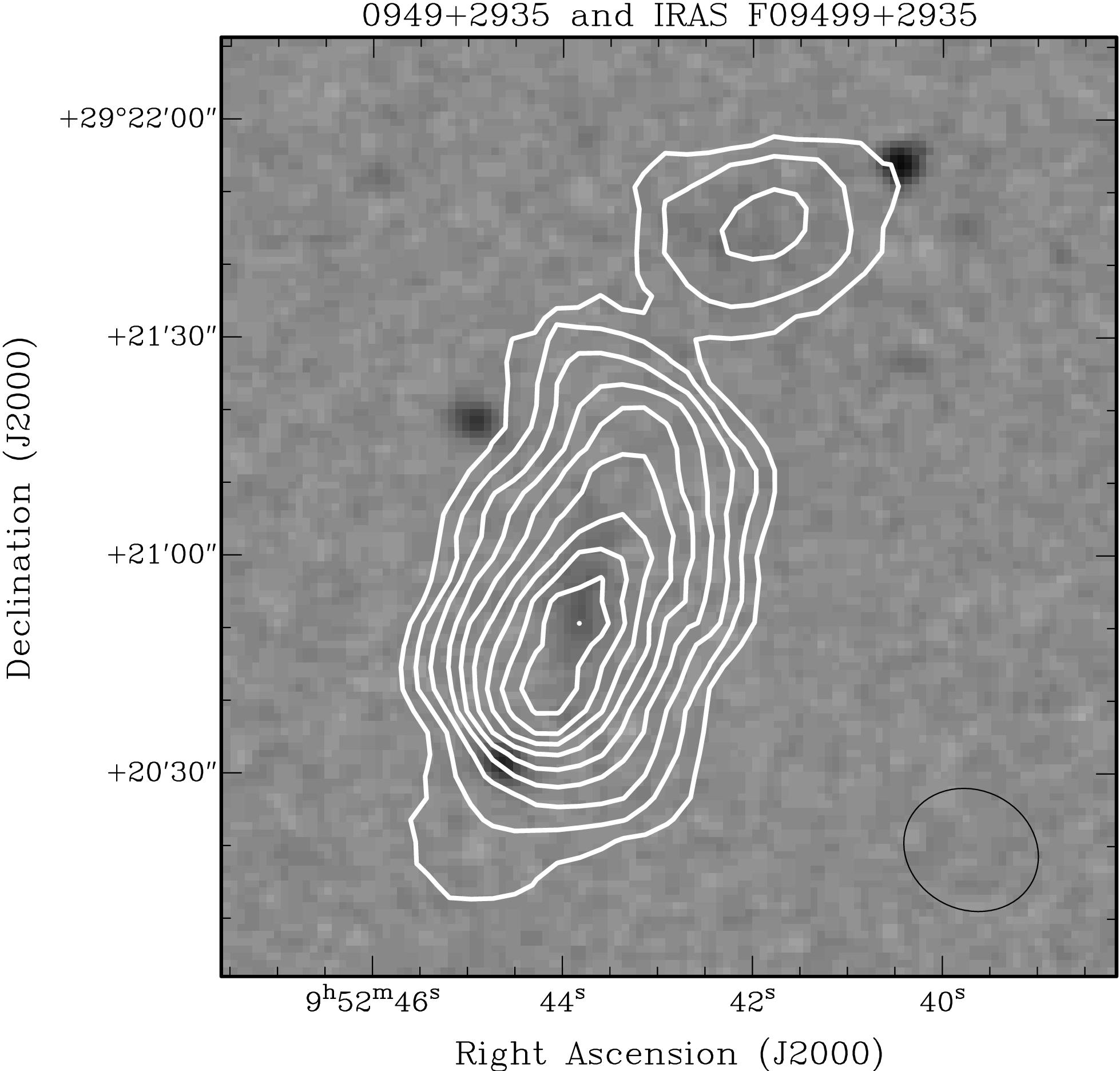}\\
\epsscale{0.8}
\plottwo{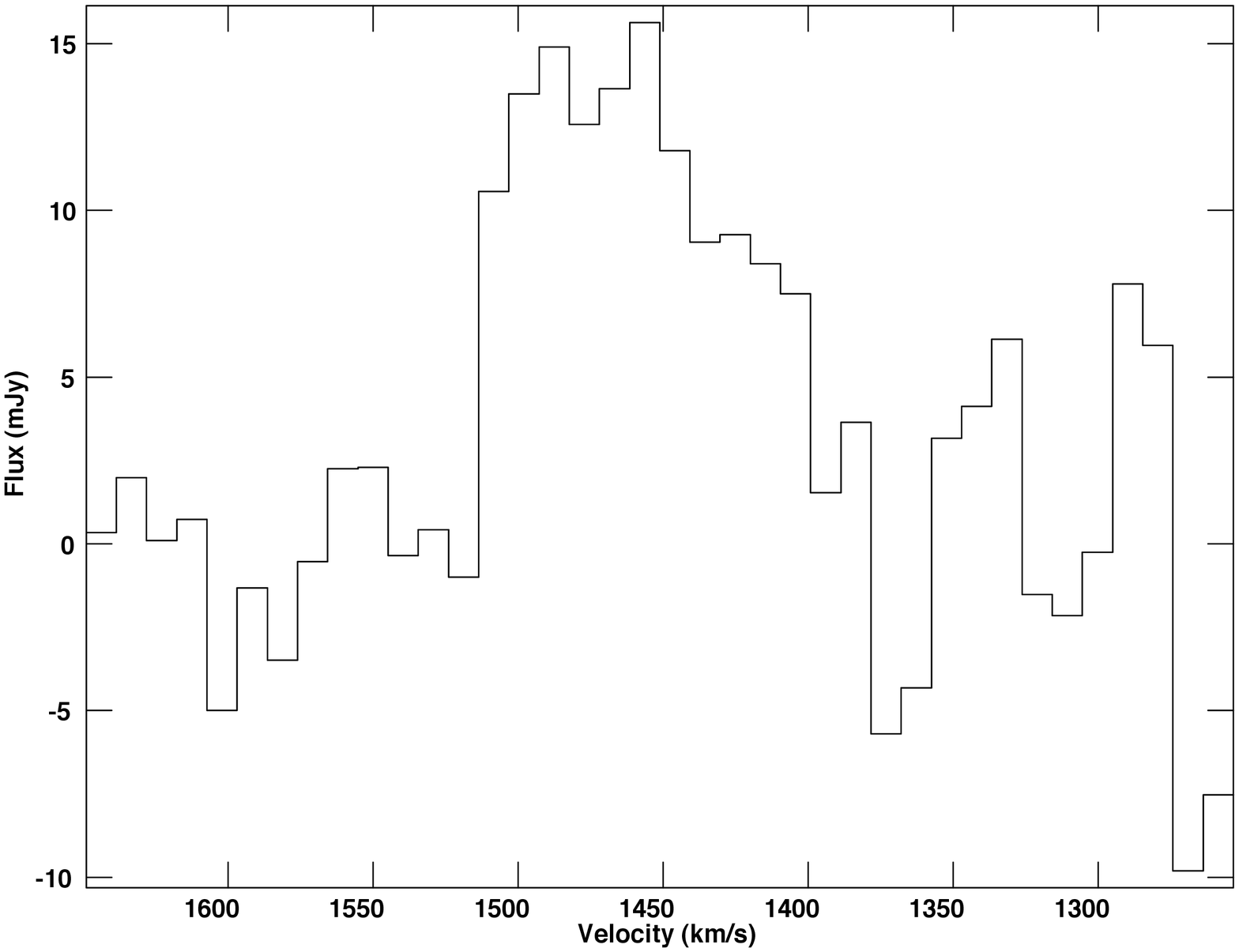}{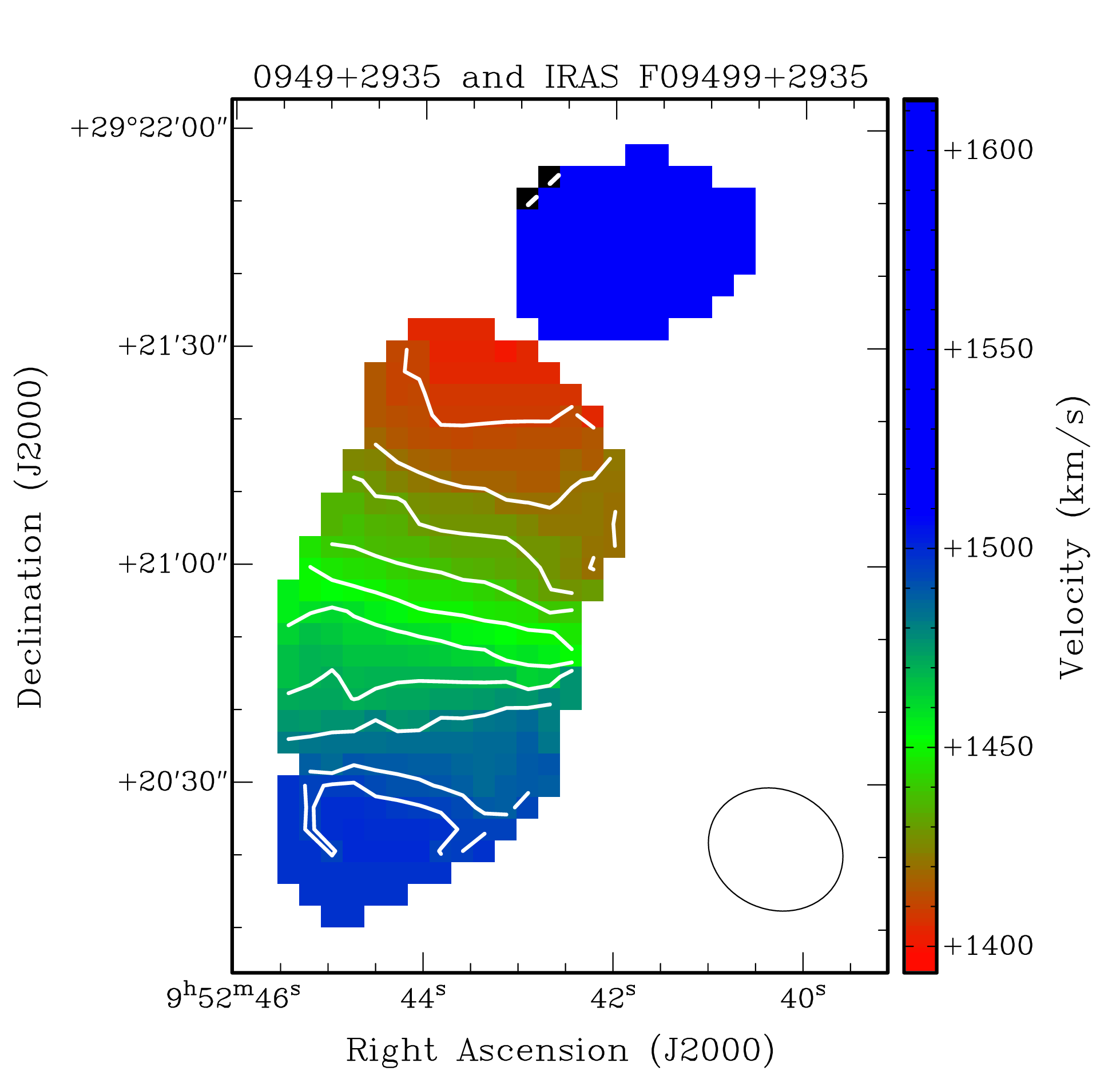}
\caption{\scriptsize  DS96 0949$+$2935 and an unknown source.  {\bf Top}: Solid black (positive) and solid grey (negative) contours show the HI integrated intensity in units of $-20\%$, $-10\%$, $-$1$\%$, 1$\%$, 10$\%$, 20$\%$, 30$\%$, 40$\%$, 50$\%$, 60$\%$, 70$\%$, 80$\%$ and 90$\%$ of the peak (0.29 Jy beam$^{-1}$ km s$^{-1}=~$1.0$\times$10$^{21}$ cm$^{-2}$).  The grey scale image is an SDSS2 red band image. {\bf Left bottom}: HI spectrum. Constructed in a similar manner to UGC 1503. {\bf Right bottom}: Velocity Field.  The HI intensity weighted mean velocity (moment 1) is shown in RGB color scale and in white contours from 6150 to 6450 km s$^{-1}$ in steps of 21 km s$^{-1}$.  The black ellipse shows the beam size. \label{stuff9}}
\end{figure*}
\begin{figure*}
\epsscale{0.8}
\plotone{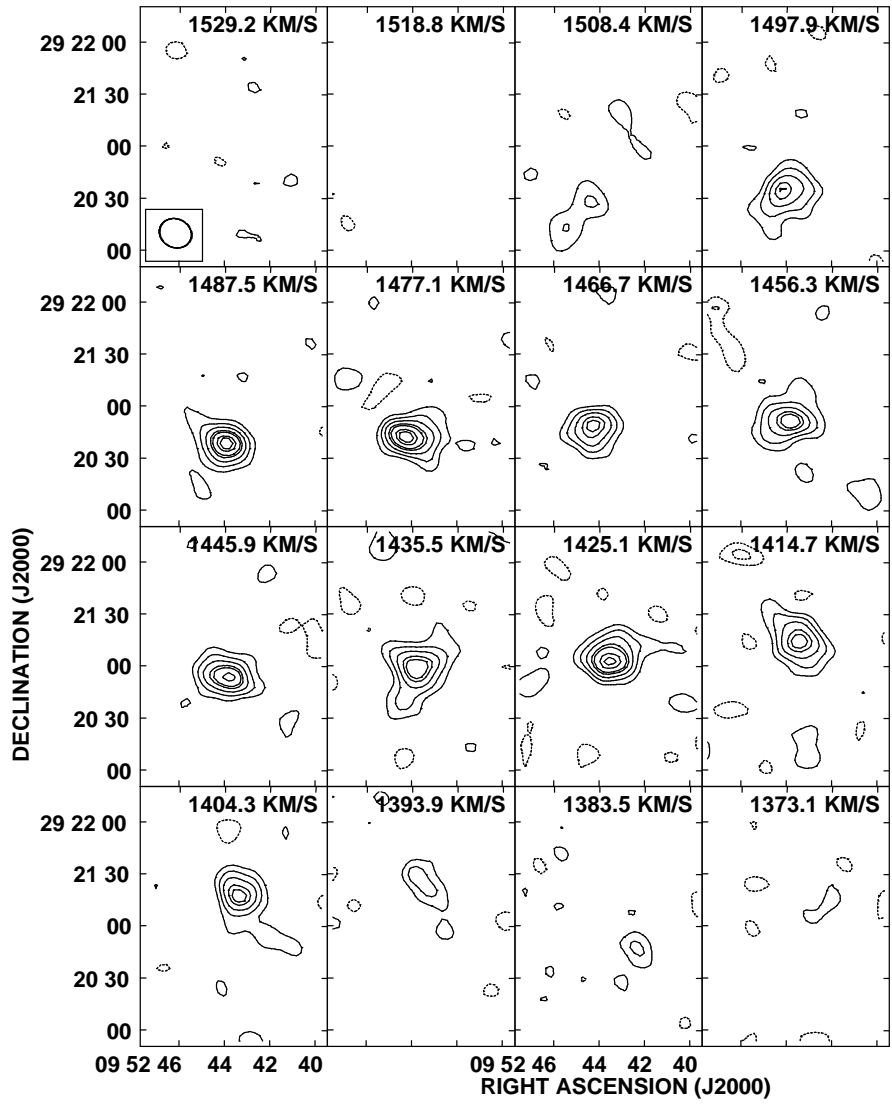}
\caption{\scriptsize   0949+2935: Individual channel maps showing HI emission.  Contour levels are -5, -3, 3, 5, 6, 6.5, 6.8, 7.2, and 7.4 times 0.6 mJy beam$^{-1}$ $\sim$1$\sigma$.  Negative contours are grey.  The velocity of each channel (in km s$^{-1}$) is indicated at the top of each panel and the beam size in the first panel in the bottom left corner.\label{stuff10}}
\end{figure*}
\begin{figure*}
\epsscale{0.5}
\plotone{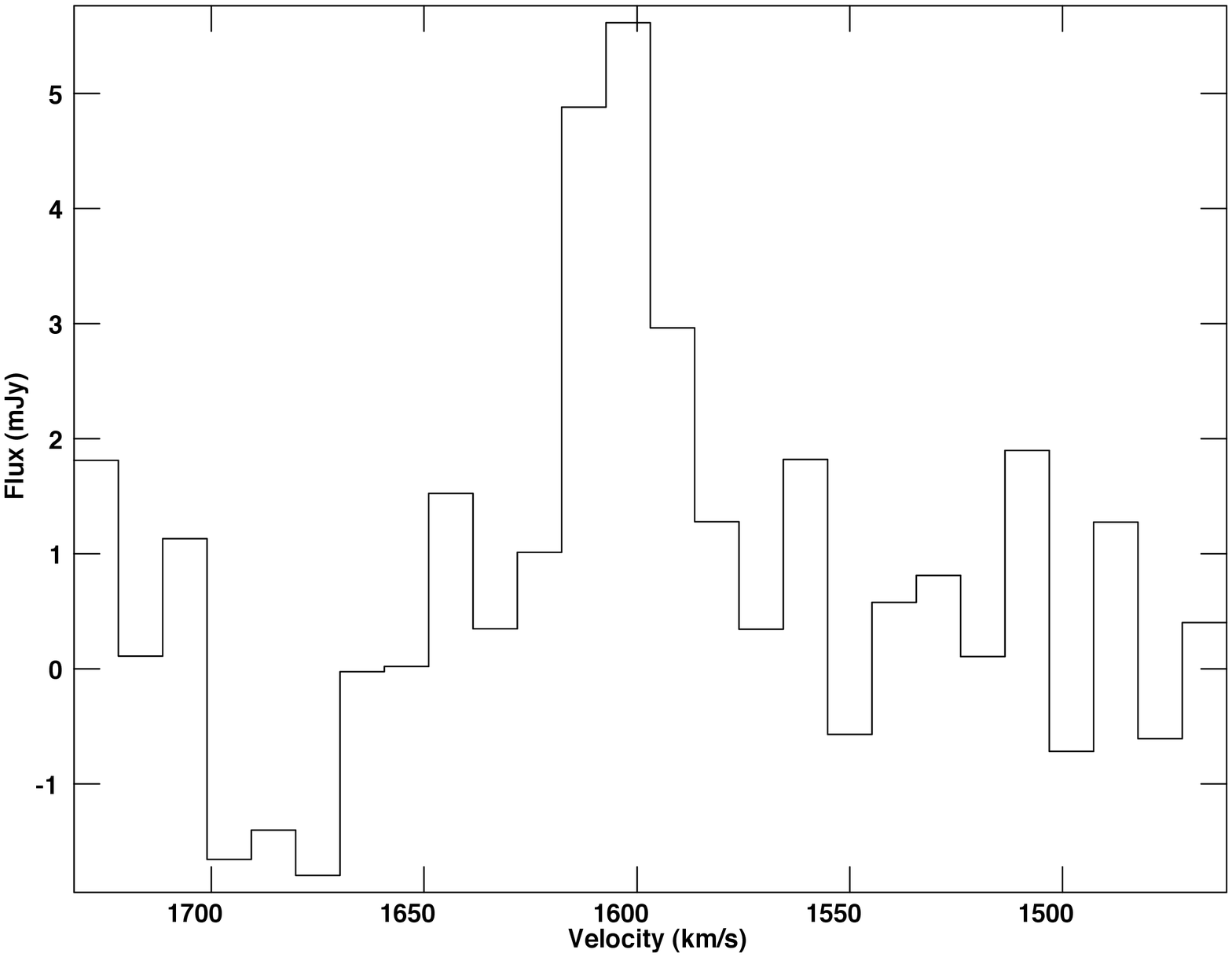}\\
\epsscale{0.5}
\plotone{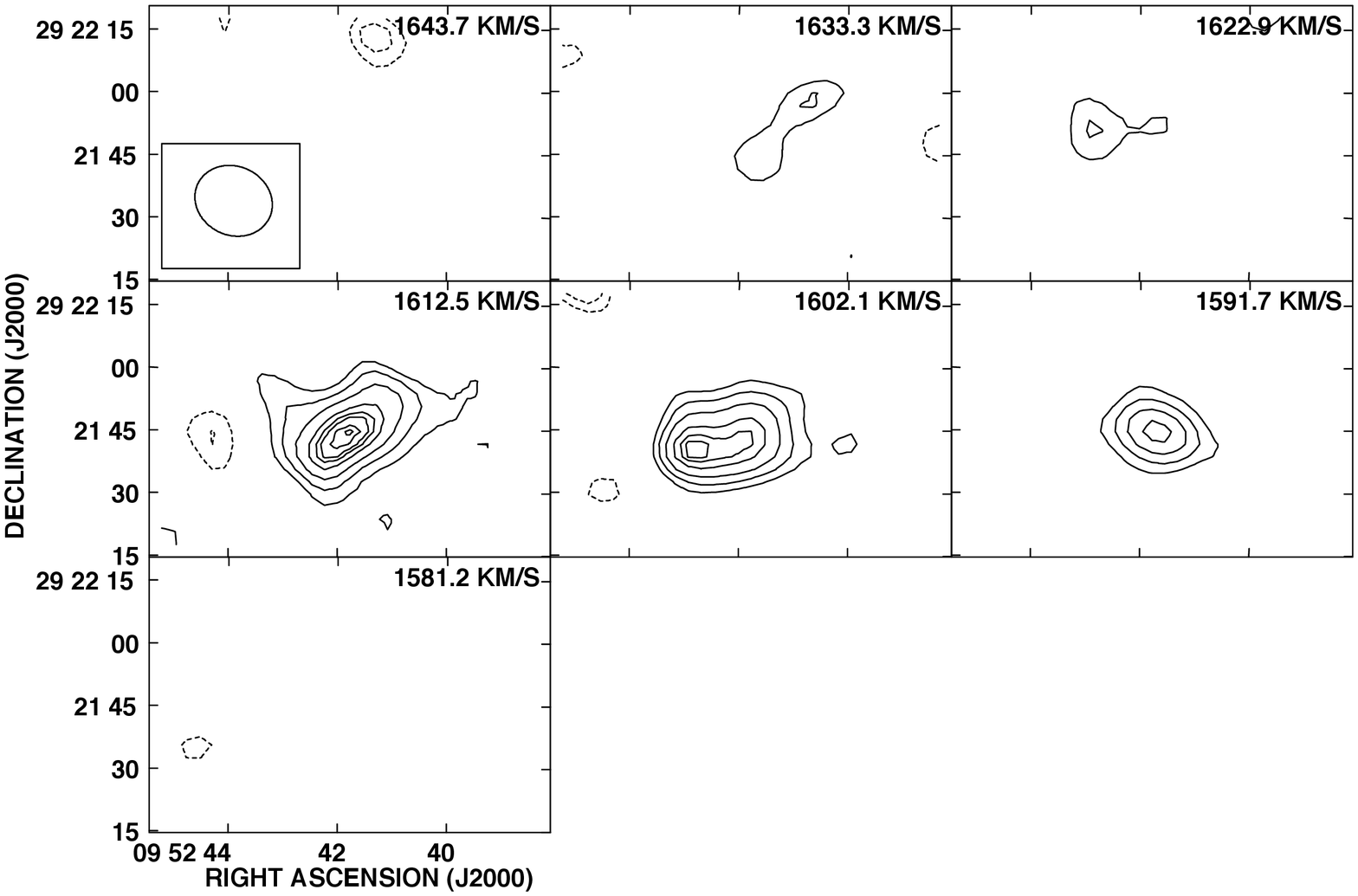}
\caption{\tiny   Unknown source.  {\bf Top}:  HI spectrum.  Constructed in a similar manner to UGC 1503.  {\bf Bottom}: Individual channel maps showing HI emission.  Contour levels are -4, -3, 3, 4, 5, 7, 9, 10, 12, and 13 times 0.6 mJy beam$^{-1}$ $\sim$1$\sigma$.  Negative contours are grey.  The velocity of each channel (in km s$^{-1}$) is indicated at the top of each panel and the beam size in the first panel in the bottom left corner.\label{stuff11}}
\end{figure*}
We detect three distinct sources of HI emission about 12\arcmin\ to the Northeast of NGC 3032 in projection.  These include:

{\bf KUG 0950+295}- Classified as an irregular galaxy in the RC3 catalogue, it is located 13\arcmin\ to the northeast of NGC 3032 in projection.  The total HI flux is 0.94$\pm$0.10 Jy km s$^{-1}$, and it is detected over 100$\pm$11 km s$^{-1}$ centered on a systemic velocity of 1650$\pm$11 km s$^{-1}$ which gives a circular velocity of V$_{circ}$$\sin{i}=$50 km s$^{-1}$.  This flux is consistent with single dish HI observations made with the 100-m radio telescope at Effelsberg which quote an HI flux of 1.2 Jy km s$^{-1}$ detected over 90 km s$^{-1}$ centered on a systemic velocity of 1638$\pm$8 km s$^{-1}$ (Huchtmeier \etal\ 2000).  We derive a distance of 22 Mpc using the systemic velocity derived from the 100-m observations and H$_0=$73 km s$^{-1}$ Mpc$^{-1}$.  Using this distance and our flux we calculate an HI mass of $1.1\times10^8$ M$_\odot$.  The HI in this galaxy is centrally peaked.  As in NGC 3032 the HI distribution is only a few beams across and so is not well resolved.  An inspection of the HI velocity field shows that the emission exhibits solid body rotation with only a slight suggestion of becoming flat near the outer parts of the HI distribution.  

{\bf DS96 0949$+$2935}- The HI emission associated with DS96 0949+2935 has a double horned HI profile indicative of a rotating disk galaxy, but only a few faint features are visible in the POSS2 images which make a clear classification difficult (Duprie \& Schneider 1996).  The HI distribution is fairly well resolved and the velocity field clearly turns over at the edge of the disk suggesting the HI is is a rotating disk.  The total HI flux of DS96 0949+2935 is 1.2$\pm$0.1 Jy km s$^{-1}$, and is detected over 100$\pm$11 km s$^{-1}$ centered on a systemic velocity of 1450$\pm$11 km s$^{-1}$ which gives a circular velocity of V$_{circ}$$\sin{i}=$50 km s$^{-1}$.  This is consistent with a Arecibo 300 meter telescope flux of 1.3 Jy km s$^{-1}$ detected over 106 km s$^{-1}$ centered on a systemic velocity of 1456 km s$^{-1}$ (Duprie \& Schneider 1996). 

{\bf Unknown source}- A faint source of HI emission located  $\sim$1\arcmin\ away from DS96 0949$+$295 in projection at RA 09h 52m 41.78s and Dec 29d 21m 44.75s.  This faint source is probably associated with a low surface brightness dwarf galaxy that appears as a faint smudge in the POSS2 image.  The HI flux of the fainter source has a flux of 0.2$\pm$0.02 Jy km s$^{-1}$, and is detected over 33$\pm$11 km s$^{-1}$ centered on a systemic velocity of 1602 km s$^{-1}$.  The systemic velocity of this faint object puts it 2 Mpc behind DS96 0949+2935.  There is no clear indication from the VLA data or in the optical images that might suggest that this unidentified source is interacting with DS96 0949$+$2935.  There are no other HI detections of this faint source mentioned in the literature.
\nocite{*}
\clearpage
\bibliographystyle{/Users/daniellelucero/Desktop/diss/apj3}
{
 \bibliography{/Users/daniellelucero/Desktop/diss/FINAL_PUB/dlucero.ms.2}
}
\end{document}